\documentclass[12pt,preprint]{article}
\usepackage{epsfig}
\usepackage{psfig}
\normalsize
\newcommand{\eqb}{\begin{equation}}
\newcommand{\eqe}{\end{equation}}
\newcommand{\dmb}{\begin{displaymath}}
\newcommand{\dme}{\end{displaymath}}
\newcommand{\pd}{\partial}

\newcommand{\eab}{\begin{eqnarray}}
\newcommand{\eae}{\end{eqnarray}}
\newcommand{\ra}{\right\rangle}
\newcommand{\la}{\left\langle}

\newcommand{\be}{\begin{equation}}
\newcommand{\ee}{\end{equation}}

\newcommand{\La}{\Lambda}

\begin{document}
\begin{titlepage}
\begin{flushright}
2004-09 \\
HD-THEP-04-19\\ 
\end{flushright}
\vspace{0.6cm}

\begin{center}
\Large{{\bf Nonperturbative approach to 
Yang-Mills thermodynamics}}

\vspace{1cm}

Ralf Hofmann

\end{center}
\vspace{0.3cm}

\begin{center}
{\em 
Institut f\"ur Theoretische Physik\\ 
Universit\"at Heidelberg\\ 
Philosophenweg 16, 69120 Heidelberg, Germany}\vspace{0.5cm}

\end{center}
\vspace{0.5cm}

\begin{abstract}

An analytical, macroscopic approach to SU(N) Yang-Mills thermodynamics is 
developed. This approach self-consistently assumes that at a 
temperature much larger than the Yang-Mills scale $\Lambda_{\tiny\mbox{YM,N}}$ (embedded and noninteracting) SU(2) 
calorons of trivial holonomy form an adjoint 
Higgs field (electric phase). Macroscopically, this field turns out to be thermodynamically and 
quantum mechanically stabilized. As a consequence, 
the problem of the infrared instability in the perturbative loop expansion of thermodynamical potentials, 
generated by the soft magnetic modes, is resolved. 
An evolution equation with two fixed points follows for the effective gauge coupling $e(T)$ 
from self-consistent thermodynamics involving the 
ground-state and its quasiparticle excitations. A plateau value of $e(T)$, which is an 
attractor of the evolution, is consistent 
with the existence of isolated magnetic monopoles of conserved charge 
being generated by dissociating calorons of nontrivial holonomy. The 
(up to negligible corrections exact) one-loop and downward evolution of 
$e(T)$ predicts the condensation of magnetic monopoles in a $2^{\tiny\mbox{nd}}$-order phase transition 
at a critical temperature $T_{E,c}$. At $T_{E,c}$ tree-level 
massive gauge modes decouple thermodynamically. This is the confinement phase transition identified in lattice simulations. 
For N=2 we compute 
the critical exponent taking the mass of 
the dual photon as an order parameter. For arbitrary N we show the 
restoration of the global electric $Z_{\tiny\mbox{N}}$ 
symmetry in the monopole condensed (magnetic) phase by investigating 
the Polyakov loop in the effective theory. The magnetic gauge coupling $g(T)$ starts its downward 
evolution from zero at 
$T_{E,c}$ and runs into a logarithmic pole at $T_{M,c}<T_{E,c}$. 
At $T_{M,c}$ center-vortex loops condense, the abelian gauge modes decouple 
thermodynamically, and 
the equation of state is $\rho=-P$ (zero entropy density). The Hagedorn transition 
to the vortex condensing phase (center phase) goes with a complete breakdown of the local 
magnetic $Z_{\tiny\mbox{N}}$ symmetry. After a rapid 
reheating in terms of (intersecting) center-vortex loops has taken place 
the ground-state pressure vanishes {\sl identically} on tree level. 
This result is protected against radiative corrections. Throughout the electric 
and the magnetic phase and for N=2,3 we compute the 
temperature evolution of the (infrared sensitive) pressure and 
energy density and for the (infrared insensitive) entropy density and 
compare our results with lattice data. We show that the 
disagreement for the two former quantities at low temperature 
(negative pressure) originates from severe finite-size artefacts in lattice 
simulations. For the entropy density we 
obtain excellent agreement with lattice results. The implications of our results for 
particle physics and cosmology are discussed.

\end{abstract} 

\end{titlepage} 

\section{Introduction}

The beauty and usefulness of the gauge principle for local field theories is 
generally appreciated. Yet, in a perturbative approach to gauge theories 
like the SM and its (non)supersymmetric extensions it is hard if not impossible 
to convincingly address a number of recent experimental and 
observational results in particle physics and 
cosmology: nondetection of the Higgs particle at LEP \cite{Higgs2000}, 
indications for a rapid thermalization and strong collective behavior 
in the early stage of an ultra-relativistic heavy-ion collision 
\cite{RHIC2003,ShuryakTeaney2001}, dark energy and dark matter at present, a 
strongly favored epoch of cosmological inflation in the early Universe 
\cite{Linde1982,Guth1982,Starobinsky1982,WMAP2003}, 
and the likely existence of intergalactic magnetic 
fields \cite{Dai2002,IMF}. An analytical and 
nonperturbative approach to strongly interacting gauge 
theories may further our understanding of these phenomena. 

It is difficult to gain insights in the dynamics of a strongly interacting 
field theory by analytical means. 
We conjecture with Ref.\,\cite{Hagedorn1965} that a thermodynamical approach 
is an appropriate starting point for such an endeavor. 
On the one hand, this conjecture is reasonable since a strongly interacting 
system being in equilibrium communicates perturbations almost instantaneously due to 
rigid correlations. Thus equilibrium is restored very rapidly. 
On the other hand, the requirement of 
thermalization poses strong constraints on the {\sl construction} 
of a macroscopic, effective theory for the ground state and 
its (quasiparticle) excitations. 
The objective of the present paper is the thermodynamics of 
SU(N) Yang-Mills theories in four dimensions. 

Let us very briefly recall some aspects of the analytical 
approaches to thermal SU(N) Yang-Mills theory as they are 
discussed in the literature. Because of asymptotic 
freedom \cite{GrossWilczek1973,Politzer1973} one would naively 
expect thermal perturbation theory to work well for temperatures 
$T$ much larger than the Yang-Mills scale $\La_{\tiny\mbox{YM,N}}$ since the gauge coupling 
constant $\bar{g}(T)$ logarithmically approaches zero 
for $\frac{T}{\La_{\tiny\mbox{YM,N}}}\to\infty$. 
It is known for a long time that this expectation is too optimistic since at 
any temperature perturbation theory 
is plagued by instabilities arising from the infrared sector (weakly screened, 
soft magnetic modes \cite{Linde1980}). As a consequence, the pressure $P$ can be computed 
perturbatively only up 
to (and including) order $\bar{g}^5$. The effects of resummations of 
one-loop diagrams (hard thermal loops), which rely on a 
scale separation in terms of the small value of 
the coupling constant $\bar{g}$, are summarized in terms of a 
nonlocal effective theory for soft and 
semi-hard modes \cite{BraatenPisarski1990}. In the computation of radiative 
effects based on this effective theory 
infrared effects due to soft modes still appear in an 
uncontrolled manner. This has 
lead to the construction of an effective theory where soft modes are collectively described in 
terms of classical fields whose dynamics is influenced by integrated 
semi-hard and hard modes \cite{Bodeker1998,Bodapps1998}. 
In Quantum Chromodynamics (QCD) a perturbative calculation of $P$ was pushed up to order 
$\bar{g}^6\log\,\bar{g}$, and an additive `nonperturbative' 
term at this order was fitted to lattice results \cite{Kajantie2002}. Within the 
perturbative orders a poor convergence of the expansion is observed for 
temperatures not much larger than the $\overline{\mbox{MS}}$ scale. While 
the work in \cite{Kajantie2002} is a computational masterpiece it could, by definition, 
not shed light on the nonperturbative physics of 
the infrared sector. Screened perturbation theory, which relies 
on a separation of the tree-level 
Yang-Mills action using variational parameters, 
is a very interesting idea. Unfortunately, this approach generates 
temperature dependent ultraviolet divergences in its presently used 
form, see \cite{Blaizot2003} for a recent review. 

The purpose of this paper is to report in a detailed way\footnote{Some 
aspects of the low-temperature physics 
are revised in the present paper as compared to \cite{Hofmann2003F}.} on a nonperturbative and 
analytical approach to the thermodynamics of SU(N) Yang-Mills theory 
(see \cite{Hofmann2000t2003} for intermediate stages). 
Conceptually, this approach is similar to the 
macroscopic Landau-Ginzburg-Abrikosov (LGA) theory for superconductivity in metals 
\cite{GinzburgLandau1950,Abrikosov1957}. Recall, that this theory does not derive 
the condensation of Cooper pairs from first principles but rather describes 
the condensate by a nonvanishing amplitude of a complex scalar field (local order parameter) 
which is charged under the electromagnetic gauge group U(1). This 
nonvanishing amplitude is driven by a phenomenologically introduced potential $V$. 
As a consequence, a (macroscopic) U(1) gauge field $a_\rho$, which is deprived of the 
microscopic gauge-field fluctuations associated with 
the formation of Cooper pairs and their subsequent condensation, acquires mass, the U(1) symmetry is 
spontaneously broken, and physical phenomena originating from 
this breakdown can be explored in dependence of the parameters appearing in the effective action, 
and in dependence of an external magnetic field and/or temperature. 

When applying this idea to the construction of a macroscopic 
theory for SU(N) Yang-Mills thermodynamics (YMTD) one is in a much better 
position as far as the uniqueness of the stabilizing potentials 
in each phase of the theory is concerned. These potentials are 
determined by thermodynamics and the requirement 
that, in a first step of the construction, they admit 
energy- and pressure-free macroscopic configurations describing the collective effects in 
an ensemble of energy- and pressure-free, noninteracting, and self-dual topological field 
configurations in the underlying theory. 
If a particular phase supports propagating gauge modes then, in a second step, 
the interactions between these topological defects 
are treated by solving the macroscopic gauge-field equations in terms of a pure 
gauge configuration in the background of the (inert) energy- and pressure-free 
scalar field.  

More specifically, we assume that at a large temperature a macroscopic 
adjoint scalar field $\phi$ 
is generated by a dilute gas of trivial-holonomy calorons 
\footnote{We discuss in Sec.\,\ref{MPlanck} why the critical 
temperature $T_P$ for the onset of the formation of $\phi$ should be comparable to 
the cutoff-scale for the local field theory in four dimensions.} 
\cite{HarrigtonShepard1977}. Calorons are Bogomolnyi-Prasad-Sommerfield (BPS) 
saturated (or self-dual) solutions \cite{PrasadSommerfield1975}
to the classical Yang-Mills equations of motion in four-dimensional Euclidean spacetime\footnote{Whenever we speak of a topological soliton 
this automatically includes the antisoliton.} (time coordinate $\tau$ is 
compactified on a circle, $0\le\tau\le\frac{1}{T}$) with varying 
topological charge and embedding in SU(N). Calorons are topologically 
nontrivial, saturate the lowest possible value of the Euclidean action 
in a given topological sector, and thus are energy- and pressure-free configurations. 
Calorons with nontrivial holonomy have BPS magnetic monopole constituents 
\cite{Nahm1984,KraanVanBaalNPB1998,vanBaalKraalPLB1998}. Their one-loop effective 
action scales with the three volume of the system \cite{GrossPisarskiYaffe1981}, and 
thus they should play no role in the thermodynamic limit. 
This conclusion, however, is no longer valid if the 
system generates domains of large but finite volume whose boundaries 
are generated by discontinuous changes of the color orientation of 
the field $\phi$. Microscopically, nontrivial-holonomy 
calorons can be dynamically generated out 
of trivial-holonomy calorons by macroscopic domain collisions. These calorons dissociate into 
their magnetic monopole constituents subsequently, see \cite{Diakonov} for a 
discussion of the destabilizing effects of quantum fluctuations in the case of 
nontrivial holonomy. We thus anticipate the occurence of 
isolated magnetic charge whose 
abundance is governed by the ($T$ dependent) 
typical volume of a domain. 

The property of vanishing energy and pressure of a caloron derives from its self-duality, 
that is, the kinetic and the interaction part in the Euclidean energy-momentum tensor 
precisely cancel when evaluated on a caloron. A potential $V_E$ 
(the subscript $E$ stands for electric phase) is constructed which stabilizes 
the modulus $|\phi|$ for given $T$ quantum mechanically and thermodynamically 
and which reflects the assumption that $\phi$ is composed of noninteracting, trivial-holonomy 
calorons. We would like to stress at this point that the effects of calorons are reflected  
as a $\frac{1}{\sqrt{T}}$ dependence of the modulus of $\phi$. 
As a consequence, the nontrivial-topology sector of the 
theory, indeed, is irrelevant at 
asymptotically large temperatures.  

A unique decomposition of each 
gauge-field configuration $A_\rho$ contributing to the partition function of the 
fundamental Yang-Mills theory is 
\eqb
\label{decfl}
A_\rho=A_\rho^{\tiny\mbox{top}}+a_\rho\,.
\eqe
In Eq.\,(\ref{decfl}) $A_\rho^{\tiny\mbox{top}}$ is a minimally (that is, BPS saturated) 
topological part, represented by calorons, 
and $a_\rho$ denotes a remainder which has trivial topology. The 
configurations in $A_\rho^{\tiny\mbox{top}}$ having trivial holonomy would build up 
the ground state described by $\phi$ if no holonomy-changing interactions 
between them were allowed for. A change in holonomy by 
interactions, mediated by the topologically trivial sector, will macroscopically manifest itself in terms of a 
{\sl finite}, pure-gauge background $a_\rho^{bg}$. A fluctuation 
$\delta a_\rho$ about this background 
acquires mass by the adjoint Higgs mechanism if 
$[\phi,\delta a_\rho]\not=0$ and thus the underlying gauge symmetry SU(N) 
is spontaneously broken to U(1)$^{\tiny\mbox{N-1}}$ at most. The degree of gauge 
symmetry breaking by calorons is a boundary 
condition set at an asymptotically high temperature $T_P$ 
where the effect of $\phi\propto T^{-1/2}$ on the Yang-Mills spectrum 
and its pressure $\sim T^4$ is very small since the ground state pressure scales as 
$\propto T$. On the one hand, Higgs-mechanism 
induced masses provide infrared cutoffs in the 
loop expansions of thermodynamical quantities 
which resolves the problem of the infrared 
instability entcountered in perturbation theory. 
On the other hand, the compositeness scale $|\phi|$ constrains the 
hardness of quantum fluctuations, and so the usual 
renormalization program needed to address ultra-violet divergences in perturbation theory is superfluous 
in the effective theory. Notice that this way of introducing 
a composite field $\phi$ in an effective description differs from the usual implementation of a 
Wilsonian flow, where {\sl high-momentum} modes are successively 
integrated out \cite{BraatenPisarski1990,Wetterich1996}, 
since $\phi$ is built of calorons with an `instanton' radius $\rho$ not being smaller than $|\phi|^{-1}$. 
At the present stage the description of the ground state of an SU(N) Yang-Mills 
theory at high temperatures in terms of 
the field $\phi$ is self-consistent. The phase and the modulus of the field $\phi$ are derived 
from a microscopic definition in \cite{HerbstHofmann2004}.

The nonperturbative approach to SU(N) YMTD proposed here 
implies the existence of three rather 
than two phases: an electric phase at high temperatures, a magnetic 
phase for a small range of temperatures comparable to the scale 
$\La_{\tiny\mbox{YM,N}}$, and a center phase for 
low temperatures. The ground state in the magnetic phase 
confines fundamental, static test charges 
but allows for the propagation of 
massive, dual gauge bosons. 
The center phase is thermodynamically disconnected 
from the magnetic and the electric phase. 
In the electric phase an evolution equation for the 
effective gauge coupling constant $e$, which follows from the requirement of thermodynamical
self-consistency of the one-loop expression for the pressure, has two fixed points associated with 
a highest and a lowest attainable temperature $T_P$ and $T_{E,c}$. It turns out that practically all strong-interaction
effects of the theory are described by a temperature dependent ground-state pressure and 
tree-level masses for thermal quasiparticles such that higher loop corrections to 
thermodynamical quantities are tiny. 

At $T_{E,c}$ the effective coupling $e(T)$ exhibits a thin divergence of the form 
\eqb
\label{coupldiv}
e(T)\sim -\log(T-T_c)\,,
\eqe
and the theory undergoes a 2$^{\mbox{\tiny{nd}}}$ 
order like phase transition to a magnetic phase which is driven by the 
condensation of some of the magnetic monopoles 
residing inside dissociating nontrivial-holonomy calorons. In this transition a part of the 
continuous gauge symmetry, which survived the formation of the adjoint Higgs field $\phi$ 
in the electric phase, is 
broken spontaneously and the tree-level 
massive gauge modes of the electric phase decouple thermodynamically. In the case of submaximal gauge-symmetry 
breaking by $\phi$ in the electric phase condensates of magnetic and 
color-magnetic monopoles occur in the magnetic phase. The former are described by 
complex scalar and the latter by adjoint Higgs fields. In the case of maximal 
gauge symmetry breaking to U(1)$^{\tiny\mbox{N-1}}$, which we will only investigate in this paper, 
the (local) magnetic center symmetry $Z_{\tiny\mbox{N/2,mag}}$ and the continuous gauge symmetry U(1)$^{\tiny\mbox{N/2-1}}$ 
survive the transition to the magnetic phase, the (global) electric center symmetry 
$Z_{\tiny\mbox{N/2,elec}}$ is fully restored. An approach to the 
thermodynamics in the magnetic phase, which is conceptually analoguous 
to the one in the electric phase, yields an evolution equation 
for the magnetic gauge coupling $g$ which has two 
fixed points at $T_{E,c}$ and 
$T_{M,c}$ (highest and lowest attainable temperature). Approaching $T_{M,c}$ from above, 
the equation of state is increasingly dominated by the ground state contributions. 
At $T_{M,c}$ we have 
\eqb
\label{eosintro}
\rho\sim -P\,.
\eqe
The theory undergoes a phase transition to a phase whose ground-state is 
a condensate of center-vortex loops. In this phase 
$Z_{\tiny\mbox{N,mag}}$ is entirely broken, and all gauge boson 
excitations are thermodynamically decoupled. Once each of the vortex-loop condensates, 
described by nonlocally defined 
complex scalar fields, has relaxed to the one of the N degenerate minima 
of its potential, the energy density and the pressure of the 
ground state are precisely zero (no radiative corrections), and the system has created particles 
by local $Z_{\tiny\mbox{N}}$ phase shifts of each vortex-condensate field which 
are associated with localized (intersecting) center fluxes forming closed loops. 
The corresponding density of states is over-exponentially 
rising implying that the magnetic-center transiton is of the 
Hagedorn type and thus nonthermal.   

There are many claims in the scenario outlined above. We will, step by step, 
verify them as we proceed. The paper is organized as follows: 

In Sec.\,\ref{EP} we explain our approach 
to the electric phase. We start with the basic assumption that it is noninteracting 
trivial-holonomy calorons that form a macroscopic adjoint Higgs field $\phi$ at 
high temperatures (electric phase) 
and explore its consequences. A nonlocal definition for $\phi$ is given. 
We then elucidate the details of the ground-state dynamics and the properties 
of topology-free gauge modes. Subsequently, 
an evolution equation for the effective gauge coupling constant $e$ is 
derived and solved, interpretations of the solution are given, and an 
argument is provided why the temperature $T_P$ for the onset of caloron 'condensation' 
has to be comparable to the cutoff-scale for the 
local field-theory description in four dimensions. In a next step, we perform the 
counting of isolated magnetic 
monopoles species in the effective theory for the electric phase 
when assuming maximal gauge symmetry breaking 
by $\phi$. The next part of Sec.\,\ref{EP} is devoted to a discussion of two-loop corrections to 
thermodynamical quantities. For the SU(2) case we investigate 
the simplest one-loop contribution to the `photon' polarization 
and perform formal weak and strong coupling limits of this expression. We also discuss the 
implementation of thermodynamical self-consistency when higher loop corrections to the pressure 
are taken into account. 

In Sec.\,\ref{MP} we investigate the magnetic phase, again 
assuming maximal gauge symmetry by caloron 'condensation': the pattern of gauge symmetry breaking 
by monopole condensation is explored, the thermodynamics of the ground state and 
its excitations is elucidated, an evolution equation for the magnetic gauge coupling constant 
$g$ is derived. Solutions to this equations are obtained numerically and their 
implications are discussed. Finally, we discuss the Polyakov loop 
in the electric and the magnetic phase and compute the critical exponent of the phase transition for SU(2).

In Sec.\,\ref{CVCM} we investigate the center phase. A nonlocal definition 
for the local fields describing the condensed center-vortex loops is given, their transformation properties 
under magnetic center rotations are determined, and their dynamics is 
discussed. 

In Sec.\,\ref{MCC} we derive a matching condition for 
the mass scales $\La_E$ and $\La_M$ which appear in the respective potentials for the caloron and magnetic 
monopole condensates. 

In Sec.\,\ref{PEEN} we compute the temperature evolution of 
the thermodynamical potentials pressure, energy density, and entropy density throughout the electric and 
the magnetic phases at one loop for N=2,3 and compare our 
results with lattice data. 

A conclusion and a discussion of likely implications of our results for 
particle physics and cosmology are given in Sec.\,\ref{CO}.    

\section{The electric phase\label{EP}}

\subsection{Conceptual framework\label{PR}}

Our analysis is based on the following assumption about the ground-state physics 
characterizing SU(N) YMTD at high temperatures.
\vspace{0.4cm}\\ 
{\sl At a temperature 
$T_P\gg\Lambda_{\tiny\mbox{YM,N}}$ SU(N) YMTD, defined on a Euclidean, four-dimensional, and 
flat spacetime, generates an adjoint Higgs field $\phi$ out of noninteracting (dilute), trivial-holonomy 
SU(2) calorons.} \vspace{0.4cm}\\  
\noindent Calorons are BPS saturated solutions 
to the Euclidean equations of motion of SU(N) Yang-Mills theory at finite temperature 
\cite{HarrigtonShepard1977,Nahm1984,KraanVanBaalNPB1998,vanBaalKraalPLB1998}. 
One distinguishes SU(2) calorons according to their holonomy, that is, the behavior of the 
Polyakov loop 
\eqb
\label{polloop}
{\bf P}={\cal P}\exp\left[i\bar{g}\int_0^{1/T}d\tau A_4(\vec{x}, \tau)\right]\,
\eqe
at $|{\vec x}|\to \infty$ when evaluated on 
the solution. In Eq.\,(\ref{polloop}) ${\cal P}$ denotes the
path-ordering symbol and $\bar{g}$ the gauge coupling constant 
of the SU(2) Yang-Mills theory. 
Trivial (nontrivial) holonomy means that 
have ${\bf P}_{\tiny{|{\vec x}|\to \infty}}={\bf 1}\ \ \ (\not={\bf 1})$. In the former 
case the SU(2) caloron has no isolated magnetic-monopole constituents, in the 
latter case it exhibits a monopole and its 
antimonopole. The masses of these constituents are determined 
by the value of $A_4(|{\vec x}|\to \infty$. In the following 
we only consider SU(2) nontrivial-holonomy calorons with no net magnetic charge. 
Analytical expression for SU(2) caloron solutions  
of trivial (nontrivial) holonomy can be found in \cite{HarrigtonShepard1977} (\cite{Nahm1984,KraanVanBaalNPB1998,Brower1998}), 
see also Fig.\,1.. 
\begin{figure}
\begin{center}
\leavevmode
\leavevmode
\vspace{4.3cm}
\includegraphics{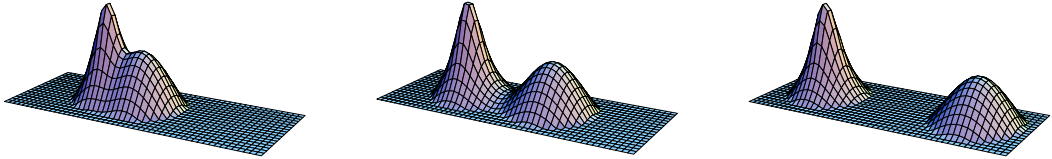}
\end{center}
\caption{Action density of SU(2) nontrivial-holonomy calorons with 
increasing 'instanton radius' (left to right) at fixed temperature. Figures are taken 
from a paper by Kraan and van Baal. The peaks of the action density 
coincide with the positions of constituent BPS monopoles.\label{instrad}}      
\end{figure}
Since calorons are BPS saturated or self-dual their 
energy-momentum tensor vanishes identically. 

An SU(2) caloron of topological charge $k$ has a classical 
Euclidean action $S_{\mbox\tiny{cal},2}=\frac{8\pi^2}{\bar{g}^2}k$. 
For trivial holonomy the one-loop effective action of a charge-one caloron 
is given as \cite{GrossPisarskiYaffe1981}
\eqb
\label{GPYSeffth}
S_{\tiny\mbox{eff}}=\frac{8\pi^2}{\bar{g}^2}+\frac{4}{3}(\pi\rho T)^2
\eqe
where $\rho$ and $T$ denote the 'instanton' radius 
and temperature, respectively. For large $\bar{g}$ and small 
enough $\rho$ the trivial-holonomy caloron 
thus sizably contributes to the partition function of the theory. 
The one-loop effective 
action of a nontrivial-holonomy caloron is
\eqb
\label{GPYSeffnth}
S_{\tiny\mbox{eff}}\propto T^3 V
\eqe
where $V$ denotes the spatial volume of the system. 
In the thermodynamic limit $V\to\infty$ nontrivial-holonomy 
calorons thus do not contribute to the partition function. 
As we will show below, however, the thermodynamic limit is 
not physical due to a domanization of the ground state 
of the theory. The suppression of nontrivial-holonomy 
calorons in the partition function is thus governed by the size 
of a typical domain. 

It was shown in \cite{Diakonov} that SU(2) calorons are 
unstable under one-loop quantum fluctuations. 
Namely, for a holonomy close to trivial there is an 
attractive potential between constituent monopole and antimonopole. In the opposite case 
the potential is repulsive implying the dissociation of the caloron 
into a monopole-antimonopole pair. In a mesoscopic level, an isolated 
monopole arises at a point in space 
where four or more Higgs-field domains meet \cite{Kibble1976}.

Since the action {\sl density} of a caloron 
is $T$-dependent the action density of the 
macroscopic, adjoint Higgs field $\phi$ should be $T$-dependent through 
the $T$-dependence of the configuration $\phi$. 

The effective 
theory describing the (electric) phase macroscopically is an adjoint Higgs model:
\eqb
\label{actE}
S_E=\int_0^{1/T}d\tau\hspace{-0.1cm}\int d^3x\,\left(\frac{1}{2}\,\mbox{tr}_{\tiny\mbox{N}}
\,G_{\mu\nu}G_{\mu\nu}+\mbox{tr}_{\tiny\mbox{N}}\,{\cal D}_\mu\phi{\cal D}_\mu\phi+
V_E(\phi)\right)\,.
\eqe
In Eq.\,(\ref{actE}) $V_E$ denotes the potential responsible for the stabilization of 
$\phi$. The covariant derivative is 
defined as ${\cal D}_\rho\phi=\pd_\rho+ie[\phi,a_\rho]$, the field 
strength as $G_{\mu\nu}=G^a_{\mu\nu}t^a$, where 
$G^a_{\mu\nu}=\pd_\mu a^a_\nu-\pd_\nu a^a_\mu-ef^{abc}a^b_\mu a^c_\nu$, 
$e$ denotes the effective gauge coupling constant, and $\mbox{tr}_{\tiny\mbox{N}}\,t^a t^b\equiv 
1/2\,\delta^{ab}$. While the effect of nontrivial topology is described by the scalar 
sector of the effective theory the curvature $G_{\mu\nu}$ is 
generated by the topologically trivial fluctuations. 

For future work \cite{HerbstHofmann2004} 
we propose the following nonlocal definition for the phase of a given SU(2) block 
$\tilde{\phi}$. 
\eqb
\label{locdefphi}
\frac{\tilde{\phi}^a(\tau)}{|\tilde{\phi}|}\sim \mbox{tr}_{2} \int d\rho\, d^3x\, 
\lambda^a\,F_{\mu\nu}((\tau,0))\,[(\tau,0),(\tau,\vec{x})]
F_{\mu\nu}((\tau,\vec{x}))[(\tau,\vec{x}),(\tau,0)]+\cdots\,.
\,
\eqe
The dots in Eq.\,(\ref{locdefphi}) denote the contributions of higher $n$-point functions, and 
the $\sim$ sign indicates that this expansion very likely is asymptotic 
at best as a powers series in a dimensionless parameter 
$\xi$. This, however, is not an obstacle to determining $\tilde{\phi}'s$ phase and 
modulus \cite{HerbstHofmann2004}. 
Each block $\tilde{\phi}$ receives a nontrivial phase by the corresponding SU(2)-embedded trivial-holonomy 
caloron $A^C_\beta$ (or anticaloron $A^A_\beta$) over which the correlator 
in Eq.\,(\ref{locdefphi}) is evaluated\footnote{The topological 
charges of $A^C_\beta$ or $A^A_\beta$ may, in principle, vary 
from block to block.}. In the definition Eq.\,(\ref{locdefphi}) 
$[(\tau,0),(\tau,\vec{x})]$ denotes a Wilson line in the fundamental representation 
which is taken to be along a
straight path connecting the two points $(\tau,0)$ and $(\tau,\vec{x})$:
\eqb
\label{Wilsonline}
[(\tau,0),(\tau,\vec{x})]
\equiv {\cal P}\,\exp\left[i\int_{(\tau,0)}^{(\tau,\vec{x})}dy_\beta\,A_\beta(y)\right]\,.
\eqe
In a lattice simulation at finite temperature $T$ the average in Eq.\,(\ref{locdefphi}) 
can be computed by using an ensemble of cooled configurations whose action  
is an integer multiple of $8\pi^2/{\bar g}^2$\footnote{The nontrivial-holonomy part is then cooled away.}. 
Local gauge-singlet composites such as the gluon condensate 
\eqb
\label{gluoncondensate}
\la \mbox{tr}_{\tiny\mbox{N}} 
F_{\mu\nu}(x)\,F_{\mu\nu}(x)\ra\,
\eqe
are thermodynamically irrelevant for the following reasons: 
Since they do not couple to the 
topologically trivial sector they do not 
influence the mass spectrum of fluctuations $\delta a_\rho$. Moreover, a singlet 
composite, arising from noninteracting trivial-holonomy calorons, 
would have zero energy density and pressure 
because of the BPS saturation: a situation which cannot be changed by 
interactions mediated by the trivial sector due to the missing 
gauge charge. The situation is different though 
if an axial anomaly, arising from integrated-over chiral fermions, 
is operative. In this case the composite in 
Eq.\,(\ref{gluoncondensate}) determines the 
mass of the axion, and thus it is visible.       
  
The key question now is whether the potential $V_E$ in Eq.\,(\ref{actE}) is uniquely 
determined by our basic assumption. 
What are the properties of the field $\phi$ that can be deduced? 
In thermal equilibrium $\phi$ must be periodic in 
Euclidean time $\tau$ ($0\le\tau\le 1/T$). 
Since $\phi$ describes the ground state of the thermal system its 
modulus $|\phi|$ must not depend on $\tau,\vec{x}$ but should depend 
on $T$. Since $\phi$ is built 
of noninteracting, self-dual configurations (zero energy density and pressure) 
it must also be pressure - and energy-free. This is the case 
if and only if $\phi(\tau)$ (not its modulus!) is BPS saturated, that is, 
it solves the following equation
\eqb
\label{BPSEP}
\pd_{\tau}\phi=v_E\,
\eqe
where $v_E$ is a `square root' of the potential $V_E$: 
\eqb
\label{squareroot}
V_E(\phi)=\mbox{tr}_{\tiny\mbox{N}}v_E^\dagger v_E\,.
\eqe
The above properties fix the potential uniquely 
to be $V_E(\phi)\mbox{tr}_{\tiny\mbox{N}}=\Lambda_E^6\mbox{tr}_{\tiny\mbox{N}}(\phi^{2})^{-1}$. 
As it turns out, a (winding) solution to Eq.\,(\ref{BPSEP}) 
is quantum mechanically and thermodynamically 
inert and thus can be used as a background to the macroscopic equation of 
motion for the trivial-topology sector of the theory.  

The equation of motion 
\eqb
\label{GF}
{\cal D}_\mu G_{\mu\nu}=2ie[\phi,{\cal D}_\nu\phi]\,,
\eqe
which follows from the effective action (\ref{actE}), 
determines a configuration $a^{bg}_\rho$. For $a^{bg}_\rho$ to describe 
the ground state of the theory it 
needs to be pure gauge, that is, $G_{\mu\nu}[a^{bg}_\rho]\equiv 0$. Otherwise 
the invariance of the thermal system under spatial rotations would be spontaneously broken. 
It will turn out that such a pure-gauge solution $a^{bg}_\rho\propto \frac{T}{e}$ 
exists for ${\cal D}_\nu\phi\equiv 0$. 
As a consequence, the action density in Eq.\,(\ref{actE}) when 
evaluated on $\phi,a^{bg}_\rho$ reduces to the potential $V_E$. We thus describe 
on a macroscopic level interactions between trivial-holonomy 
calorons as mediated by the topologically trivial sector. 
Namely, the vanishing ground-state energy density (pressure) of 
noninteracting trivial-holonomy calorons is shifted from 
zero to $V_E(\phi)$ ($-V_E(\phi)$). Moreover, a macroscopic 
holonomy arises which indicates that (unstable) nontrivial-holonomy 
calorons are generated by gluon exchange and, as a consequence, that isolated 
magnetic monopoles occur. This precludes our conceptual 
discussion of the ground-state physics.

An adjoint Higgs field $\phi$ breaks the SU(N) gauge 
symmetry down to U(1)$^{\tiny\mbox{N-1}}$ at most. Whether SU(N) gauge 
symmetry is broken maximally or submaximally is decided by a boundary 
condition to the BPS equation (\ref{BPSEP}) set at an 
asymptotically high temperature. Interacting calorons emit and absorb 
gauge-field fluctuations $\delta a_\rho$. To discuss their 
quasiparticle mass spectrum a gauge transformation to $a^{bg}_\rho\equiv 0$ (unitary gauge) 
must be performed. We will show explicitly for the SU(2) case that such a 
transformation is on the one hand nonperiodic but on the other hand a 
symmetry transformation for all thermodynamical quantities. This is true since the transformation does 
not affect the periodicity 
of the fluctuations $\delta a_\rho$ (no Hosotani mechanism \cite{Hosotani1983}). The nonperiodic gauge transformation maps the 
Polyakov loop from ${\bf -1}$ to ${\bf 1}$, therefore generates a 
global electric center transformation and thus interpolates between the 
two physically equivalent ground states of the theory. As a consequence, 
the global symmetry $Z_{\tiny\mbox{2,elec}}$ is spontaneously broken and 
hence the electric phase is deconfining. The generalization to arbitrary 
N is straight forward.

There are tree-level heavy 
(TLH) and tree-level massless (TLM) modes in $\delta a_\rho$. 
Due to the $T$ dependence of $\phi$ 
on-shell TLH modes are thermal quasiparticles. 

Due to the $T$ dependent Higgs mechanism and the $T$ dependent 
ground-state energy, which are 
both generated by the macroscopic field $\phi$, implicit temperature 
dependences arise in a loop expansion of thermodynamical quantities. 

To guarantee in the effective theory the validity of the Legendre transformations between  
thermodynamical quantities, as they can be derived from the partition function of the underlying 
theory, thermodynamical self-consistency has to be demanded. This condition determines the temperature 
evolution of the effective gauge coupling constant $e$ with temperature. 
As we will see, there is an attractor to this evolution which is the constancy of 
$e$ except for a logarithmic pole at a temperature $T_{c,E}$. We thus recover in the effective theory 
the ultraviolet-infrared decoupling that follows from the 
renormalizability of the underlying theory. 

The approach to the ground-state dynamics is an inductive one. 
Namely, we first define a potential $V_E(\phi)$ and subsequently show that this definition 
implies the above properties of $\phi$, a small action for 
calorons at the temperature where they are assumed to first form the field $\phi$, and the 
existence of a pure-gauge solution $a^{bg}_\rho$ to Eq.\,(\ref{GF}). The thermal system 
decomposes into a ground state and a part represented by very weakly interacting 
quasiparticle fluctuations \footnote{We compute two-loop correction to the pressure in 
\cite{HerbstHofmannRohrer2004}.}. One can consider the former as a heat bath for the latter at low temperatures 
and vice versa at high temperatures.

\subsection{Caloron `condensate', macroscopically \label{CCM}}

\subsubsection{The case of even N: Ground-state physics}

We first address the macroscopic dynamics 
of the adjoint Higgs field $\phi$ when N is even. 
The case of odd N is discussed in Sec.\,\ref{oddN}. 
We may always work in a gauge where $\phi$ is SU(2) block diagonal:
\eqb
\label{SU2diag}
\phi=\left(\begin{array}{cccc}
\tilde{\phi}_1 & {\bf 0} & {\bf 0} &\cdots\\ 
{\bf 0}&\tilde{\phi}_2& {\bf 0} &\cdots\\ 
{\bf 0}&{\bf 0}& \ddots & \\ 
\vdots& \vdots& & 
\end{array}\right)\,.
\eqe
In Eq.\,(\ref{SU2diag}) each field 
$\tilde{\phi}_l,\,(l=1,\cdots,\mbox{N}/2)$, 
lives in an independent SU(2) subalgebra of SU(N), and we define the SU(2) modulus as
\eqb
\label{modulus}
|\tilde{\phi}_l|^2\equiv \frac{1}{2}\,\mbox{tr}_2\,\tilde{\phi_l}^2\,.
\eqe
The potential $V_E$ in Eq.\,(\ref{actE}) is defined as 
\eqb
\label{potentialE}
V_E=\mbox{tr}_{\tiny\mbox{N}} v^\dagger_E v_E\equiv\Lambda_E^6\,
\mbox{tr}_{\tiny\mbox{N}}\,(\phi^2)^{-1}\,
\eqe
where $\La_E$ is a fixed mass scale generated by dimensional transmutation. 
It is important to note already at this point 
that there is only one independent mass scale describing the 
thermodynamics in {\sl all} phases of the theory.  

We define $v_E$ as follows:
\eqb
\label{sqpotentialE}
v_E\equiv i\La_E^3\left(\begin{array}{cccc}
\lambda_1\tilde{\phi}_1/|\tilde{\phi}_1|^2& {\bf 0} & {\bf 0} &\cdots\\ 
{\bf 0}&\lambda_1\tilde{\phi}_2/|\tilde{\phi}_2|^2& {\bf 0} &\cdots\\ 
{\bf 0}& {\bf 0}& \ddots & \\ 
\vdots& \vdots& & 
\end{array}\right)\,
\eqe
where $\lambda_i,\,(i=1,2,3)$, denote the Pauli matrices. This definition is modulo 
global SU(2)-block gauge transformations. This 
global symmetry (the `direction' of winding along a U(1) circle around the group manifold $S^3$ od SU(2)) 
will translate into a gauge symmetry once the theory condenses magnetic monopoles, 
see Sec.\,\ref{windingmag}. 

In SU(2) decomposition 
the solution $\tilde{\phi}_l$ to the BPS equation 
(\ref{BPSEP}) reads 
\eqb
\label{solBPSE}
\hspace{-0.5cm}\tilde{\phi}_l(\tau)=\sqrt{\frac{\Lambda_E^3}{2\pi T |K(l)|}}\,\lambda_3
\exp(-2\pi i T K(l)\lambda_1\tau)\,
\eqe
where $K(l)$ is a non-zero integer. The solution in Eq.\,(\ref{solBPSE}) 
is periodic in $\tau$ and depends on $T$. 
The set $\{K(1),\dots,K(\mbox{N}/2)\}$, which is a boundary condition to 
Eq.\,(\ref{BPSEP}) at the large temperature $T_P$, determines the value of the potential 
at a given temperature. It also specifies to what extent the 
SU(N) gauge symmetry is spontaneously broken by caloron 'condensation'. For example, 
the sets $\{1,1\}$ and $\{1,2\}$
break SU(4) down to SU(2)$\times$SU(2)$\times$U(1) 
and U(1)$^3$, respectively. Out of 15 gauge-field 
modes 7 modes remain massless in the former and 3 modes in the latter case. 
For a description in terms of a given SU(N) Yang-Mills theory the 
set $\{K(1),\dots,K(\mbox{N}/2)\}$ has to be measured, 
see also the discussion in Sec.\,\ref{MPlanck}. For definiteness and simplicity we 
assume in the following that the gauge symmetry breaking 
is maximal in such a way that the 
potential $V_E(\phi)$ is minimal (MGSB). 
This corresponds to the boundary condition $\{K(1),\dots,K(\mbox{N}/2)\}=\{1,2,\dots,\mbox{N}/2\}$ 
or a (local) permutation thereof. 

Let us now verify that the solution in Eq.\,(\ref{solBPSE}) is quantum mechanically and thermodynamically 
stabilized. Assuming MGSB, the following ratios are obtained 
\eqb
\label{ratios}
\frac{\pd^2_{|\tilde{\phi}_l|}V_E}{T^2}=12\pi^2\,l^2\,, \ \ \ \ \ \ 
\frac{\pd^2_{|\tilde{\phi}_l|}V_E}{|\tilde{\phi}_l|^2}=3l^3\,\lambda_E^3\,
\eqe
where the dimensionless temperature $\lambda_E$ is defined 
as $\lambda_E\equiv 2\pi T/\La_E$. For N not too large we have 
$\lambda_E\gg 1$, see Sec.\,\ref{TSCE}. As one can infer from Eq.\,(\ref{ratios}), 
the mass $m_l^2\equiv \pd^2_{|\tilde{\phi}_l|}V_E$ of collective caloron fluctuations is much larger than $T$ and 
the compositeness scale $|\tilde{\phi}_l|$. The off-shellness of 
quantum fluctuations of the field $\tilde{\phi}_l$ is 
cut off at this scale in Minkowskian or Euclidean signature as
\eqb
\label{Mineuosn}
|p^2-m_l^2|\le |\tilde{\phi}_l|^2\,,\ \ \ \ \ \ \mbox{or} \ \ \ \ \ 
\ p_e^2+m_l^2\le |\tilde{\phi}_l|^2\,,\ \ \ \ (p_e^2\ge 0)\,.
\eqe
So if no off-shellness in Minkowskian or Euclidean signature is 
allowed for on the one hand. On the other hand, statistical fluctuations of 
on-shell $\phi$-particles are strongly Boltzmann suppressed and thus negligible. 
We conclude 
that the solution $\phi$ in Eq.\,(\ref{solBPSE}) is 
stabilized against fluctuations $\delta\phi$ and the potential $V_E$ in Eq.\,(\ref{potentialE}) 
is a truly effective one. Thus $\phi$ is nothing but a background for the 
thermodynamics of the topologically trivial sector of the theory. 
As we will see in Sec.\,\ref{TSCE}, topologically trivial quantum fluctuations $\delta a_\rho$ 
generate only a tiny correction to the tree-level value $V_E(\phi)$.    

Before we investigate the properties of the fluctuations $\delta a_\rho$ 
let us complete our construction of the ground state. 
The ground-state configurations $a_\rho^{bg}$ 
needs to be a pure-gauge solution to the classical equation of motion Eq.\,(\ref{GF}). In order for 
$a_\rho^{bg}$ not to break the rotational invariance of the system 
it needs to be pure gauge. Inserting the background (\ref{solBPSE}) 
(winding numbers: $\{l=1,\dots,l=N/2\}$ for MGSB) into Eq.\,(\ref{GF}), 
we obtain the following pure-gauge solution: 
\eqb
\label{curvfree}
a^{bg}_\rho=\frac{\pi}{e}\,T \delta_{\rho 4}\left(\begin{array}{cccc}
\lambda_1& {\bf 0} & {\bf 0} &\cdots\\ 
{\bf 0}&2\,\lambda_1& {\bf 0} &\cdots\\ 
{\bf 0}& {\bf 0}& \ddots & \\ 
\vdots& \vdots& & 
\end{array}\right)\,.
\eqe
Moreover, we have
\eqb
\label{covder0}
{\cal D}_\rho\phi=0\,
\eqe
on $\phi, a^{bg}_\rho$. A remarkable thing has happened: 
On a macroscopic level we describe the generation 
of a nontrivial holonomy by interactions 
between trivial holonomy calorons, mediated by trivial-topology fluctuations, 
in terms of a macroscopic holonomy associated with $a^{bg}_\rho$! 
For the microscopic physics this implies the generation of (rare) nontrivial-holonomy calorons 
and their subsequent dissociation into magnetic monopoles. On a mesoscopic level, 
this is nothing but the Kibble mechanism for monopole creation \cite{Kibble1976} arising from the 
domanization of color orientations of $\phi$. 
Moreover, the vanishing pressure and energy density of a hypothetical 
ground state composed of noninteracting trivial-holonomy calorons, 
is shifted to $\mp V_E(\phi)$ with 
\eqb
\label{gspot}
V_E(\phi)=\frac{\pi}{2}\,\Lambda_E^3 T \mbox{N(N+2)}\,
\eqe
by gluon exchange (insert Eqs.\,(\ref{curvfree}) and (\ref{covder0}) into Eq.\,(\ref{actE})). 

Let us now split the topologically trivial part in Eq.\,(\ref{decfl}) 
further into the ground-state part $a^{bg}_\rho$ and fluctuations $\delta a_\rho$:
\eqb
\label{gffluct}
a_\rho=a^{bg}_\rho+\delta a_\rho\,.
\eqe
To make the mass spectrum of the fluctuations $\delta a_\rho$ 
visible it would be desirable to work in unitary 
gauge where $a^{bg}_\rho\equiv 0$ and thus no coupling 
of $\delta a_\rho$ to the background $a^{bg}_\rho$ takes place. 

The gauge rotation $\Omega\equiv e^{i\theta}$, 
which transforms $\phi$ and $a_\rho$ according to
\eab
\label{gurot}
\phi &\to& \Omega^\dagger\,\phi\,\Omega\nonumber\\ 
a_\rho&\to& \Omega^\dagger\,a_\rho\,\Omega+\frac{i}{e}\,(\pd_\rho\Omega^\dagger)\,\Omega
\eae
from winding gauge to unitary gauge is for MGSB  given as 
\eqb
\label{thetamat}
\theta=\left(\begin{array}{cccc}
-\pi\lambda_1T\tau & {\bf 0 }& {\bf 0} &\cdots\\ 
{\bf 0}&-2\pi\lambda_1T\tau & {\bf 0} &\cdots\\ 
{\bf 0}&{\bf 0}& \ddots & \\ 
\vdots& \vdots& & 
\end{array}\right)\equiv \left(\begin{array}{cccc}
\theta_1 & {\bf 0 }& {\bf 0} &\cdots\\ 
{\bf 0}&\theta_2& {\bf 0} &\cdots\\ 
{\bf 0}&{\bf 0}& \ddots & \\ 
\vdots& \vdots& & 
\end{array}\right)\,.
\eqe 
Notice that the gauge transfromation $\Omega$ as parametrized by Eq.\,(\ref{thetamat}) is {\sl not} periodic due to its 
first, third, fifth, ... block being {\sl antiperiodic} in $\tau$. 
Is a nonperiodic 
gauge transformation physically admissible? Let us discuss this 
for the SU(2) case only. We can make $\Omega=\exp[\frac{-i\pi\lambda_1}{T}]$ 
periodic at the expense of sacrificing its smoothness at the point $\tau=\frac{1}{2T}$ 
by 
\eqb
\label{omegatrafo}
\Omega\rightarrow \tilde{\Omega}=\Omega Z(\tau)\,
\eqe
where $Z(\tau)$ is a local (electric) $Z_2$ transformation of the form
\eqb
\label{loccentertr}
Z(\tau)=2\Theta(\tau-\frac{1}{2T})-1\,,
\eqe
and $\Theta$ denotes the Heavyside step function:
\eqb
\label{Hevayside}
\Theta(x)=\left\{\begin{array}{c}
0\,,\ \ \ \ (x<0)\,,\\ 
\frac{1}{2}\,,\ \ \ \ (x=0)\,,\\ 
1\,,\ \ \ \ \ (x>0)\,.
\end{array}\right.\,.
\eqe
Applying $\Omega^\prime$ 
to $a_\mu=a^{bg}_\rho+\delta a_\rho$, 
where $\delta a_\rho$ is a periodic fluctuation in winding gauge, we have
\eab
\label{trafiwtoug}
a_\rho&\rightarrow &\tilde{\Omega}^\dagger(a^{bg}_\rho+\delta a_\rho)\tilde{\Omega}+
\frac{i}{e}\pd_\rho \tilde{\Omega}^\dagger\tilde{\Omega}\nonumber\\ 
&=& \Omega^\dagger(a^{bg}_\rho+\delta a_\rho)\Omega+
\frac{i}{e}\left((\pd_\rho\Omega^\dagger)\Omega+(\pd_\rho Z(\tau))Z(\tau)\right)\nonumber\\ 
&=&\Omega^\dagger\delta a_\rho\Omega+\frac{2i}{e}\delta(\tau-\frac{1}{2T})Z(\tau)\nonumber\\ 
&=&\Omega^\dagger\delta a_\rho\Omega\,.
\eae
Since $\Omega^\dagger(0)=-\Omega^\dagger(\frac{1}{T})=\Omega(0)=-\Omega(\frac{1}{T})$ 
we conclude that if the fluctuation $\delta a_\rho$ 
is periodic in winding gauge it is also periodic in unitary
gauge. It thus is irrelevant whether we integrate out the 
fluctuations $\delta a_\rho$ in winding or unitary gauge in a loop expansion 
of thermodynamical quantities: 
Hosotani's mechanism \cite{Hosotani1983} does not take place. What changes though 
is the Polyakov loop evaluated on 
the background configuration $a^{bg}_\rho$:
\eqb
\label{Polch}
P[a^{bg}_\rho]=-{\bf 1} \rightarrow P[0]={\bf 1}\,.
\eqe
We conclude that the theory has two equivalent ground states and 
that the global electric $Z_{\tiny\mbox{2,elec}}$ symmetry is spontaneously broken. We thus have shown that 
the elecric phase is, indeed, {\sl deconfining}. 
The generalization of this result to SU(N) with N even is straight forward.    

In unitary gauge 
we have
\eqb
\label{solBPSEug}
\hspace{-0.5cm}\tilde{\phi}_l(\tau)\equiv\sqrt{\frac{\Lambda_E^3}{2\pi T l}}\,\lambda_3\,.
\eqe
Thus the field $\phi$ is constant and diagonal. Moreover, we have
\eqb
\label{Gug}
G_{\mu\nu}^a[a_\rho]=G_{\mu\nu}^a[\delta a_\rho]\,.
\eqe
The gauge-covariant kinetic term for $\phi$ in the action Eq.\,(\ref{actE}) reduces to a sum over 
mass terms for the TLH modes contained in $\delta a_\rho$. The TLH (TLM)
modes are massive (massless) quasiparticles 
associated with three (two) polarization states. As we will show 
in \cite{HerbstHofmannRohrer2004} by computing the two-loop correction to the thermodynamical pressure for N=2 
these quasiparticles are practically noninteracting for sufficiently large temperatures. 

\subsubsection{The case of odd N\label{oddN}}

If N is odd then a decomposition of $\phi$ into SU(2) blocks only is no longer possible. One of the SU(2) blocks 
in Eq.\,(\ref{SU2diag}) is then replaced by 
an SU(3) block. Imposing MGSB and counting the number of independent 
stable and unstable magnetic monopoles microscopically on the one hand and macroscopically 
in the electric and the magnetic phase on the other hand, 
see Sec.\,(\ref{CMP}), we conclude that temporal winding should 
take place within the SU(3) block as in Eq.\,(\ref{solBPSE}) but now in 
each of the two independent SU(2) subalgebras only 
for half the time. The first SU(2) subgroup is generated by 
$\bar{\lambda}_i$, ($i=1,\cdots,3$) where
\eqb
\label{barla}
\bar{\lambda}_1=\left(\begin{array}{ccc}0&0&0\\ 
0&0&1\\ 
0&1&0\end{array}\right)\,,\ \ \ 
\bar{\lambda}_2=\left(\begin{array}{ccc}0&0&0\\ 
0&0&-i\\ 
0&i&0\end{array}\right)\,,\ \ \ 
\bar{\lambda}_3=\left(\begin{array}{ccc}0&0&0\\ 
0&1&0\\ 
0&0&-1\end{array}\right)\,.
\eqe
The second SU(2) subgroup is generated by 
$\tilde{\lambda}_i$, ($i=1,\cdots,3$) where
\eqb
\label{tildela}
\tilde{\lambda}_1=\left(\begin{array}{ccc}0&0&1\\ 
0&0&0\\ 
1&0&0\end{array}\right)\,,\ \ \ 
\tilde{\lambda}_2=\left(\begin{array}{ccc}0&0&-i\\ 
0&0&0\\ 
i&0&0\end{array}\right)\,,\ \ \ 
\tilde{\lambda}_3=\left(\begin{array}{ccc}1&0&0\\ 
0&0&0\\ 
0&0&-1\end{array}\right)\,.
\eqe
The solution of the BPS equation (\ref{BPSEP}) for the SU(3) block can be obtained by applying the following prescription to 
a single block (\ref{solBPSE}) of {\sl odd} winding number 
$K(l)$ in Eq.\,(\ref{SU2diag}): generate two configurations $\tilde{\phi}_{l,1}$ and $\tilde{\phi}_{l,2}$ 
by replacing the 
Pauli matrix $\lambda_i$ in Eq.\,(\ref{solBPSE}) by $\bar{\lambda}_i$ in the first half period, $0\le\tau\le 1/(2T)$, and 
by zero in the second  half period, $1/(2T)<\tau<1/T$, and by replacing the Pauli matrix $\lambda_i$ 
in Eq.\,(\ref{solBPSE}) by zero in the first half period, $0<\tau<1/(2T)$, and 
by $\tilde{\lambda}_i$ in the second half period, $1/(2T)\le\tau\le 1/T$. Add 
$\tilde{\phi}_{l,1}$ and $\tilde{\phi}_{l,2}$ to generate a new 
solution to the BPS equation (\ref{BPSEP}) over the entire period. Within this block a member 
of the third, dependent SU(2) algebra is generated at $\tau=0,1/(2T)$. To which block 
$l$ of odd winding number $K(l)$ 
this prescription is applied is a boundary condition to the BPS equation. 
For simplicity we will proceed in this paper by only quoting results for the case N=3.

It is clear that for even $N\ge 6$ a decomposition of $\phi$ may contain an even number of SU(3) 
blocks besides SU(2) blocks. For definiteness, 
we only consider the decomposition into SU(2) blocks as proposed in Eq.\,(\ref{SU2diag}). 

\subsection{Tree-level mass spectrum of TLH modes}

In absence of radiative corrections only N(N$-$1) TLH modes acquire 
mass by the adjoint Higgs mechanism. TLM modes aquire tiny 
screening masses radiatively. 
This effect will be discussed in Sec.\,\ref{Radcor}. 

In unitary gauge, the TLH mass spectrum calculates as 
\eqb
\label{massspectrum}
m_k^2=-2e^2\,\mbox{tr}\,[\phi,t^{k}][\phi,t^{k}]\,,\ \ \ \ \ \ (k=1,\cdots,\mbox{N(N$-$1)})\,.
\eqe
The SU(N) generators $t^k$, which are associated with the 
TLH modes, are  
\eab
\label{TLgen}
t^{IJ}_{rs}&=&1/2\,(\delta_r^I\delta_s^J+\delta_s^I\delta_r^J)\,,\ \ \ 
\bar{t}^{IJ}_{rs}=-i/2\,(\delta_r^I\delta_s^J-\delta_s^I\delta_r^J)\,,\nonumber\\ 
(I&=&1,\cdots,\mbox{N};\,J>I;\,r,s=1,\cdots,\mbox{N})\,. 
\eae
By virtue of Eqs.\,(\ref{solBPSEug}), (\ref{massspectrum}), (\ref{TLgen}) we obtain:
\eqb
\label{TLHmasses}
m_{IJ}^2=\bar{m}_{IJ}^2=e^2(\phi_I-\phi_J)^2=
e^2\,\frac{\Lambda_E^3}{2\pi T}\left\{\begin{array}{cc}
\left(\frac{1}{\sqrt{I/2}}-
\frac{1}{\sqrt{J/2}}\right)^2\,,&\hspace{0.5cm}(I\,\mbox{even}, J\,\mbox{even})\\ 
\left(\frac{1}{\sqrt{(I+1)/2}}-
\frac{1}{\sqrt{(J+1)/2}}\right)^2\,,&\hspace{0.5cm}(I\,\mbox{odd}, J\,\mbox{odd})\\ 
\left(\frac{1}{\sqrt{(I+1)/2}}+
\frac{1}{\sqrt{J/2}}\right)^2\,,&\hspace{0.5cm}(I\,\mbox{odd}, J\,\mbox{even})\\ 
\left(\frac{1}{\sqrt{I/2}}+
\frac{1}{\sqrt{(J+1)/2}}\right)^2\,,&\hspace{0.5cm}(I\,\mbox{even}, J\,\mbox{odd})\,\,.
\end{array}\right.
\eqe
For N=3 we have
\eab
\label{n3}
m_{12}^2&=&m_{13}^2=\bar{m}_{12}^2=\bar{m}_{13}^2=e^2\,\frac{\Lambda_E^3}{2\pi T}\nonumber\\ 
m_{23}^2&=&\bar{m}_{23}^2=4e^2\,\frac{\Lambda_E^3}{2\pi T}\ \ \ \ \ \mbox{or}\nonumber\\ 
m_{12}^2&=&m_{23}^2=\bar{m}_{12}^2=\bar{m}_{23}^2=e^2\,\frac{\Lambda_E^3}{2\pi T}\nonumber\\ 
m_{13}^2&=&\bar{m}_{13}^2=4e^2\,\frac{\Lambda_E^3}{2\pi T}\,.
\eae

\subsection{Thermodynamical self-consistency and 
$e(T)$ at one loop\label{TSCE}}

The TLH modes $\delta a^k_\rho$ are thermal quasiparticle fluctuations on tree-level 
since their masses $m_{IJ}$ and $\bar{m}_{IJ}$, given in Eqs.\,(\ref{TLHmasses}) and (\ref{n3}), 
are $T$ dependent. Moreover, the ground-state pressure, given by $-V_E$, is linearly 
dependent on $T$, see Eq.\,(\ref{gspot}). Thermodynamical quantities such as the pressure, 
the energy density, or the
entropy density are interrelated by Legendre transformations as 
derived from the partition function associated with 
the underlying SU(N) Yang-Mills Lagrangian. To assure 
that the same Legendre transformations are valid in the effective electric theory, where ground-state pressure 
and particle masses are temperature dependent, a condition for thermodynamic self-consistency needs 
to be imposed. In general, this condition assures that $T$-derivatives of quantities that enter the 
action {\sl density} of the effective theory (in our case the TLH masses and the ground-state pressure) 
cancel one another. 

\begin{figure}
\begin{center}
\leavevmode
\leavevmode
\vspace{2.5cm}
\includegraphics{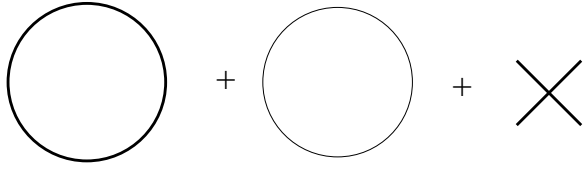}
\end{center}
\caption{Diagrams contributing to the pressure when radiative corrections are 
ignored. A thick line corresponds to TLH modes and a 
thin one to TLM modes. The cross depicts the ground-state contribution arising 
from caloron `condensation'.  
\label{diagpress}}      
\end{figure}

Let us formulate this condition at one-loop accuracy. It is convenient to 
work with dimensionless quantities. The quantity $a_k$ (mass over temperature) 
is defined as
\eqb
\label{dimlessdef}
a_k\equiv c_k a\,\ \ \ \ \ \mbox{where} \ \ \ \ \  a\equiv e\sqrt{\frac{\Lambda_E^3}{2\pi T^3}}\,,
\ \ \ \ (k=1,\cdots,\mbox{N}(\mbox{N}-1))\,,
\eqe
and the coefficient $c_k$ is equal to the square root of one of the numbers 
appearing to the right of the curly bracket in Eq.\,(\ref{TLHmasses}) or, for N=3, to 
1 or 2, compare with Eq.\,(\ref{n3}). Recalling 
our definition of a dimensionless temperature
\eqb
\label{lambdaE}
\lambda_E\equiv \frac{2\pi T}{\La_E}\,,
\eqe
we have
\eqb
\label{massP}
a=2\pi e\lambda_E^{-3/2}\,.
\eqe
Let us also define the (negative definite) function $\bar{P}(a)$ as
\eqb
\label{P(a)}
\bar{P}(a)\equiv\int_0^\infty dx \,x^2\,\log[1-\exp(-\sqrt{x^2+a^2})]\,.
\eqe
Ignoring higher loop corrections, the total thermodynamical pressure 
$P(\lambda_E)$ associated with the diagrams in 
Fig.\,(\ref{diagpress}) calculates as
\eqb
\label{treelP}
P(\lambda_E)=-\La_E^4\left\{\frac{2\lambda_E^4}{(2\pi)^6}\left[2(\mbox{N}-1)\bar{P}(0)+
3\sum_{k=1}^{\tiny\mbox{N(N-1)}}\bar{P}(a_k)\right]+
\frac{\lambda_E}{2}\left(\frac{\mbox{N}}{2}+1\right)\mbox{N}\right\}\,\ \ (\mbox{N\ \ even})\,.
\eqe
In Eq.\,(\ref{treelP}) we have neglected the `nonthermal' contribution $-\Delta V_E$ to the 
one-loop bubbles in Fig.\,\ref{diagpress} which is estimated as (quantum fluctuations are cut 
off at the compositeness scale $|\phi|$)
\eqb
\label{1loopvacbubble}
\Delta V_E<(\mbox{N}^2-1)\frac{3}{8\pi^2}\int_0^{|\phi|}dp\, p^3\log\left(\frac{p}{|\phi|}\right)=
-3(\mbox{N}^2-1)\frac{|\phi|^4}{128\pi^2}\,.
\eqe
To neglect $-\Delta V_E$ is well justified for sufficiently small N, say N$<$20, since we have
\eqb
\label{DeltaV}
\left|\frac{\Delta V_E}{V_E}\right|<
(\mbox{N}^2-1)\frac{3}{128\pi^2}\left(\frac{|\phi|}{\La_E}\right)^6\sim 
(\mbox{N}^2-1)\frac{3}{128\pi^2}\,\lambda_E^{-3}\,,
\eqe
and we will show later that the minimal temperature in the electric phase 
$\lambda_{c,E}$ is much larger than unity. 

\begin{figure}
\begin{center}
\leavevmode
\leavevmode
\vspace{4.5cm}
\includegraphics{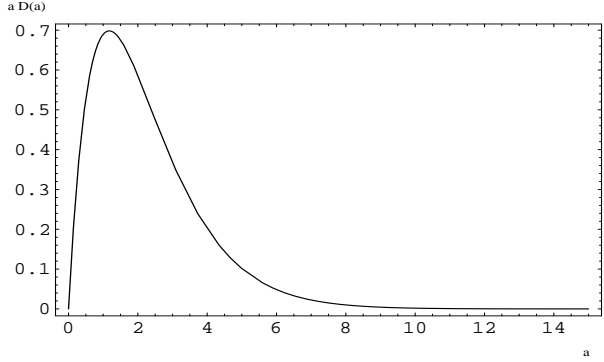}
\end{center}
\caption{The function $a\,D(a)$.\label{DAf}}      
\end{figure}

\begin{figure}
\begin{center}
\leavevmode
\leavevmode
\vspace{4.5cm}
\includegraphics{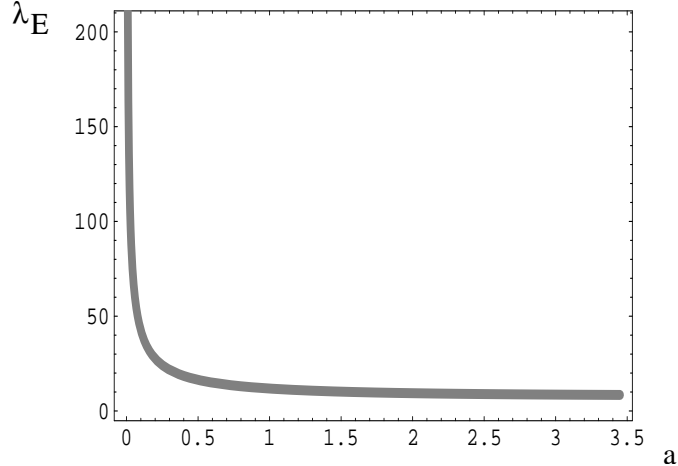}
\end{center}
\caption{The solution $\lambda_E(a)$ to Eq.\,(48) for N=2 subject to the 
boundary condition $\lambda_{E,P}\equiv\lambda_E(a=0)=10^3$.\label{laofa}}      
\end{figure}

For N=3 we obtain 
\eqb
\label{treelP3}
P(\lambda_E)=-\La_E^4\left\{\frac{2\lambda_E^4}{(2\pi)^6}\left[4\bar{P}(0)+3\left(
4\bar{P}(a)+2\bar{P}(2a)\right)\right]+
2\lambda_E\right\}\,\ \ \ (\mbox{N=3})\,.
\eqe
A particular Legendre transformation following from the partition function of the underlying 
theory maps pressure into energy density as
\eqb
\label{rhoPree}
\rho=T\frac{dP}{dT}-P\,.
\eqe
For Eq.\,(\ref{rhoPree}) to hold also in the effective electric theory the following situation 
has to be {\sl arranged for}: only the explicit 
T-dependence in $P$, arising from the explicit T-dependence of the 
Boltzmann weight, should contribute to the derivative $\frac{dP}{dT}$ while implicit $T$ dependences 
of gauge-boson masses and the ground-state pressure ought to cancel one other. 
This condition is expressed as \cite{Gorenstein1995}
\eqb
\label{TSC}
\pd_a P=0\,.
\eqe
Before we proceed let us recall that the $\lambda_E$ dependence of the ground-state pressure, as indicated
in Eqs.\,(\ref{gspot}) and (\ref{treelP}), could be expressed in terms of a dependence on the 
mass parameter $a$ by virtue of Eq.\,(\ref{massP}) if the $\lambda_E$ dependence of the gauge 
coupling constant $e$ was known. Keeping this in mind, we derive from Eqs.\,(\ref{TSC}), (\ref{treelP}), and 
(\ref{treelP3}) the following evolution equation 
\eqb
\label{eeq}
\pd_a \lambda_E=-\frac{24\,\lambda_E^4\,a}{(2\pi)^6 
\mbox{N(N+2)}}\sum_{k=1}^{\tiny\mbox{N(N-1)}}c_k^2 D(a_k)\,,\ \ (\mbox{N\ \ even})\,.
\eqe
For N=3 we have
\eqb
\label{eeq3}
\pd_a \lambda_E=-\frac{12\,\lambda_E^4\,a}{(2\pi)^6} 
\left(D(a)+2D(2a)\right)\,,\ \ (\mbox{N=3})\,.
\eqe
The function $D(a)$ is defined as
\eqb
\label{DA}
D(a)\equiv \int_0^{\infty} dx\,
\frac{x^2}{\sqrt{x^2+a^2}}\frac{1}{\exp(\sqrt{x^2+a^2})-1}\,.
\eqe
Eqs.\,(\ref{eeq}) and (\ref{eeq3}) describe the evolution of temperature as a 
function of tree-level gauge boson mass. The right-hand sides of these equations 
are negative definite since the function $D(a)$ in Eq.\,(\ref{DA}) 
is positive definite, see Fig.\,\ref{DAf}.
As a consequence, the solutions $\lambda_E(a)$ 
to Eqs.\,(\ref{eeq}) and (\ref{eeq3}) can be inverted to $a(\lambda_E)$. In Fig.\,\ref{laofa} a solution 
for N=2 subject to the initial condition $\lambda_{E,P}\equiv\lambda_E(a=0)=10^3$ is shown. 
We have noticed numerically that the low-temperature behavior of $\lambda_E(a)$ 
is practically independent of the value $\lambda_{E,P}$ as long as $\lambda_{E,P}$ is 
sufficiently large. Let us show this analytically. For $a$ sufficiently smaller than unity we may 
expand the right-hand side of Eq.\,(\ref{eeq}) only taking the 
linear term in $a$ into account. The inverse of the solution is then of the following form
\eqb
\label{anasolv}
a\propto \lambda_E^{-3/2}\sqrt{1-\left(\frac{\lambda_E}{\lambda_{E,P}}\right)^3}\,.
\eqe
For $\lambda_E$ sufficiently smaller than $\lambda_{E,P}$ this can be approximated as
\eqb
\label{anasolapp}
a\propto\lambda_E^{-3/2}\,.
\eqe
The dependence in Eq.\,(\ref{anasolapp}) thus 
is an {\sl attractor}. So at whatever asymptotically high temperature the formation of the 
adjoint Higgs field $\phi$ out of noninteracting trivial-holonomy calorons 
is assumed does not influence the behavior of the theory at much 
lower temperatures. This result is reminiscent of the 
ultraviolet-infrared decoupling property of the renormalizable, 
underlying theory.  

Notice that Fig.\,(\ref{DAf}) and Eq.\,(\ref{eeq}) imply that there are 
fixed points of the evolution $\lambda_E(a)$ at $a=0$ and $a=\infty$. 
The points $\lambda_{E,P}\equiv\lambda_E(a=0)$ and $\lambda_{E,c}\equiv\lambda_E(a=\infty)$ are 
associated with the highest and the lowest attainable temperaturesin the electric phase, respectively. 
In a bottom-up evolution no information can be obtained about $\lambda_{E,P}$ if the temperature that is maximally 
reached is sufficiently smaller than $\lambda_{E,P}$.  
In a top-down evolution, where $\lambda_{E,P}$ is set as a boundary value, the prediction of 
$\lambda_{E,c}$ is independent of $\lambda_{E,P}$. These two 
statements are immediate consequences of the existence of an 
attractor in the thermodynamical evolution. 

\begin{figure}
\begin{center}
\leavevmode
\leavevmode
\vspace{5.5cm}
\includegraphics{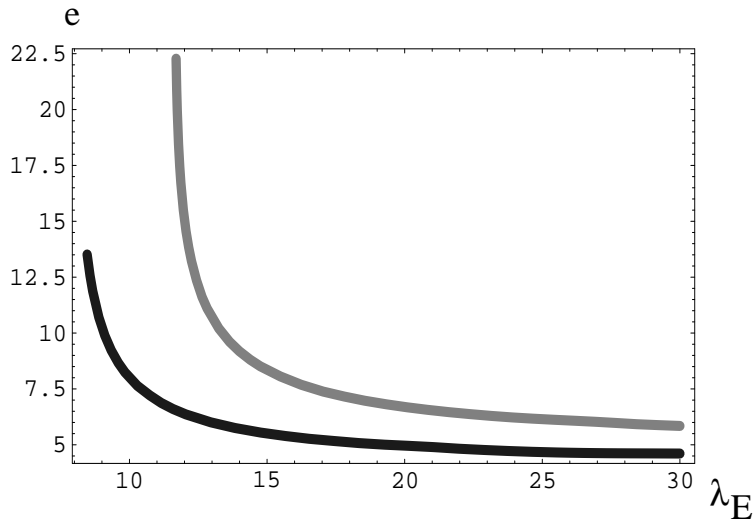}
\end{center}
\caption{The low-temperature evolution of the gauge 
coupling $e$ in the electric phase for 
N=2 (grey line) and N=3 (black line). The gauge coupling 
diverges logarithmically, $e\propto -\log(\lambda_{E}-\lambda_{c,E})$, at 
$\lambda_{E,c}=11.65\ (\mbox{N=2})$ and $\lambda_{E,c}=8.08\ (\mbox{N=3})$. 
The plateau values are $e=5.1\ (\mbox{N=2})$ and $e=4.2\ (\mbox{N=3})$.\label{eoflam}}      
\end{figure}

Numerically inverting the solution $\lambda_E(a)$ to $a(\lambda_E)$, the evolution of the gauge coupling constant 
$e$ can be computed using Eq.\,(\ref{massP}):
\eqb
\label{coupP}
e(\lambda_E)=\frac{1}{2\pi}a(\lambda_E)\lambda_E^{3/2}\,.
\eqe
We show the result in Fig.\,\ref{eoflam} for N=2,3. Before we interprete this result 
a remark on the interpretation of the effective gauge coupling constant $e$ for $T\sim T_P$ is in order. 
Since $e$ determines the 
strength of the interaction between nontopological gauge field fluctuations $\delta a_\rho$ 
and the {\sl coherent} caloron state $\phi$ there is, in general, no reason for it 
to be equal to the gauge coupling constant $\bar{g}$ of the fundamental Yang-Mills theory. 
However, for temperatures very close to $T_P$, where $\phi$ is assumed to form 
(see also Sec.\,\ref{MPlanck}), the coupling constant $e$ should be roughly equal to $\bar{g}$. 
From Fig.\,\ref{eoflam} we see that the effective gauge coupling constant $e\sim\bar{g}$ 
evolves to values larger than unity 
shortly below the initial temperature $\lambda_{E,P}$. This is in agreement 
with our assumption that trivial-holonmy SU(2) {\sl calorons} (of sufficiently small `instanton' radius) 
have a large action and thus contribute sizably to the partion function of the underlying theory, 
compare with Eq.\,(\ref{GPYSeffth}). For a grandly unifying SU(N) gauge theory we argue 
in Sec.\,\ref{MPlanck} that $\lambda_{E,P}\sim \frac{M_P}{\La_E}$ where $M_P\gg \La_E$ denotes the
cutoff scale for the description in terms of a local, four-dimensional field theory. 

There is no handle on the relation between the fundamental gauge coupling 
$\bar{g}$ and $e$ after the condensate has formed and is sustained by interactions 
between the trivial-holonomy calorons. Since collective effects are strong 
they can make up for a large value of the action of an isolated caloron generated by a 
small value of $\bar{g}$. This may justify the use of universal 
perturbative expressions for the beta function in lattice 
simulation at low temperatures.

Notice that the dependence of $a$ on $\lambda_E$ in 
Eq.\,(\ref{anasolapp}) is canceled in the dependence of $e$ on $\lambda_E$ 
such that a plateau is reached quickly in Eq.\,(\ref{coupP}). We interprete the fact that the 
gauge coupling constant $e$ remains constant for a large range of temperatures 
as another indication for the existence of spatially isolated and conserved 
magnetic charges in the system, see also Sec.\,\ref{CMP}. 
During the relaxation of $e$ to its plateau value 
constituent BPS magnetic monopoles residing in dissociating nontrivial-holonomy 
calorons form as isolated defects 
\cite{KraanVanBaalNPB1998,vanBaalKraalPLB1998,Diakonov}. Since the 
interaction between a monopole and an antimonopole, as mediated by TLM modes, 
is screened \cite{KorthalsAltes,HoelbingRebbiRubakov2001} these defects need 
not be considered explicitly in the effective, thermal theory discussed in the present work. Implicitly, 
their presence is accounted for here by the holonomy of the 
background field $a^{bg}_\rho$.  
 
The effective gauge coupling constant 
$e$ runs into a logarithmic (needle) pole at $\lambda_{E,c}$ of the form
\eqb
\label{logpolee}
e(\lambda_E)\propto-\log(\lambda_E-\lambda_{E,c})\,,
\eqe
compare with Fig.\,\ref{laofa}. 

The plateau values for $e$ are $e\sim 5.1$ and $e\sim 4.2$ 
for N=2 and N=3, respectively. For a given N they do not depend on where the 
boundary condition $\lambda_{E,P}\equiv\lambda_E(a=0)$ is set if $\lambda_{E,P}$ is sufficiently larger than
$\lambda_{E,c}$. For N=2 we have $\lambda_{E,c}=11.65$ and for N=3 we have $\lambda_{E,c}=8.08$. 
It thus is self-consistently justified to neglect the one-loop quantum corrections 
to $V_E(\phi)$ as they are estimated in Eq.\,(\ref{DeltaV}).
Naively, one would conclude that the large plateau values would render the 
fluctuations $\delta a_\rho$ to be very strongly coupled and that radiative 
corrections to the thermodynamical potentials would thus be uncontrolled. This, however, 
does not happen due to the fact that the TLH modes acquire masses by the Higgs mechanism 
which are proportional to $e$ and due to the existence of a compsiteness scale 
$|\phi|$. The latter constrains the momentum $p$ of quantum fluctuations 
in $\delta a_\rho$ as
\eab
\label{qfconstr}
|p^2-m^2_k|&\le &|\tilde{\phi}_l|^2\,,\ \ (\mbox{TLH, Minkowskian})\,,\ \ 
|p^2|\le |\tilde{\phi}_l|^2\,, \ \ (\mbox{TLM, Minkowskian})\,,\nonumber\\ 
\ p_e^2+m_k^2&\le&|\tilde{\phi}_l|^2\,,\ \  (\mbox{TLH, Euclidean})\,, \ \ 
p_e^2\le |\tilde{\phi}_l|^2\,, \ \  (\mbox{TLM, Euclidean})\,.
\eae
Since in nonlocal two-loop contributions, see Fig.\,\ref{looppress}, 
each TLH line is on-shell for $e>1$ the effect 
of a strongly coupled vertex is compensated by the very 
small phase space that is allowed for the progagation of 
a TLM mode coupling to the TLH mode. The center-of-mass energy flowing into or out of a four-vertex 
is constrained in addition to Eq.\,(\ref{qfconstr}) 
to be smaller than $|\tilde{\phi}_l|$. We show in \cite{HerbstHofmannRohrer2004} 
that the two-loop contributions to 
the pressure for SU(2) are, depending on temperature, at most $\sim 0.1$\% 
of the one-loop result.

\subsection{What is $T_P$?\label{MPlanck}}  

At temperatures 
larger than the highest attainable temperature $T_P$ in the electric phase (corresponding to $a=0$) 
a grandly unifying SU(N) gauge symmetry, which generates all matter and its (nongravitational) 
interactions in the Universe at lower temperatures, would be unbroken. We assume here 
that gravity is a perfectly classical theory up to the Planck 
mass $M_P$ \footnote{This assumption is usually made 
in field-theory models of cosmological 
inflation.}. The perturbative phase of the SU(N) Yang-Mills theory at 
$T>T_P$ would have a trivial vacuum state represented by weakly interacting 
quantum fluctuations. The momenta associated with these 
fluctuations can be maximally as hard as some cutoff scale at which the 
four-dimensional setup ceases to be reliable. Common belief is that this cutoff scale is 
$M_P$. 

The highest temperature $T_{\tiny\mbox{cutoff}}$ 
attainable in the perturbative phase thus is comparable to $M_P$, $T_{\tiny\mbox{cutoff}}\sim M_P$. 
At temperatures rnaging between $T_{\tiny\mbox{cutoff}}$ and $T_P$ perturbative 
vacuum fluctuations would generate a cosmological 
constant $\La_{\tiny\mbox{cosmo}}$ given as
\eqb
\label{wbLambdacosmo}
\La_{\tiny\mbox{cosmo}}\sim M_P^4\,.
\eqe
At $T_{\tiny\mbox{cutoff}}$ the vacuum energy density $M_P^4$ 
would be comparable to the thermal energy density of on-shell fluctuations 
$\sim T_{\tiny\mbox{cutoff}}^4$. While the former is constant 
the later dies off quickly as the Universe cools down. Thus for $T$ 
slightly smaller than $T_{\tiny\mbox{cutoff}}$ the vacuum energy density dominate the expansion hence the 
Universe would rapidly decrease its temperature as 
\eqb
\label{decT}
T\sim \exp[-M_P \Delta t] T_{\tiny\mbox{cutoff}}\,
\eqe
where $\Delta t\equiv t-t_P$ and $t_P\sim M^{-1}_P$. A sudden termination 
of this Planck-scale inflation would occur at $T=T_P$ where 
the field $\phi$ comes into existence. While the {\sl radiation} component of the total 
energy density is continuous across the phase boundary 
at $T_P$ the energy density of the {\sl ground-state} 
would be discontinuously reduced from $M_P^4$ to 
$\sim T_P \La_{YM,\tiny\mbox{N}}^3$, see Eq.(\ref{gspot}). On the one hand, 
this release of latent heat is a characteristic for a (strong if $T_P\ll T_{\tiny\mbox{cutoff}}$) 
1$^{\tiny\mbox{st}}$ order transition. On the other hand, the order parameter $a$ for the onset of the electric
phase is continuous, see Fig.\,(\ref{eoflam}) (screening masses are comparable 
on both sides of the phase boundary, see Eq.\,(\ref{thermPiweakcoupl})). 
But this is signalling a 2$^{\tiny\mbox{nd}}$ order phase transition. There is only one way 
to avoid this 
contradiction: The phase boundary at 
$T_P$ needs to be hidden beyond the point 
$T_{\tiny\mbox{cutoff}}\sim M_P$, that is, $T_P\ge T_{\tiny\mbox{cutoff}}$. 

The reader may 
object that our conclusion about the 2$^{\tiny\mbox{nd}}$ order 
of the phase transition (order parameter $a$) is resting on a 
one-loop analysis of the gauge-coupling evolution. Usually, 
it is understood that such a mean-field treatment breaks down close 
to a 2$^{\tiny\mbox{nd}}$ order transition due to fact that long range 
correlations mediated by low-momentum quantum fluctuations become important. 
In the electric phase these long-range correlations are, however, 
contained in the field $\phi$ which does not fluctuate at 
any temperature $T_{E,c}<T\le T_P$. Due to $|\phi|$ being a 
cutoff for the quantum fluctuations of the TLM 
and TLH modes and due to the fact that $|\phi|$ dies off as $T^{-1/2}$ the long-range correlating effects of 
TLM and TLH quantum fluctuations can safely be 
neglected if $T_P\gg\La_E$. The above discussion and conclusion thus are valid.

So far we had in mind the simplified case of MGSB at $T_P$ in 
grandly unifying SU(N) Yang-Mills theory. We know from experiment, however,  
that gauge-symmetry breakdown at $T_P$ is {\sl not} maximal in Nature. If the gauge-symmetry breaking 
by $\phi$ at $T_P$ is submaximal, however, the 
same contradiction between the orders of the
phase transition arises for $T_P<M_P$. In this case a 
fundamental gauge symmetry SU(N) would be broken 
to a product of group with factors SU(M) M$<$N or U(1). Recall, 
that submaximal gauge-symmetry breaking in the electric phase takes
place if SU(2) blocks in $\phi$ with equal winding number 
are generated at $T_P$, see Eq.\,(\ref{SU2diag}).

The theory would then condense magnetic SU(M) color and magnetic U(1) monopoles at the 
temperature $T_{E,c}$. While condensates of the latter are described by 
complex scalar fields, see Sec.\,\ref{magPT}, condensates of the
former are, again, described by adjoint Higgs fields. Maximal or submaximal 
breakings of the residual SU(M) gauge symmetries would be possible at the electric-magnetic 
transition of the fundamental SU(N) theory. For the effective 
description of SU(M) thermodynamics at $T<T_{E,c}$ this 
boundary condition effectively is set during the phase transition at $T_{E,c}$ and not at $T_P$. 
By matching the thermodynamical pressures at $T_{E,c}$ the scale $\La_E^\prime$ of the 
effective theory SU(M) is determined in terms of the scale $\La_E$ 
of the fundamental SU(N) theory and the pattern of symmetry breaking at $T_P$ and $T_{E,c}$. 
After a sequence of such matching procedures has taken place (in which residual U(1) 
factors have thermodynamically decoupled) a hierarchy 
between $\La_E$ and the scale $\La_E^{\prime\cdots\prime}$ of an effective 
SU(L) theory ($L\ll N$), seen experimentally at low energies (or local 
temperatures for that matter), is generated. 

To summarize, 
we provided an argument that in a grandly unifying and four-dimensional SU(N) Yang-Mills theory the dynamical 
generation of the adjoint background field $\phi$ must 
take place at a temperature $T_{\tiny\mbox{cutoff}}\sim M_P$ where the local 
field-theory description breaks down. Any lower SU(L) gauge symmetry, which is 
generated from the SU(N) theory by a sequence of condensations of (color) monopoles and confining
transitions, is matched to its `predessesor' theory in 
2$^{\tiny\mbox{nd}}$ order like phase transitions. The scale of this SU(L) gauge theory can be much 
lower than the scale $\Lambda_E$ of the fundamental SU(N) theory.

\subsection{Stable und unstable magnetic monopoles\label{CMP}}

Due to the presence of an adjoint Higgs field $\phi$ 
in the electric phase there are 't Hooft-Polyakov magnetic monopoles 
\cite{'tHooft1974,Polayakov1974} which are centered at 
the isolated zeros of $\phi$. On a mesoscopic level, these zeros occur at points in space 
where four or more color-orientation domains of a 
given block $\tilde{\phi}_l$ meet \cite{Kibble1976}. Microscopically, 
BPS monopoles \cite{PrasadSommerfield1974} are contained within 
decaying nontrivial-holonomy calorons \footnote{A perturbative analysis of 1-loop radiative corrections to an isolated 
instanton were performed in \cite{'tHooft1976}. In \cite{Diakonov} this was done for 
the nontrivial-holonomy caloron. As a result a repulsive potential for the constituent monopole and 
antimonopole was obtained for a sufficiently large holonomy.}. 

Close to $T_P$ we have $\bar{g}\sim e$, and the 
following processes take place: 
Trivial-holonomy calorons grow rapidly in size, 
start to overlap, and thus generate calorons with holonomy by their 
interactions. If this holonomy is sufficiently large then two following processes take place: (i) 
SU(2) nontrivial-holonomy calorons of the same embedding in SU(N) decay independently into 
their constituent magnetic monopoles and antimonopoles, and (ii) SU(2) 
nontrivial-holonomy calorons generated from trivial-holonomy calorons 
of different SU(2) embeddings\footnote{Recall, that we have assumed that these calorons 
come with different topological charge to 
obtain maximal symmetry breaking.} in SU(N) do not decay into constituent 
monopoles and antimonopoles since they would have to live in instable 
superpositions of the embeddings of the asymptotic 
trivial-holonomy calorons. While the former process generates stable magnetic 
dipoles the latter generates instable monopoles and antimonopoles.
 
In our macroscopic approach it is hard to see how the size of a typical trivial-holonomy 
caloron changes with temperature after $e$ has reached its plateau value 
since $e$ plays a different role than the fundamental coupling constant
$\bar{g}$. We may, however, infer from lattice simulations that trivial-holonomy calorons 
are large enough to not generate a topological susceptibility on lattices of 
presently feasible sizes. 

A magnetic monopole-antimonopole pair, which is connected by a magnetic flux line, becomes a stable dipole 
if the pair has a sufficiently large spatial separation. 
In this case the monopole and the antimonopole are practically 
noninteracting \cite{HoelbingRebbiRubakov2001} and 
thus are stable defects. If a single monopole is produced then it 
is unstable unless it connects with its antimonopole, produced in 
an independent collision. N$-$1 independent SU(2) subgroups exist and so 
N$-$1 independent magnetic monopoles may occur in the case of MGSB. 
Since the monopole constituents in a caloron are BPS 
saturated we also expect an isolated monopole in a stable monopole-antimonopole pair 
to be BPS saturated. The analytical expression 
for an SU(2) charge-one BPS monopole in a gauge where the Higgs field $\phi$ 
winds around the group manifold at spatial infinity \cite{PrasadSommerfield1974} is given as
\eqb
\label{BPSmonop}
A^a_0=0\,,\ \ \ \ A^a_i=\epsilon_{aij}\hat{r}_j\frac{1-K(r)}{er}\,,\ \ \ \ \phi^a=\hat{r}_a \frac{H(r)}{er}\,,
\eqe
where $r\equiv \sqrt{\vec{r}^2}$, and $\hat{r}$ is a spatial unit vector. 
The form of the functions $K(r)$ and $H(r)$ is
\eqb
\label{KandH}
K(r)=\frac{Cr}{\sinh(Cr)}\,,\ \ \ \ \ \ \ \ \ H(r)=Cr\coth(Cr)-1\,.
\eqe
In Eqs.\,(\ref{KandH}) the mass scale $C$ is proportional to the asymptotic Higgs 
modulus $|\phi(|\vec x|\to\infty)|$ and the gauge coupling constant $e$. 
The mass of a BPS monopole is given as \cite{PrasadSommerfield1974}
\eqb
\label{monopmass}
M=\frac{8\pi}{\sqrt{2}\,e} |\phi(|\vec x|\to\infty)|\,.
\eqe
A dual, abelian field strength $\tilde{G}_{\mu\nu}$ can be defined as 
\cite{'tHooft1974}
\eqb
\label{tT}
\tilde{G}_{\mu\nu}=\frac{\phi^a G_{\mu\nu}^a}{|\phi|}-
\frac{\epsilon_{abc}}{e|\phi|^3}\phi_a ({\cal D}_\mu \phi)_b ({\cal D}_\mu \phi)_c\,.
\eqe
The expression in Eq.\,(\ref{tT}) reduces to $\tilde{G}_{\mu\nu}=\pd_\mu a^3_\nu-\pd_\nu a^3_\mu$ 
in unitary gauge $\phi^a=\delta^{a3}|\phi|$. Eq.\,(\ref{tT}) defines the field strength of a 
dual photon which couples 
to the magnetic charge $4\pi/e$ of the monopole. 
Both the gauge dynamics involving only dual photons and magnetic monopoles and 
the entire gauge dynamics in the electric phase are blind with respect to the 
magnetic (local in space) center symmetry $Z_{\tiny\mbox{N,mag}}$ 
(the field strength ${G}_{\mu\nu}$, the field $\phi$ and the covariant derivative 
${\cal D}_\mu \phi$ are invariant under center transformations and, 
as a consequence of Eq.\,(\ref{tT}) so is the dual field strength 
$\tilde{G}_{\mu\nu}$). However, a local-in-time 
transformation $\in Z_{\tiny\mbox{N,elec}}$ may transform the Dirac string between 
a static monopole and a static antimonopole in a given SU(2) embedding 
into a Dirac string belonging to a dipole in a different SU(2) 
embedding. This does not violate the 
conservation of total magnetic charge and certainly has no effect 
on any gauge invariant quantity. In an effective theory, 
where monopoles are condensed degrees of freedom, a local $Z_{\tiny\mbox{N,elec}}$ 
transformation should thus be represented by a local permutation of the 
fields describing the monopole condensates and the gauge-field fluctuations 
which couple to them. In such an effective theory the action thus ought to be 
invariant under these local permutations.

How can we see the occurrence of stable and unstable magnetic monopoles in the macroscopic, 
effective theory for the electric phase? In winding gauge the temporal winding 
of the $l^{\tiny\mbox{th}}$ SU(2) block in $\phi$ 
is complemented by spatial winding at isolated points in 3D space. By a large, $\tau$ 
dependent gauge transformation the monopole's spatially asymptotic SU(2) 
Higgs field is rotated to spatial constancy and into the direction given 
by the temporal winding of the unperturbed block $\tilde{\phi}_l$. Its 
Dirac string rotates as a function of Euclidean time. Since this monopole is stable, 
a correlated antimonopole must exist. We arrive at a dipole rotating 
about its center of mass at an 
angular frequency $2\pi l T$. This rotation is an 
artifact of our choice of gauge. Rotating the dipole to unitary gauge by the gauge 
function $\theta_l$ in Eq.\,(\ref{thetamat}), we arrive 
at a (quasi)static dipole. There are isolated 
coincidence points (CPs) in time where the lower right (upper left) corner 
of the $l^{\tiny\mbox{th}}$ ($(l+1)^{\tiny\mbox{th}}$) 
SU(2) block (now ($l=1,\cdots,\mbox{N}/2-1$))
together with the number zero of its 
right-hand (left-hand) neighbour are proportional to the generator $\lambda_3$. 
Coincidence also takes place between the first and last diagonal entry in $\phi$. 
At a CP, spatial winding may take place at isolated points in 3D space. Moreover, 
coincidence also takes place between nonadjacent 
SU(2) blocks and the first and last diagonal entry in $\phi$. The associated 
monopoles are, however, not independent. The spatial winding associated with the additional 
SU(2) generators `flashing out' at the CPs corresponds to unstable magnetic monopoles.  

Summing up all independent monopole species, we have: 
\eqb
\label{counting}
\frac{\mbox{N}}{2}+\frac{\mbox{N}}{2}-1=\mbox{N}-1\,.
\eqe
In the case N=3 the field $\phi$ winds with winding number one in each of the 
two independent SU(2) subalgebras for half the time, see Sec\,\ref{oddN}. The match between these 
subalgebras happens at the CPs $\tau_{CP}=0,1/(2T)$ where an element of the 
third, dependent SU(2) algebra is generated, see Sec.\,\ref{oddN}. Due to these CPs we have 
unstable monopoles.

\subsection{Outlook on radiative corrections\label{Radcor}}

\subsubsection{Contributions to the TLM self-energy}

Let us now investigate for N=2 and at one loop the simplest contribution to 
the polarization tensor for the TLM mode. A complete investigation of 
two-loop contributions to the pressure is the objective of \cite{HerbstHofmannRohrer2004}. We work in unitary 
gauge $\phi=\mbox{diag}(\phi_1,\phi_2)$, $a_\rho^{bg}=0$. 
This condition fixes the gauge up to U(1) rotations 
generated by $\lambda_3$. This remaining gauge freedom can be used to gauge the TLM mode to 
transversality: $\pd_i \delta a^{\tiny\mbox{TLM}}_i=0$ (radiation or Coulomb gauge). 
No ghost fields need to be introduced 
in unitary-Coulomb gauge. 

After an analytical continuation to Minkowskian signature\footnote{For the purpose of the present 
work we do not need the matrix formulation of the real-time propagators.} 
the asymptotic propagator of a free TLM mode is given as \cite{LandsmanWeert1987}
\eqb
\label{TLMprop}
D^{\tiny\mbox{TLM},0}_{\mu\nu,ab}(k,T)=-\delta_{ab}P^T_{\mu\nu}
\left(\frac{i}{k^2}+2\pi\delta(k^2)n_B(|k_0|/T)\right)\,
\eqe
where 
\eab
\label{PT}
P^T_{00}&=&P^T_{0i}=P^T_{i0}=0\,,\nonumber\\ 
P^T_{ij}&=&\delta_{ij}-\frac{k_ik_j}{\vec{k}^2}\,,
\eae
and $n_B(x)\equiv\frac{1}{\exp[x]-1}$ denotes the Bose distribution function. 
The analytically continued asymptotic propagator of 
a free TLH mode $D^{\tiny\mbox{TLH},IJ,0}_{\mu\nu,ab}(k,T)$ 
is that of a massive vector boson
\eqb
\label{TLHprop} 
D^{\tiny\mbox{TLH},IJ,0}_{\mu\nu,ab}(k,T)=-\delta_{ab}
\left(g_{\mu\nu}-\frac{k_\mu k_\nu}{m_{IJ}^2}\right)
\left[\frac{i}{k^2-m_{IJ}^2}+2\pi\delta(k^2-m_{IJ}^2)n_B(|k_0|/T)\right]\,.
\eqe
The vertices for the interactions of TLH and TLM modes are the usual ones. 
In unitary-Coulomb gauge the 4D loop integrals over quantum fluctuations 
are cut off at the compositeness scale $|\phi|(T)$ of the effective theory. 
Thermal fluctuations, associated with 3D loop integrals, are automatically cut off 
by the distribution function $n_B$.
\begin{figure}
\begin{center}
\leavevmode
\leavevmode
\vspace{3.5cm}
\includegraphics{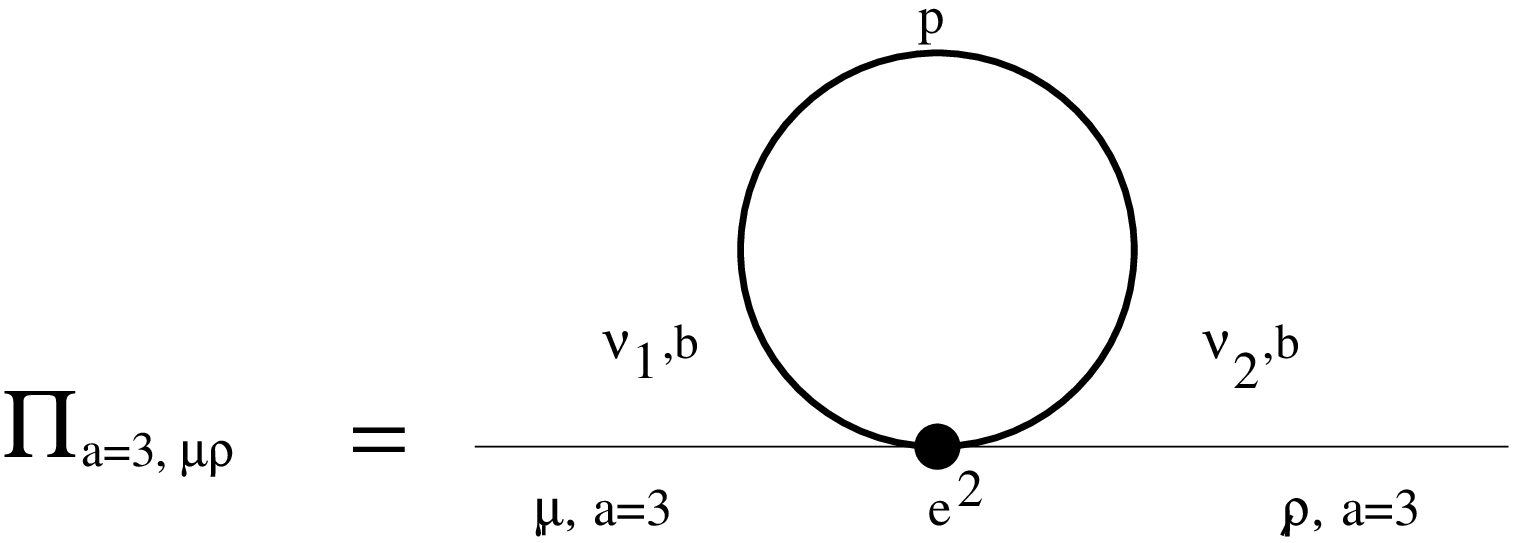}
\end{center}
\caption{A tadpole contribution to the self-energy of the TLH mode.\label{tadpole}}      
\end{figure}
Let us now look at the tadpole contribution to $\Pi_{a=3,\mu\rho}$ as shown in Fig.\,\ref{tadpole}. 
This diagram decomposes into a part for the vacuum fluctuations in the loop, 
which has a $T$ dependence only due to the $T$ dependence of TLH masses, 
and a thermal part. Contracting the Lorentz indices, the former can be calculated as
\eqb
\label{vacPi}
\Pi^{\tiny\mbox{vac},\mu}_{a=3,\mu}=-\frac{3(e|\tilde{\phi}|)^2}{2\pi^2}\int_0^{\sqrt{1-(2e)^2}}dx\,
\frac{x^3(4+\frac{x^2}{(2e)^2})}{x^2+(2e)^2}\,
\eqe
while the thermal part reads
\eqb
\label{thermPi}
\Pi^{\tiny\mbox{therm},\mu}_{a=3,\mu}=\frac{18}{\pi^2}(e|\tilde{\phi}|)^2\int_0^\infty dy\,
\frac{y^2}{\sqrt{y^2+(2e)^2}}\,\,\frac{1}{\exp\left[2\pi\lambda_E^{-3/2}\sqrt{y^2+(2e)^2}\right]-1}\,.
\eqe
It is instructive to perform the weak and strong coupling limits 
in Eqs.\,(\ref{vacPi}) and (\ref{thermPi}). 

\noindent For $e<\frac{1}{\sqrt{2}}$, we obtain 
\eqb
\label{vacPiweakcoupl}
\Pi^{\tiny\mbox{vac},\mu}_{a=3,\mu}=-\frac{3|\tilde{\phi}|^2}{32\pi^2}\left(1+16\,e^2-
80\left(1-\frac{12}{5}\log(2e)\right)\,e^4\right)\,
\eqe
and for $e\ll 1$
\eqb
\label{thermPiweakcoupl}
\Pi^{\tiny\mbox{therm},\mu}_{a=3,\mu}\to 3\pi^2\,(eT)^2+O(e^4)\,.
\eqe
For $e\gg \frac{1}{\sqrt{2}}$ there is no vacuum contribution, $\Pi^{\tiny\mbox{vac},\mu}_{a=3,\mu}=0$.
The thermal part reads 
\eqb
\label{thermstrongcoupl}
\Pi^{\tiny\mbox{therm},\mu}_{a=3,\mu}\to\frac{18}{\pi^3}\,e^3\,\Lambda_E^2\,\lambda_E^{1/2}\,
K_1(4\pi\,e\,\lambda_E^{-3/2})\,
\eqe
where $K_1(x)$ denotes a modified Bessel function. 
The weak coupling result for the thermal part in Eq.\,(\ref{thermPiweakcoupl}) coincides, 
up to a numerical factor, with the perturbative expression for the electric 
screening (or Debye) mass-squared, as it should. In the limit of infinite coupling, 
which is reached due to the logarithmic pole for 
$\lambda_E\searrow\lambda_{E,c}$, see Eq.\,(\ref{logpolee}), 
the thermal part in Eq.\,(\ref{thermstrongcoupl}) vanishes. 
This agrees qualitatively with results obtained in 
thermal quasiparticle models fitted to lattice data 
\cite{Biro1990,Peshier1995,LevaiHeinz1998}. It was found in these models that the Debye mass 
vanishes for $T\searrow T_{E,c}$ \footnote{We foretake at this point that the deconfinement phase transiiton
seen on the lattice is the electric-magnetic transition at $T_{E,c}$. We will discuss 
in Sec.\,\ref{complat} why the lattice is not capable of measuring infrared 
sensitive quantities such as the pressure at temperatures below $T_{E,c}$.}. 
A large and {\sl constant} value of $e$, as it is generated 
by one-loop evolution (compare with Eqs.\,(\ref{anasolapp}) and (\ref{coupP})), 
implies that the approximation leading to Eq.\,(\ref{thermstrongcoupl}) breaks down for 
high temperatures. It is, however, clear from Eq.\,(\ref{thermPi}) that the 
weak coupling result at $O(e^2)$ in Eq.\,(\ref{thermPiweakcoupl}) 
gives an upper bound on the contribution to the screening mass-squared at any 
temperature and any value of the 
coupling constant.

We expect that the situation is similar for the nonlocal one-loop diagrams. 
As for the tadpole correction in 
the polarization operator of a TLH
mode there is a contribution $\propto e^2$ for strong coupling which arises from the 
vacuum part with the TLM mode in the loop. This contribution is, however, 
suppressed due to the constraint that the center-of-mass energy flowing into or out of 
the vertex must be smaller than $|\phi|$ \cite{HerbstHofmannRohrer2004}.

\subsubsection{Loop expansion of the pressure}

\begin{figure}
\begin{center}
\leavevmode
\leavevmode
\vspace{3.5cm}
\includegraphics{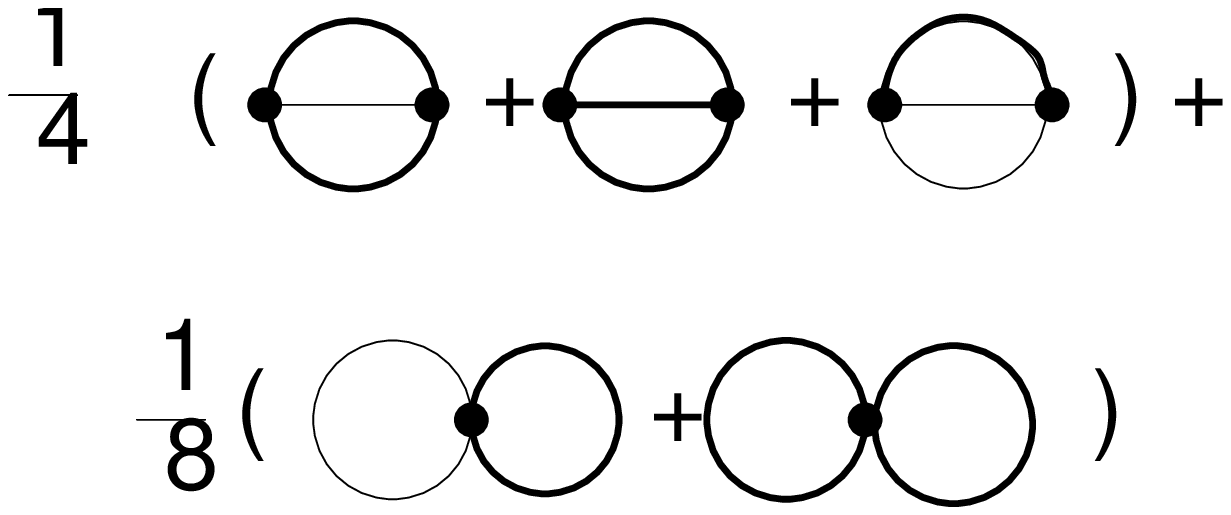}
\end{center}
\caption{Two-loop diagrams contributing to the pressure. Thick lines denote propagators of TLH modes, 
thin lines those of TLM modes.\label{looppress}}      
\end{figure}

Two-loop diagrams contributing to the pressure in a real-time formulation 
are indicated in Fig.\,\ref{looppress}. We do not compute them here but in 
\cite{HerbstHofmannRohrer2004} for N=2. 
A general remark concerning thermodynamical self-consistency 
is in order already here. Recall, that on one-loop level 
we have obtained an evolution equation from the requirement of thermal 
self-consistency $\pd_a P=0$. This gave a functional relation between temperature and mass  
which could be inverted for all temperatures in the electric phase. 
After the relation Eq.\,(\ref{coupP}) between 
coupling constant $e$ and mass $a$ was exploited we obtained 
a functional dependence of the effective gauge coupling constant 
$e$ on temperature. Equivalently, we could have 
demanded $\pd_e P=0$ since $e$ is the only variable parameter of our 
effective theory for the electric phase. This would have {\sl directly} 
generated an evolution equation for temperature 
as a function of $e$. 

Radiative corrections $\Delta P$ to the 
pressure have a separate dependence on $a$ and $e$, 
\eqb
\label{radpressure}
\Delta P=T^4 \Delta\tilde{P}(e,a,\lambda_E)\,,
\eqe
where $\Delta\tilde{P}$ is a dimensionless function of its dimensionless arguments. 
To implement thermodynamical self-consistency by demanding $\pd_a P=0$ one has to 
express the explicitly appearing $e$ in Eq.\,(\ref{radpressure}) 
in terms of $a$ by means of Eq.\,(\ref{coupP}) and distinguish temperature 
dependences arising from a simple rescaling and those 
arising from the $T$ dependent ground-state physics. For SU(2) we have 
\eqb
\label{e(a)}
\frac{m^2}{|\phi|^2}\equiv e^2(a,\lambda_E)=\frac{T^2}{2}\,\times\frac{a^2}{|\phi|^2}=
\frac{\lambda_E^2}{8\pi^2}\times a^2\lambda_E\,.
\eqe
The first factor on the right-hand sides of Eq.\,(\ref{e(a)}) arises 
from rescaling, so only the second factor needs to be differentiated: 
\eqb
\label{e(a)D}
\pd_a e(a,\lambda_E)=\frac{\lambda_E^2}{8\pi^2}\times\left(2a\lambda_E+a^2\pd_a\lambda_E\right)\,.
\eqe
After solving $\pd_a P=0$ for the term $\pd_a\lambda_E$ 
we obtain a modified right-hand side of the 
evolution equation Eq.\,(\ref{eeq}). The inverted solution to this 
evolution equation describes the dependence of mass on temperature or, 
after applying Eq.\,(\ref{massP}), the dependence of $e$ on temperature 
when two-loop diagrams for the pressure are taken into account. 
The Euclidean momenta $p_e$ of off-shell fluctuations of the TLH-modes are constrained by the condition 
\eqb
\label{osmomTLH}
p^2_e\le|\phi|^2(1-c_k^2\,e^2\,\frac{|\tilde{\phi}_1|}{|\phi|^2})\,,\ \ \ \ 
(k=1,\cdots,\mbox{N}(\mbox{N}-1))\,,
\eqe
and by the requirement that the total momentum squared 
flowing into or our of a four-vertex cannot be larger than $|\phi|^2$, s
ee Eq.\,(\ref{Mineuosn}) and Eq.\,(\ref{dimlessdef}). Since at one loop $e$ is smaller than unity 
for $T\sim T_P$ only, we expect TLH-mode quantum 
fluctuations to be absent at temperatures lower than $T_P$ also at 
higher-loop accuracy.

\section{The magnetic phase\label{MP}}

\subsection{The electric-magnetic phase transition\label{magPT}}

In Sec.\,\ref{CMP} we have discussed how stable and unstable BPS monopoles 
are generated as isolated objects in the electric phase. For definiteness we have assumed MGSB 
by the adjoint scalar $\phi$. The mass of the N/2 stable BPS monopole species is given as in 
Eq.\,(\ref{monopmass}) when replacing $\phi\to\tilde{\phi}_l\,,\ (l=1,\cdots,\mbox{N}/2)$. 
As a consequence of the evolution of the gauge 
coupling $e(\lambda_E)$ following from Eq.\,(\ref{eeq}) the mass of a stable 
monopole vanishes at $\lambda_E=\lambda_{E,c}$ due 
the logarithmic pole of $e$. Stable monopoles do not 
carry any Euclidean action at 
this point, and thus they condense. 

TLM modes, which couple to isolated 
monopoles with strength $g=\frac{4\pi}{e}$, become dual gauge bosons. They couple to the 
monopole {\sl condensates} with a strength $g$, which may now 
continuously vary with temperature, starting with $g=0$ at $\lambda_E=\lambda_{E,c}$. 
The TLH modes of the electric phase decouple kinematically at 
$\lambda_E=\lambda_{E,c}$ since their masses, $\propto e|\phi|$, diverge. 
At the onset of the {\sl magnetic} phase, 
where N/2 species of stable monopoles are condensed, we are thus left with an 
effective Abelian theory of $\mbox{N}-1$ dual gauge fields $a^D_{\mu,k}$, $(k=1,\cdots,\mbox{N}-1)$ 
and N/2 condensates of stable monopoles described by complex scalar fields 
$\varphi_l\,,\ (l=1,\cdots,\mbox{N}/2)$. The temporal winding of these fields is the 
same as that of the associated SU(2) blocks $\tilde{\phi}_l$ in the electric phase. 
For N=3 there are two independent condensates of stable monopoles. 

What happens to the N/2$-$1 independent unstable monopoles (N even, $\mbox{N}>3$? 
Unstable monopoles are generated by gluon exchanges 
between trivial-holonomy calorons in different SU(2) embeddings. We conclude, 
that at $\lambda_E=\lambda_{E,c}$ the intact continuous gauge symmetry 
U(1)$^{\tiny\mbox{N-1}}$ of the electric phase becomes a gauge symmetry 
U(1)$_D^{\tiny\mbox{N-1}}$ which is spontaneously 
broken as 
\eqb
\label{breaking}
\mbox{U}(1)_D^{\tiny\mbox{N-1}}\to \mbox{U}(1)_D^{\tiny\mbox{N/2-1}} 
\ (\mbox{N}>3)\,
\eqe
for $\lambda_E<\lambda_{E,c}$ in the magnetic phase. For $N=2,3$ the spontaneous 
breakdown of continuous gauge symmetry is maximal. The monopole condensate 
$\bar{\varphi}_k$, which is associated with the 
dual gauge-field fluctuation $\delta a^D_{\mu,k}$, is  
defined as
\eqb
\label{monopcondsit}
\bar{\varphi}_k=\left\{\begin{array}{c} \hspace{-1.1cm}\varphi_i\,,\ \ \ (k=1,\cdots,\mbox{N}/2)\,,\nonumber\\ 
 0\,,\ \ \ (k=\mbox{N}/2+1,\cdots,\mbox{N}-1)\,,\end{array}\right.\,.
\eqe
The local $Z_{\tiny\mbox{N,elec}}$ symmetry acts 
on $\delta a_{\mu,k}^D, \bar{\varphi}_k$ as a 
local-in-time permutation (see the discussion in Sec.\,\ref{CMP})
\eqb
\label{localcenter}
(\delta a^D_{\mu,k},\bar{\varphi}_k)\to 
(\delta a^D_{\mu,(k+j(\tau))\tiny{\mbox{mod\, (N-1)}}},\bar{\varphi}_{(k+j(\tau))
\tiny{\mbox{mod\, (N-1)}}})\,,\ \ \ \ (j\in \bf{Z})\,.
\eqe
In Eq.\,(\ref{localcenter}) the integer-valued functions $j$ are piecewise constant 
on extended regions of Euclidean spacetime. The symmetry defined in Eq.\,(\ref{localcenter}) 
leaves the ground state of the system invariant, and thus 
the discrete, local symmetry $Z_{\tiny\mbox{N},elec}$ is unbroken in 
the magnetic phase. As a consequence the {\sl global} 
$Z_{\tiny\mbox{N},elec}$ associated with the Polyakov loop as an 
order parameter is also unbroken, see also Sec.\,\ref{polyaloop}.

\subsection{Monopole condensates, macroscopically\label{windingmag}}

The effective theory describing the magnetic phase is 
constructed in close analogy to the effective theory 
describing the electric phase. Recall, that we assume 
MGSB in the electric phase. Since the condensation of monopoles is driven by their 
masslessness the complex scalar fields $\varphi_l$, 
which describe the monopole condensates, 
are energy- and pressure-free in the absence of 
monopole interactions mediated by dual gauge-field fluctuations in the topologically trivial sector 
of the theory. For the N=2 case the exponent of the phase of the local field $\varphi$ is defined as
\eqb
\label{defvarphiloc}
i\log\left[\frac{\varphi}{|\varphi|}\right]=\la \int d\Sigma_{\mu\nu} \tilde{G}_{\mu\nu}
\ra_{\tiny\mbox{z. m. of n.i. magn.
monop.}}\,.
\eqe
In Eq.\,(\ref{defvarphiloc}) the dual field strength $\tilde{G}_{\mu\nu}$ is the 't Hooft tensor of 
Eq.\,(\ref{tT}), the (surface-) integal is over a spatial 
2-sphere of infinite radius, and the average is over the zero-mode deformations of a 
noninteracting magnetic monopole. Again, Eq.\,(\ref{defvarphiloc}) defines a dimensionless 
entity in accord with the fact that the Yang-Mills scale is a parameter to be 
measured and not to be calculated. The right-hand side of Eq.\,(\ref{defvarphiloc}) measures the magnetic 
flux. If monopoles are point-like, that is, if 
they are massive, then the right-hand side of 
Eq.\,(\ref{defvarphiloc}) vanishes identically due to cancellation of ingoing and outgoing 
fluxes. If monopoles are massless (condensed), that is, if their charges are 
spread over the entire Universe, 
then this cancellation does not take place, see \cite{HerbstHofmann2004} for a more 
detailed investigation. It is clear that definition (\ref{defvarphiloc}) relies on 
definitions (\ref{locdefphi}) and (\ref{tT}). So when expressed in terms of 
fundamental caloron and topologically trivial fields it looks quite involved. 
No (nonlocally defined) lattice operator of this type has ever been constructed. 
One more point needs to be discussed: The phase in Eq.\,(\ref{defvarphiloc}) should be a function of Euclidean time $\tau$, 
see Eq.\,(\ref{BPSsolM}). From the definition of the 't Hooft tensor, Eq.\,(\ref{tT}), we 
see that such a time dependence manifests itself in terms a $\tau$ dependent 
angle in adjoint color space between 
the fields $\phi^a$ and $G_{\mu\nu}^a$. Deep in the electric phase, where monopoles are 
isolated defects, this angle is subject to a 
global gauge choice for the direction of winding of the field $\phi^a$.  In the magnetic phase, where 
monopoles are condensed, the global gauge choice in the electric phase 
is promoted to a {\sl local} gauge 
choice for the composite field $\varphi$. This situation is reminiscent of 
Kaluza-Klein like gemoetarical compactifications where 
global space-time symmetries along `extra' dimensions become gauge symmetries upon 
compactification \cite{Kaluza1921,Klein1926} within a low-energy formulation of the theory. 
This also seems to happens in the low-energy formulation of the Yang-Mills 
theory being in its magnetic phase.    

The fields $\varphi_l$ are 
energy- and pressure-free if and only if their Euclidean time dependence is BPS saturated. 
Moreover, the $\varphi_l$ must be periodic in time, and their gauge invariant 
modulus must not depend on spacetime. Since BPS monopoles have resonant 
excitations \cite{ForgasVolkov2003}, which are activated by the exchanges of 
dual gauge bosons, one expects the ground-state energy 
of the system and the tree-level mass spectrum of dual 
gauge-field fluctuations to be $T$ dependent. 
The local permutation symmetry discussed in Secs.\,\ref{CMP} and 
\ref{magPT} is respected by the effective potential 
$\tilde{V}_{M}(\varphi_1,\cdots,\varphi_{\tiny\mbox{N}/2})$ if it 
decomposes into a {\sl sum over potentials} $V_{M}(\varphi_l)$:
\eqb
\label{potmagn}
\tilde{V}_M(\varphi_1,\cdots,\varphi_{\tiny\mbox{N}/2})\equiv\sum_{l=1}^{\tiny\mbox{N}/2} 
V_{M}(\varphi_l)\,.
\eqe
The potential $V_{M}$ is uniquely determined by the above conditions. We have
\eqb
\label{singlepotM}
V_{M}(\varphi_l)\equiv\overline{v_{M}(\varphi_l)}v_{M}(\varphi_l)
\, \ \ \ \ \mbox{and} \ \ \ \ v_{M}(\varphi_l)=i\La_M^3/\varphi_l\,.
\eqe
In Eq.\,(\ref{singlepotM}) $\La_M$ denotes a mass scale which is 
related to the mass scale $\La_E$ in the electric phase by a matching 
condition, see Sec.\,\ref{MCC}. The effective action for the magnetic phase 
reads
\eqb
\label{effactM}
S_M=\int_0^{1/T}
d\tau\int d^3x\,\left[\frac{1}{4}\sum_{k=1}^{\tiny\mbox{N}-1}\,
\tilde{G}_{\mu\nu,k}\tilde{G}_{\mu\nu,k}+
\frac{1}{2}\left(\sum_{l=1}^{\tiny\mbox{N}/2}\overline{\tilde{{\cal D}}_{\mu,l}\varphi_l}
\tilde{{\cal D}}_{\mu,l}\varphi_l+\tilde{V}_M(\varphi_1,\cdots,\varphi_{\tiny\mbox{N}/2})\right)\right]\,.
\eqe
In Eq.\,(\ref{effactM}) $\tilde{G}_{\mu\nu,l}$ 
denotes the Abelian field strength of the dual field $a^D_{\mu,l}$, $\tilde{G}_{\mu\nu,l}\equiv 
\pd_\mu a^D_{\nu,l}-\pd_\nu a^D_{\mu,l}$, the covariant derivative is defined as 
$\tilde{{\cal D}}_{\mu,l}\equiv \pd_\mu+ig\,a^D_{\mu,l}$, and $g$ denotes 
the magnetic gauge coupling constant. One remark concerning the normalization of the kinetic 
terms for the fields $\varphi_l$ is in order. The ratio between the gauge kinetic term $\frac{1}{4}\,
\tilde{G}_{\mu\nu,l}\tilde{G}_{\mu\nu,l}$ and the kinetic term 
$\frac{1}{2}\,\tilde{{\cal D}}_{\mu,l}\varphi_l\tilde{{\cal D}}_{\mu,l}\varphi_l$ 
defines the mass spectrum of the gauge-field fluctuations $\delta a^D_{\mu,l}$ in unitary gauge. 
A redefinition of the factor in front of 
$\tilde{{\cal D}}_{\mu,l}\varphi_l\tilde{{\cal D}}_{\mu,l}\varphi_l$ (and $\tilde{V}_M$) 
changes this ratio. At the same time, however, the scale $\La_M$ is changed because the 
matching condition - equality of the pressure in the magnetic and electric phase at the phase boundary - 
is unchanged. 
The canonical normalization used in Eq.\,(\ref{effactM}) 
is thus nothing but a convention for defining the scale $\La_M$ in terms of $\La_E$.  

\noindent The solutions to the BPS equations
\eqb
\label{BPSM}
\pd_\tau \varphi_l=\bar{v}_{M}(\bar{\varphi}_l)
\eqe
read 
\eqb
\label{BPSsolM}
\varphi_l=\sqrt{\frac{\La_M^3}{2\pi T K(l)}}\exp[-2\pi i T K(l)\tau]
\eqe
where $K(l)$ is an integer. The condition of MGSB by 
minimal ground-state energy in the electric phase translated into  
\eqb
\label{maxsymbr}
K(l)=l\,.
\eqe
Since the $l^{\tiny\mbox{th}}$ stable monopole species in the electric phase is 
associated with zeros of the $l^{\tiny\mbox{th}}$ SU(2) block in $\phi$ the temporal 
winding of its {\sl condensate} $\varphi_l$ is accordingly. 
From Eqs.\,(\ref{BPSsolM}) and (\ref{maxsymbr}) we derive a potential 
\eqb
\label{potMsol}
1/2\,\tilde{V}_M=\frac{\pi\mbox{N(N+2)}}{8}\,T\Lambda_M^3 
\eqe
for even N. For N=3 we obtain
\eqb
\label{potMsolN3}
1/2\,\tilde{V}_M=\pi\,T\Lambda_M^3\,. 
\eqe
Again, statistical fluctuations of the fields $\varphi_l$ are negligible and quantum fluctuations 
are absent:
\eqb
\label{qsflphiM}
\frac{\pd^2_{|\varphi_l|} V_{M}(\varphi_l)}{T^2}=24\pi^2l^2\,,\ \ \ \ \ 
\frac{\pd^2_{|\varphi_l|} V_{M}(\varphi_l)}{|\varphi_l|^2}=6\,l^3\lambda_M^3\,.
\eqe
In Eq.\,(\ref{qsflphiM}) we have defined $\lambda_M\equiv \frac{2\pi T}{\La_M}$. For $\mbox{N}=2,3$ 
$\lambda_M$ is larger than unity throughout the magnetic phase, see Sec.\,\ref{TSCM}. 
Since the fields $\varphi_l$ do not fluctuate they are a background 
to the macroscopic gauge-field equations of
motion 
\eqb
\label{eomdualG}
\pd_\mu \tilde{G}_{\mu\nu,l}=ig\left[\overline{\tilde{{\cal D}}_{\nu,l}\varphi_l}\varphi_l-\bar{\varphi_l}
\tilde{{\cal D}}_{\nu,l}\varphi_l\right]\,.
\eqe
There exist pure-gauge solutions to Eq.\,(\ref{eomdualG}) given as
\eqb
\label{pgsolM}
a^{D,bg}_{\mu,l}=\delta_{\mu 4}\frac{2\pi l}{g}\,T\,.
\eqe
Again, a macroscopic `holonomy' is created by interacting monopoles which can be 
related to the existence of isolated loops of magnetic flux: ANO vortices. 
We have $\tilde{{\cal D}}_{\nu,l}\varphi_l=0$. 
As a consequence, the action in Eq.\,(\ref{effactM}) reduces to the potential term 
on the ground-state solutions $\varphi_l, a^{D,bg}_{\mu,l}$. This results in a shift of 
the ground-state energy density and pressure to $\pm 1/2\,\tilde{V}_M$, respectively.

\noindent Nonperiodic gauge functions  
\eqb
\label{grM}
\theta_l=2\pi l T\tau
\eqe
transform each pair $\varphi_l,\, a^{D,bg}_{\mu,l}$ to unitary gauge
\eqb
\label{unitgaugeM}
\varphi_l=|\varphi_l|\,,\ \ \ \ \  a^{D,bg}_{\mu,l}=0\,.
\eqe
In analogy to the electric phase one shows that 
the gauge rotations $\Omega_l=e^{\theta_l}$ leave intact the 
periodicity of the fluctuations $\delta a^D_{\mu,l}$. These gauge rotations thus 
do not change the physics upon integrating out the fields $\delta a^D_{\mu,l}$ 
at one loop. Due to the effective theory being Abelian and the monopole condensates $\varphi_l$ 
inert the one-loop calculation is exact. There is, however, a pronounced difference 
to the electric phase. In winding as well as in unitary gauge the 
Polyakov loop evaluated on each of the background fields $a^{D,bg}_{\mu,l}$ 
is {\sl unity}. We conclude that the ground state of the system is unique, 
much in contast to the electric phase, and thus that the global 
$Z_{\tiny\mbox{N,elec}}$ symmetry is {\sl restored}. For a discussion of the 
full Polyakov loop see Sec.\,\ref{polyaloop}. Hence the magnetic phase confines 
fundamental test charges at the same time as it allows for the propagation of 
massive, Abelian gauge modes!

\subsection{Gauge-field excitations and \\ thermodynamical self-consistency\label{TSCM}}

The Abelian Higgs mechanism generates a tree-level
mass spectrum for the fluctuations $\delta a^D_{\mu,l}$. It is given as
\eqb
\label{massspecM}
m_l=g|\varphi_l|\equiv a_l\,T\,.
\eqe
Due to the noncondensation of unstable monopoles there are N/2$-1$ 
dual gauge field fluctuations 
\eqb
\label{masslessM}
\delta a^D_{\mu,i}\,,\ \ (i=\mbox{N}/2+1,\cdots,\mbox{N})\,
\eqe
which are massless. In analogy to the electric phase we derive 
an evolution equation for temperature as a function of mass from the 
requirement of thermodynamical self-consistency, $\pd_a P=0$ 
(for the definition of $a$ in the magnetic phase
see Eq.\,(\ref{defM})). Notice that the 
one-loop expression for thermodynamical quantities and the tree-level masses of
the dual gauge boson fluctuations $\delta a^D_{\mu,k}\,,\ (k=1,\mbox{N}-1)$ are exact due to 
the effective theory being Abelian. We obtain  
\eqb
\label{evolMeven}
\pd_a\lambda_M=-\frac{96}{(2\pi)^6 \mbox{N(N+2)}}\lambda_M^4\, a
\sum_{l=1}^{\tiny{\mbox{N/2}}}c_l^2 D(a_l)\,,
\ \ \ (\mbox{N\ \ even})\,.
\eqe
For N=3 we have
\eqb
\label{evolM3}
\pd_a\lambda_M=-\frac{12}{(2\pi)^6}\lambda_M^4\, a D(a)\,.
\eqe
In Eqs.\,(\ref{evolMeven}) and (\ref{evolM3}) we have defined:
\eab
\label{defM}
c_l&\equiv&\frac{1}{\sqrt{l}}\,,\ \ \ \ 
\lambda_M\equiv\frac{2\pi T}{\La_M}\,,\nonumber\\  
a&\equiv& \frac{g}{T}|\varphi_1|=2\pi\,g\,\lambda_M^{-3/2}\,,\ \ \ \ a_l\equiv c_l a\,.
\eae
In deriving Eqs.\,(\ref{evolMeven}) and (\ref{evolM3}) 
we have neglected the `nonthermal' contribution $-\Delta \tilde{V}_M$ to the pressure which is 
very small for sufficiently small N (quantum fluctuations are cut 
off at the compositeness scales $|\varphi_l|$):
\eqb
\label{DeltaVM}
\left|\frac{\Delta \tilde{V}_M}{\tilde{V}_M}\right|<
3\,\frac{\mbox{N}-1}{128\pi^2}\left(\frac{|\varphi_1|}{\La_M}\right)^6\sim 
3\,\frac{\mbox{N}-1}{128\pi^2}\,\lambda_M^{-3}\,.
\eqe
The function $D(a)$ is defined in Eq.\,(\ref{DA}). The evolution equations 
(\ref{evolMeven}) and (\ref{evolM3}) have fixed points at $a=0,\infty$ which correspond to 
the highest and lowest attainable temperatures in the magnetic phase, 
$\lambda_{E,c}$ and $\lambda_{M,c}$, respectively. 

\begin{figure}
\begin{center}
\leavevmode
\leavevmode
\vspace{4.5cm}
\includegraphics{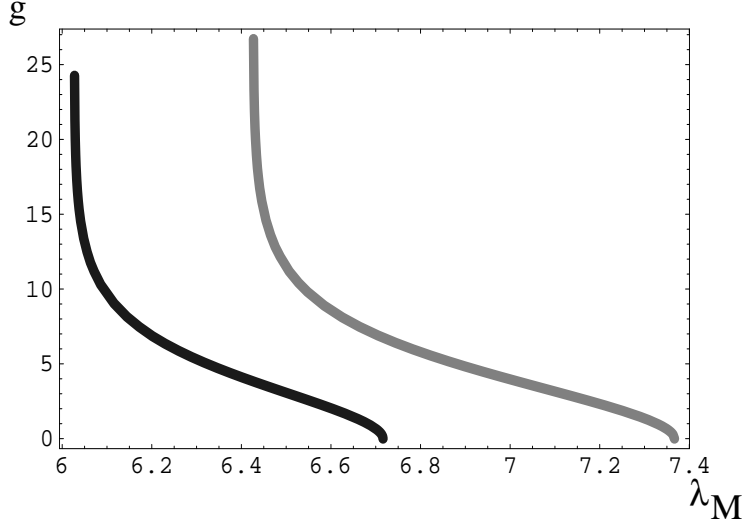}
\end{center}
\caption{The evolution of the gauge 
coupling constant $g$ in the magnetic phase for 
N=2 (thick grey line) and N=3 (thick black line). The gauge coupling constant 
diverges logarithmically, $g\propto -\log(\lambda_{M}-\lambda_{M,c})$, at 
$\lambda_{M,c}=6.41$ (N=2) and $\lambda_{M,c}=5.82$ (N=3). \label{gevol}}      
\end{figure}

The $\lambda_M$ dependence of the gauge coupling constants $g$ 
is obtained by inverting the solutions to Eqs. (\ref{evolMeven}) and 
(\ref{evolM3}) and using the relation between $g$, $\lambda_M$ and $a$ in 
Eq.\,(\ref{defM}) afterwards. Results for N=2,3 are shown in Fig.\,\ref{gevol}.
The gauge coupling constant $g$ increases continuously from $g=0$ at the electric-magnetic 
phase boundary ($\lambda_M=\lambda_{E,c}$) until it diverges logarithmically at $\lambda_{M,c}$. 
A continuous behavior of the magnetic 
coupling is not in contradiction with magnetic charge conservation 
since no isolated magnetic charges appear 
in the magnetic phase: magnetic monopoles are only present in condensed 
form, and there are no collective monopole excitations, see Eq.\,(\ref{qsflphiM}). The 
continuous rise of $g$, starting from zero at $\lambda_{M,c}$, implies a continuous behavior of the 
mass parameter $a$ in Eq.\,(\ref{defM}). Since $a$ is the measurable order parameter 
for monopole condensation this situation is reminiscent of a 
2$^{\tiny\mbox{nd}}$ order phase transition. We compute the critical exponent of this 
transition for N=2 in Sec.\,\ref{critexpo}.

Lowering $\lambda_M$ towards $\lambda_{M,c}$ the core size of ANO vortices 
becomes large, the monopole condensates are more and 
more distorted by magnetic flux lines, and thus it becomes increasingly 
irrelevant in what SU(2) embedding a particular monopole lives. Formerly 
unstable monopoles acquire longevity and thus additional monopole 
condensates form at $\lambda_{M,c}$: $\bar{\varphi}_i\not=0\,,\ (i=\mbox{N}/2+1,\cdots,\mbox{N}-1)$ for 
$\lambda_M\sim \lambda_{c,M}$. 
As a consequence, {\sl all} dual gauge-boson fluctuations $\delta a^D_{\mu,k}\,,\ (k=1,\cdots,\mbox{N}-1)$ 
are very massive close to $\lambda_{M,c}$ and decouple thermodynamically at $\lambda_{M,c}$. 
The equation of state at $\lambda_M=\lambda_{c,M}$ thus is 
\eqb
\label{eosMc}
\rho(\lambda_{M,c})=-P(\lambda_{M,c})\,.
\eqe
At $\lambda_{M,c}$, where all dual gauge-field fluctuations are very massive, the continuous 
dual gauge symmetry $U(1)_D^{\tiny\mbox{N-1}}$ is broken 
completely.

\subsection{Polyakov loop in the electric and the magnetic phase\label{polyaloop}}

In this section we show that the Polyakov loop, which is an order parameter 
for the deconfining transition, indeed is finite in the electric 
and close to zero in the magnetic phase. In each phase the Polyakov loop of the full 
effective theory (now we also consider the fluctuations $\delta a_\rho$) 
formulated in Euclidean spacetime and unitary gauge is defined as
\eab
\label{Ployeff}
{\bf P}&=&Z^{-1}\times\exp[-S_{cl}]\times
\int {\cal D}\delta b_\rho\,\exp\left[-i\gamma\int_0^{T^{-1}} d\tau\,\delta b_4^{b.g.}\right]\times\nonumber\\ 
& &\exp\left[-\int_0^{T^{-1}}d\tau d^3x\left\{\frac{1}{4}G^2[\delta b_\rho]+
\sum_k m^2_k\,\delta b_{\mu,k}\delta b_{\mu,k}\right\}\right]\,.
\eae
where $\gamma=\{e,g\}$ and $\delta b_\rho=\{\delta a_\rho,\delta a^D_\rho\}$ 
depending on whether we discuss the electric or the magnetic phase. The term 
$\exp[-S_{cl}]$ refers to the vanishing weight of the ground state in 
the partition function $Z$ in either phase. This weight, however, cancels in expectation 
values. 
\begin{figure}
\begin{center}
\leavevmode
\leavevmode
\vspace{3.5cm}
\includegraphics{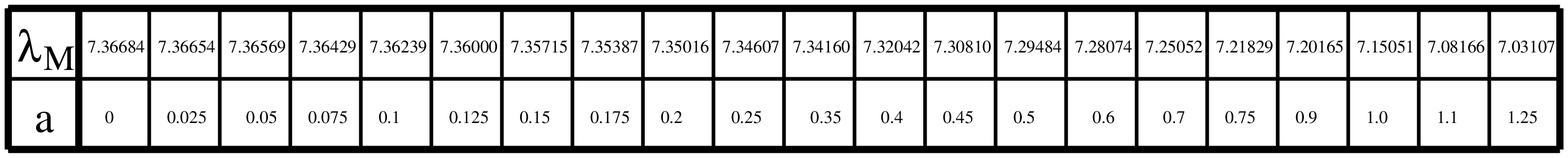}
\end{center}
\caption{The data set used for the fit of the critical exponent $\nu$.\label{nudata}}      
\end{figure}
Since the fluctuations $\delta b_\rho$ 
are periodic in time they are decomposed as
\eqb
\label{fourierb}
\delta b_\rho(\tau,{\vec x})=\sum_{n=-\infty}^{n=\infty} 
\exp\left[2\pi in \frac{\tau}{T}\right]\bar{b}_{\rho,n}({\vec x})\,.
\eqe
Modes with $n\not=0$ make the Polyakov-loop 
phase in Eq.\,(\ref{Ployeff}) unity, are action-suppressed, 
and thus they are irrelevant. Zero modes ($n=0$) make a contribution 
to the Polyakov-loop phase if they are not action-suppressed. This is the case 
if both of the following conditions are met: (i) there is no mass term for these 
fluctuations and (ii) we have $\pd_i\bar{b}_{\rho,0}({\vec x})\sim 0$ where $\pd_i$ 
denotes a spatial derivative. In the electric 
phase TLM modes are massless and thus their space-indepent zero-mode fluctuations 
generate the bulk of the (finite) Polyakov loop. For N=2,3 there are no massless 
fluctuations in the magnetic phase and thus condition (i) is violated. As a consequence, none of 
the fluctuations $\delta b_\rho$ can contribute 
to the Polyakov loop in a substantial way in the magnetic phase and thus we have 
${\bf P}\sim 0$. For N$>$3 the presence of unstable magnetic monopoles in the 
magnetic phase prevents some of the TLM modes to pick up a mass by the Abelian 
Higgs mechanism. Thus the Polyakov loop should be small but nonvanishing in 
the magnetic phase.    

\subsection{Critical exponent for the SU(2) electric-magnetic transition\label{critexpo}}

In this section we compute the critical exponent $\nu$ for the 
electric-magnetic transition for N=2. The obvious order parameter for this transition is 
the `photon' mass. 

The data set in Fig.\,\ref{nudata} is generated from an inversion of the numerical solution to the 
evolution equation Eq.\,(\ref{evolMeven}). The following model is used to fit the data
\eqb
\label{modeldata}
a(\lambda_M)=C\times |\lambda_M-\lambda_{M,em}|^\nu
\eqe
where $C$ and $\nu$ are constants, $\lambda_{M,em}$ denotes the critical 
temperature $T_{e,c}$ in units of $\frac{\Lambda_M}{2\pi}$, and $a$ is the dimensionless `photon' mass. 
Recall, that the value $\lambda_{E,c}=11.65$ is obtained from the position of the logarithmic pole 
of the coupling constant $e$ in the {\sl electric phase}. By matching the pressure at 
the electric-magnetic phase boundary, see Sec.\ref{MCC}, this translates into a 
value $\lambda_{M,em}=7.337$.  

\begin{figure}
\begin{center}
\leavevmode
\leavevmode
\vspace{3.5cm}
\includegraphics{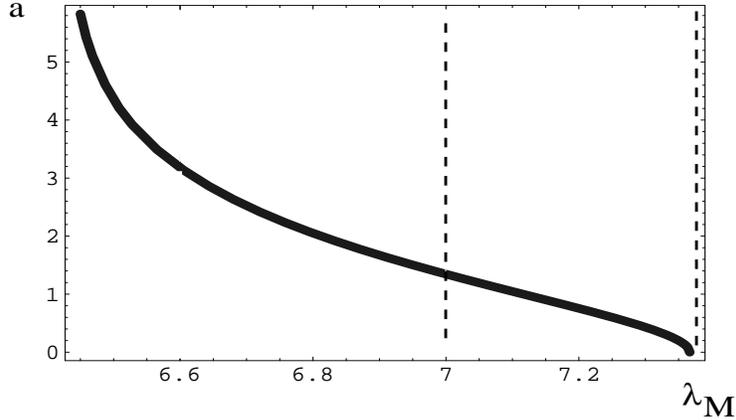}
\end{center}
\caption{The function $a(\lambda_M)$ in the vicinity of the electric-magnetic phase transition. 
The region between the dashed vertical lines corresponds to the data set in Tab.\,1 which generates the 
least sensitivity of $\nu$ on the length of the 
fitting interval $\Delta=|\lambda_{M,\tiny\mbox{min,fit}}-\lambda_{M,em}|$.\label{shapeFIT}}      
\end{figure}

\begin{figure}
\begin{center}
\leavevmode
\leavevmode
\vspace{3.5cm}
\includegraphics{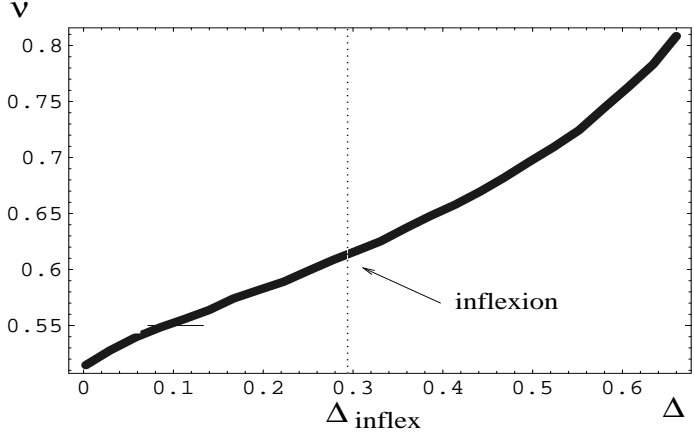}
\end{center}
\caption{Dependence of the critical exponent $\nu$ on the length of the fitting 
interval $\Delta=|\lambda_{M,\tiny\mbox{min}}-\lambda_{M,c}|$. The inflexion point 
of the curve is at $\Delta_{\tiny\mbox{inflex}}=
0.29\pm0.05$ corresponding to $\nu=0.61+0.02-0.01$.\label{inflexionpoint}}      
\end{figure}

To perform the actual fit we 
have used Mathematica's NonlinearFit function which is contained in the 
statistics package. In Fig.\,\ref{shapeFIT} the 
critical behavior of the `photon' mass $a$ is shown.
To determine the length of the fitting interval 
$\Delta=|\lambda_{M,\tiny\mbox{min,fit}}-\lambda_{M,em}|$ where $\nu$ is 
least sensitive to changes in $\lambda_{M,\tiny\mbox{min,fit}}$ we numerically 
determine the inflexion point $\Delta_{\tiny\mbox{inflex}}$ of the function $\nu=\nu(\Delta)$, 
see Fig.\,\ref{inflexionpoint}. We obtain 
\eqb
\label{inflexpo}
\Delta_{\tiny\mbox{inflex}}=0.29\pm0.05\,.
\eqe
We emphasize that this interval is well contained in the fitting 
interval used in \cite{deForcrandEliaPepe2001} where the critical exponent was determined 
from the expectation of the dual string tension. Their fit interval $0\le t\le 1$ with 
$t\equiv\frac{T_{M,em}-T}{T_{M,em}}$ correponds to an interval length 
$\Delta=\lambda_{M,em}\sim 7.34$!  

The interval of least senstivity $\Delta_{\tiny\mbox{inflex}}=0.29\pm0.05$ 
translates into 
\eqb
\label{nuFITst}
\nu=0.61+0.02-0.01\,.
\eqe
Alternatively, we can determine $\nu$ by a fit in 
very small intervals around $\lambda_{M,em}$. In this case we obtain the result expected from a naive
mean-field analysis, $\nu\to 0.5$. However, the fitted 
prefactor $C$ varies considerably for very small intervals around $\lambda_{M,em}$. 
It is worth mentioning at this point that the `would-be' critical 
exponent for N=3 has at least 
two inflexion points as a function of $\Delta$. This makes a unique determination 
of the physical value of $\nu$ 
impossible and clearly indicates that the phase transition is not second order anymore. A small latent 
heat is associated with the electric-magnetic transition for N=3, see Fig.\,\ref{rho}, 
which makes it 
weakly first order. This is seen on the lattice \cite{LuciniTeperWenger2003}.   

The critical exponent for the 3D Ising 
model (same universality class as SU(2) Yang-Mills \cite{SvetitskyYaffe1982-1,SvetitskyYaffe1982-2}) 
is $\nu_{\tiny\mbox{Ising}}\sim 0.63$. As it seems, the effective theories for the electric and the 
magnetic phases have passed an important test!

\section{The center phase\label{CVCM}} 

\subsection{Isolated center vortices in the magnetic phase} 

Away from the points $\lambda_{E,c}$ and $\lambda_{M,c}$ there 
are isolated vortices in the magnetic 
phase which form closed loops due to the 
conservation of magnetic flux. Along the 
core region of a vortex, where the monopole condensate vanishes, 
$\bar{\varphi}_k\approx 0$, monopoles and antimonopoles form a closed 
chain and move into opposite directions \cite{Olejnik1997}. 
Along a vortex loop the magnetic flux is $2\pi\,k/N\,,\ (k=1,\cdots,\mbox{N}-1)$ 
with respect to the dual gauge field $a^D_{\mu,k}$. This coins the 
name center vortex loop.
The typical action $S_{\tiny\mbox{ANO}}$ and energy $E_{\tiny\mbox{ANO}}$ of a center vortex loop can be 
estimated since an analytical solution to the Abelian Higgs 
model is known \cite{NielsenOlesen1973}:
\eqb
\label{ANOana}
S_{\tiny\mbox{ANO}}\sim\frac{1}{g^2}\,,\ \ \ \ \ \ \ E_{\tiny\mbox{ANO}}\propto \frac{1}{g}\,.
\eqe
As a consequence of Eq.\,(\ref{ANOana}) center vortex 
loops do not carry action and are 
massless at $\lambda_{M,c}$ where $g$ 
diverges. They condense to form a new ground state of 
the system: a phase transition takes place. As we will show below this phase 
transition is nonthermal and of the Hagedorn type.

\subsection{Center vortex condensates, macroscopically}

Since center vortex loops are extended, one-dimensional 
objects the local scalar fields $\Phi_k(x)\,,\ (k=1,\cdots,\mbox{N}-1)$ 
describing their respective condensates have to be 
defined in a nonlocal way. We formally define the fields $\Phi_k(x)$ 
in terms of an average over the dual Abelian gauge 
fields $A^D_{\mu,k}$ of the magnetic phase 
(the part belonging to an ANO or center vortex is included in $A^D_{\mu,k}$!) as
\eqb
\label{Wilsonloopx}
\frac{\Phi_k(x)}{|\Phi_k(x)|}=\la \exp\left[ig\oint dz_\mu A^D_{\mu,k}\right]\ra_{A^D_{\mu,k}}\,.
\eqe
In Eq.\,(\ref{Wilsonloopx}) the integration contour is spatial and circular, 
its center is at $x$, and circle's diameter is infinite. In absence of a Yang-Mills scale 
$\Lambda_C$, which is a relevant situation 
when investigating the ground-state structure in the center phase due 
to the missing gauge-mode propagation, only dimensionless quantities 
like in Eq.\,(\ref{Wilsonloopx}) can be computed. As we shall see below, the presence of 
$\Lambda_C$ can only be observed in the excitation spectrum (selfintersections of center-vortices). 
We may always chose a parametrization of the integration 
contour $z(\xi)$ in Eq.\,(\ref{Wilsonloopx}) 
such that $0\le\xi\le 1$. 

A (local) magnetic center rotation can be expressed as
\eqb
\label{centerrot}
\Omega(z)=\exp\left[\frac{2\pi i}{\mbox{N}} \sum_{k=1}^{\tiny\mbox{N-1}}\chi_k(z)\right]\,
\eqe
where $\chi_k(z)$ is either $k$ or zero. The dual gauge field $A^D_{\mu,k}$ transforms 
under $\Omega(z)$ as
\eqb
\label{trafocenterrot}
A^D_{\mu,k}\to A^D_{\mu,k}-g^{-1}\frac{2\pi}{\mbox{N}}\pd_\mu \chi_k(z)\,.
\eqe
\begin{figure}
\begin{center}
\leavevmode
\leavevmode
\vspace{4.5cm}
\includegraphics{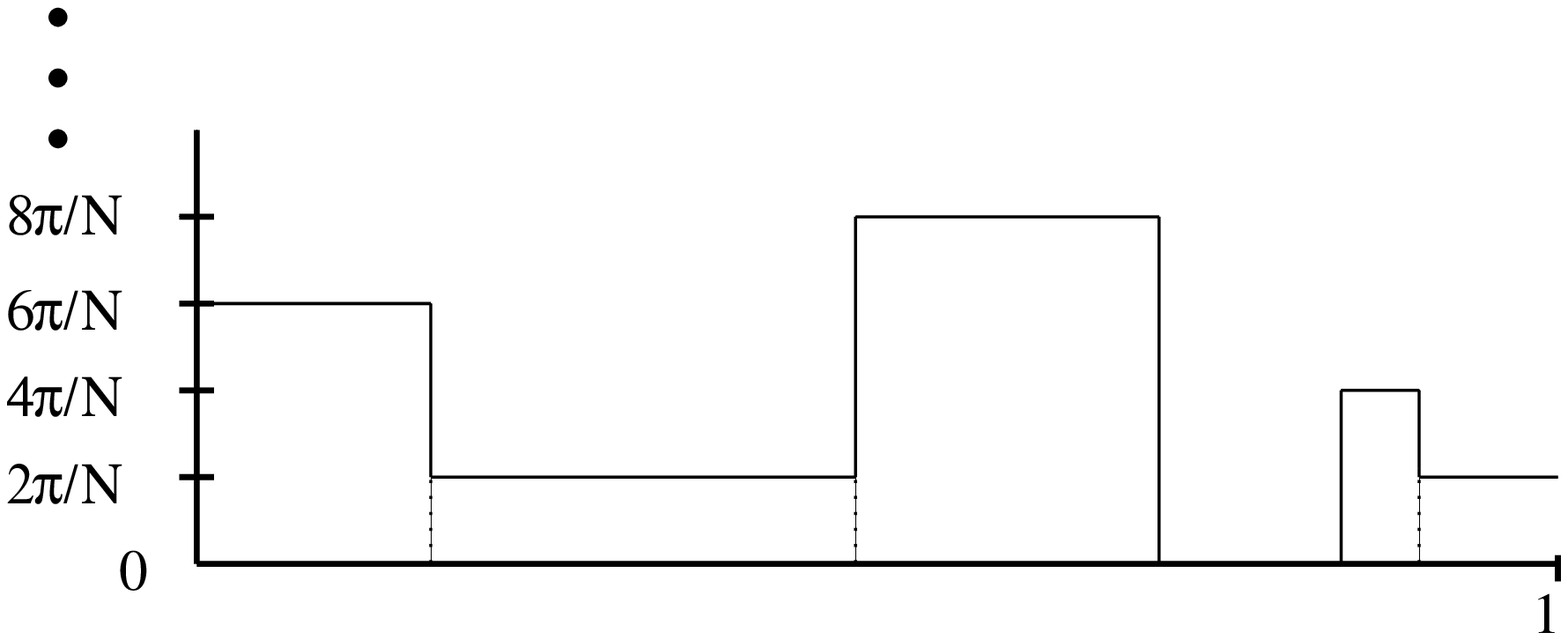}
\end{center}
\caption{The phase of a local magnetic center transformation along a 
circular contour parametrized by $0\le \xi\le 1$ and its 
decomposition into boxes. \label{locCP}}      
\end{figure}
According to the definition (\ref{Wilsonloopx}) a 
magnetic center rotation $\Omega(z)$ may locally add a 
magnetic flux quantum \cite{'tHooft1978} to the flux contained 
in $\Phi_k$. This is possible since we may have 
$\Omega(z(0))\not=\Omega(z(1))$. The action of the local magnetic 
center rotation $\Omega(z)$ on the complex field $\Phi_k(x)$ is as follows:
\eqb
\label{centerphi}
\Phi_k(x)\to\left\{\begin{array}{c}\exp[\pm\frac{2\pi i k}{\tiny\mbox{N}}]\,\Phi_k(x),\ \ \ 
(|\chi_k(z(1))-\chi_k(z(0))|=k)\,,\\ 
\,\,\,\,\,\Phi_k(x)\,,\ \ \ \ \ \ \ \ \ \ \ \ \ \ \ \mbox{otherwise} \end{array}\right.\,.
\eqe
Obviously, the action of 
$\Omega(z)$ on $\Phi_k(x)$ may locally change the phase of the field $\Phi_k$, 
and thus the ground state is not invariant under local magnetic center 
transformations: $Z_{\mbox{N,mag}}$ as a discrete gauge symmetry is spontaneously 
broken in the center phase. 

The local center transformation $\Omega$ singles out possible 
`boundaries of the circle' at the positions $z(\xi_n)$, ($\xi_0=0,\cdots$),  
where it jumps, see Fig.\,\ref{locCP}. The size and direction of an 
$\Omega$-induced magnetic flux quantum - an observable quantity - 
should not depend on the admissible re-parametrizations 
$\xi(\zeta)$, $(0\le\zeta\le 1)$ with $\xi(0)\in\{\xi_n\}$. 
For example, a translation 
\eqb
\label{zeta}
\xi(\zeta)=\zeta+\xi_1
\eqe
would have shifted the `boundary of the circle' at $z(\xi=0)$ to $z(\xi=\xi_1)$. Thus 
it would have generated a different flux quantum. To avoid such an ambiguity 
we have to impose that a 
reparametrization of the circle is compensated for by an according local 
permutation of the fields $\Phi_k(x)$: if flux quanta $\frac{2\pi k}{\mbox{N}}$ and 
$\frac{2\pi l}{\mbox{N}}$, ($l\not=k$), 
were generated by $\Omega$ in the parameterizations 
$\xi$ and $\zeta$, respectively, then a 
permutation with $\Phi_l\to\Phi_k$ needs to be 
performed {\sl after} the reparametrization $\zeta$ was implemented. 
This yields the same flux quantum as in parametrization 
$\xi$. Now, the discontinuous phase 
change in Eq.\,(\ref{centerphi}) is 
provided by the {\sl dynamics} in an effective theory 
and is not externally imposed. As a consequence, 
the demand for invariance of a generated flux quantum 
under admissible contour-reparametrizations translates 
into an invariance of the effective Lagrangian under local permutations 
of the set $\{\Phi_1(x),\cdots,\Phi_{\tiny\mbox{N-1}}(x)\}$. The following Lagrangian 
satisfies this requirement:
\eqb
\label{actcenter}
{\cal L}_C=\frac{1}{2}\sum_{k=1}^{\tiny\mbox{N-1}}
\left[\pd_\mu\Phi_k \pd_\mu\Phi_k+V_C(\Phi_k)\right]\,.
\eqe
It is clear from Eq.\,(\ref{actcenter}) that the transformation 
in Eq.\,(\ref{centerphi}) is only a symmetry of the potential term 
due to the noninvariance of the kinetic terms under the jumps in the fields $\Phi_k$. 
However, because of the conservation of magnetic flux (only closed loops of center vortex 
flux can be generated) one quantum of center flux created by a jump at one point in spacetime
is compensated by an opposite quantum of 
center flux created by the opposite jump at another point. 
The spacetime integral over ${\cal L}_C$, the action, 
is therefore invariant under local center rotations which induce magnetic 
flux in a closed flux line. Based on the continuum Lagrangian 
in Eq.\,(\ref{actcenter}) it would be interesting to perform a 
lattice simulation of the transition to the center phase taking any member of the set $\{|\Phi_k|\}$ as an 
order parameter. 
\begin{figure}
\begin{center}
\leavevmode
\leavevmode
\vspace{6.5cm}
\includegraphics{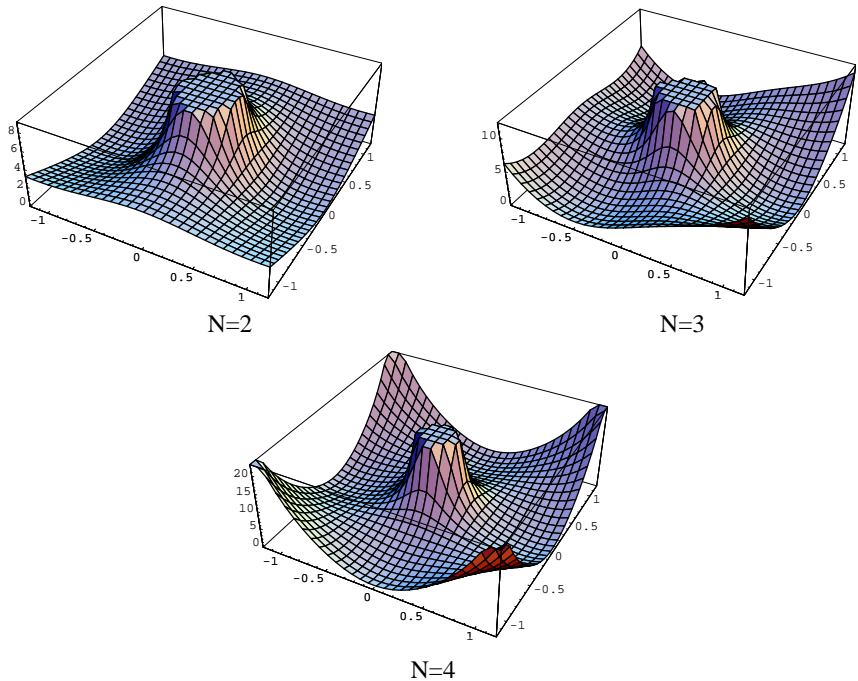}
\end{center}
\caption{The potential $V_C=\overline{v_C(\Phi)}v_C(\Phi)$ corresponding 
to the definition in Eq.\,(\ref{vCright}) 
for N=2,3,4 and $\La_C=\Lambda^\prime_C$. 
$|\Phi|$ is given in units of $\La_C$ and $V_C$ in units of $\La_C^4$. Notice the minima $V_C=0$ at 
the N$^{\tiny\mbox{th}}$ unit roots. \label{pot234}}      
\end{figure}
The function $V_C(\Phi)$ in Eq.\,(\ref{actcenter}) can be written as 
$V_C(\Phi)=\overline{v_C(\Phi)}v_C(\Phi)$. Let us now discuss the properties of 
the function $v_C$. If a matching to the magnetic phase would 
take place in thermodynamical equilibrium and a classical treatment of the ground-state dynamics could be 
justified at this point then the Euclidean time 
dependence of the (periodic) fields $\Phi_k$ would have to be BPS
saturated. Only in this case do the fields $\Phi_k$ describe the condensates of 
zero-energy center-vortex loops \footnote{Recall, that all gauge-boson fluctuations are 
decoupled in the center
phase. As a consequence, interaction between center vortices are extremely local.}. 

At first sight a candidate for $v_C$ would 
be $v_{C,\tiny\mbox{trial}}(\Phi_k)=i\Lambda_C^3/\Phi_k$. The potential $V_C$ would then not only 
be $Z_{\tiny\mbox{N,mag}}$ but also $U(1)^{\tiny\mbox{N-1}}$ symmetric. The 
latter (global) symmetry, however, does not exist in SU(N) Yang-Mills theory 
and thus $v_{C,\tiny\mbox{trial}}$ is excluded.

\begin{figure}
\begin{center}
\leavevmode
\leavevmode
\vspace{4.5cm}
\includegraphics{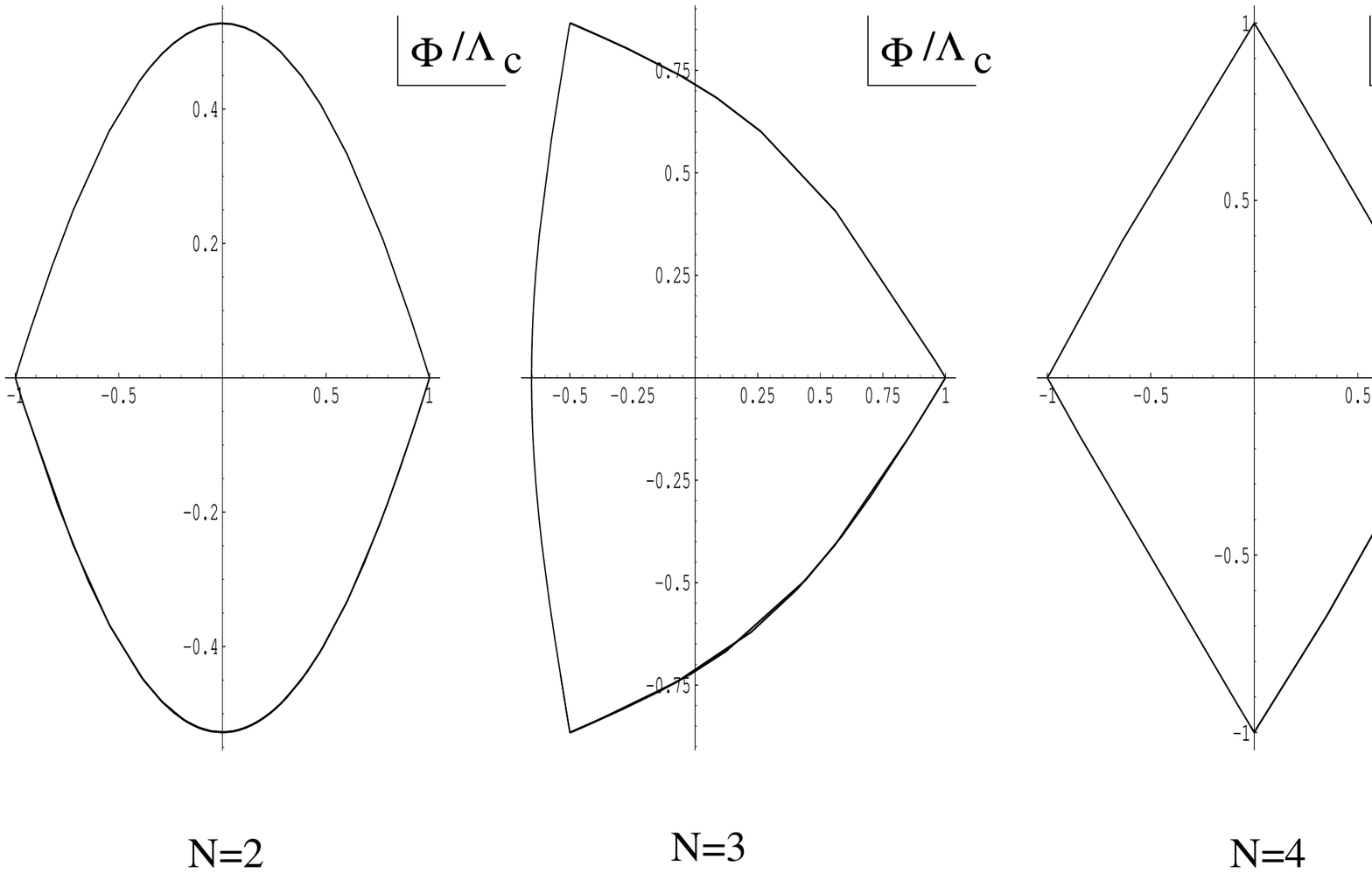}
\end{center}
\caption{Numerical solutions to the BPS equation $\pd_\tau\Phi =\overline{v_C(\Phi)}$ 
for N=2,3,4. \label{H2000}}      
\end{figure}

A function $v_C(\Phi_k)$, which is covariant {\sl only} 
under $Z_{\tiny\mbox{N}}$ transformations and, at the same time, allows for periodic and 
BPS saturated solutions along the Euclidean time coordinate $\tau$ \cite{Hofmann2002}, 
see Fig.\,\ref{H2000},
\footnote{The latter requirement derives from a consideration of 
the limit $\mbox{N}\to\infty$ where these solutions are of 
physical relevance, see below.}, is uniquely given as
\eqb
\label{vCright}
v_C(\Phi_k)=i\left(\frac{\Lambda_C^3}{\Phi_k}-\frac{\Phi_k^{\tiny\mbox{N-1}}}
{(\Lambda^\prime_C)^{\tiny\mbox{N-3}}}\right)\,
\eqe
where $\Lambda_C$ and $\Lambda^\prime_C$ denote mass scales that are a priori independent. 
The N degenerate minima of the potential $V_C(\Phi_k)$ are at
\eqb
\label{minPot}
|\Phi^{\tiny\mbox{min}}_k|=\left[\La_C^3 (\Lambda^\prime_C)^{\tiny\mbox{N-3}}\right]^{1/\tiny\mbox{N}}\,.
\eqe
At these minima the potential $V_C$ {\sl vanishes}. Periodic solutions to the BPS 
equations
\eqb
\label{PBScenter}
\pd_\tau\Phi_k=\bar{v_C(\Phi_k)}
\eqe
are parameterized by a winding number $\in\bf{Z}$ in analogy to the situation in the magnetic phase. 
The fields $\Phi_l\,,\ \ (l=1,\cdots,\mbox{N}/2)$, which 
are associated with vortices formed from the stable dipoles of the electric phase, are 
winding accordingly. For the fields $\Phi_i\,,\ \ (i=\mbox{N}/2+1,\cdots,\mbox{N}-1)$, being associated 
with vortices formed from independent monopoles, which are unstable in the electric phase, no winding numbers can be 
derived. This uncertainty in assigning winding numbers for the fields $\Phi_i$ 
is reflected in the appearance of an additional mass scale $\Lambda^\prime_C$ in 
Eq.\,(\ref{vCright}). In the cases N=2,3, however, only {\sl stable}, independent monopoles exist, 
and thus we can set $\La_C=\Lambda^\prime_C$. 

\begin{figure}
\begin{center}
\leavevmode
\leavevmode
\vspace{4.5cm}
\includegraphics{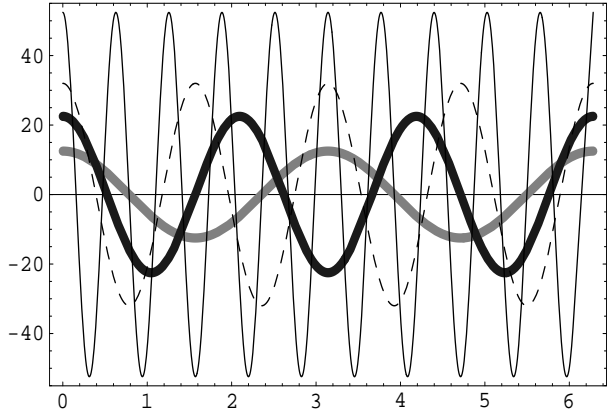}
\end{center}
\caption{The ratio $\left.\pd^2_{\theta_k} V_C(\Phi_k)/|\Phi_k|^2\right|_{|\Phi_k(0)|=0.8\La_C}$ ($V_C$ defined 
in Eq.\,(\ref{vCright}) and $\Lambda_C=\La_C^\prime$) as a function of $\frac{\theta_k}{\La_C}$ 
for N=2 (thick grey line), N=3 (thick black line), N=4 (dashed line), and N=10 
(thin solid line). \label{tac}}      
\end{figure}

In Fig.\,\ref{pot234} the 
graphs of the potential $V_C$ for N=2,3,4 and for $\La_C=\Lambda^\prime_C$ are shown. 
Notice the ridges and the valleys of 
negative and positive tangential curvature, respectively. 
Due to the BPS saturation a solution 
$\Phi_k$ to Eq.\,(\ref{PBScenter}) is guaranteed to carry no energy. This statement is, however, only then useful 
for the description of the ground state if the classical approximation can be 
justified. At finite N the modulus of BPS saturated solutions 
is no longer $\tau$ independent, see Fig.\,\ref{H2000}. 
On the one hand, this situation is in contradiction 
to thermal equilibrium. On the other hand, we can not trust the classical approximation 
since tangential fluctuations $\theta_k$, defined as
\eqb
\label{TanModes}
\Phi_k=|\Phi_k|\exp[i\frac{\theta_k}{\Lambda_C}]\,,
\eqe
can be tachyonic and therefore destabilize the classical solution to Eq.\,(\ref{PBScenter}), see 
Fig.\,\ref{tac}. In the electric and the magnetic phase tangential fluctuations of the caloron and the monopole
condensates are would-be Goldstone modes giving rise to longitudinal polarizations 
of gauge boson fluctuations. Due to the absence of a continuous gauge 
symmetry in the center phase no gauge-field fluctuations exist 
which could `eat' the tangential fluctuations. How to integrate out tachyonic tangential 
modes analytically is conceptually unclear. The only definite statement we 
can make at present is that they rapidly drive the expectation of the fields 
$\Phi_k$ towards the minima of $V_C$: $\Phi_k$ relaxes to one of the minima 
of $V_C$ along an outward directed spiral in the complex plane. During this process magnetic flux quanta 
are locally generated by discontinuous phase changes of $\Phi_k$ due the tunneling through the 
regions where the tangential fluctuations are tachyonic.   
Once the field $\Phi_k$ has settled into one of the minima 
of $V_C$ the situation is classical again. At the minima we have 
\eqb
\label{minimacur}
\left.\frac{\pd^2_{\theta_k} V_C(\Phi_k)}{|\Phi_k|^2}\right|_{\Phi^{\tiny\mbox{min}}_k}=
\left.\frac{\pd^2_{|\Phi_k|} V_C(\Phi_k)}
{|\Phi_k|^2}\right|_{\Phi^{\tiny\mbox{min}}_k}
=2\mbox{N}^2>1\,.
\eqe
The approximation $\La_C=\La_C^\prime$ used in Eq.\,(\ref{minimacur}) is exact for N=2,3. 
Since radial {\sl and} tangential quantum fluctuations are heavier than the 
compositeness scale $\Phi^{\tiny\mbox{min}}_k$ they are absent in the effective theory. 
As a consequence, the vanishing value of $V_C$ at the minima receives {\sl no} 
radiative corrections.

In the limit $\mbox{N}\to\infty$ only the pole term of the potential 
$V_C$ survives for $|\Phi_k|<|\Phi^{\tiny\mbox{min}}_k|$ and 
thus we recover the situation of a global 
$U(1)^{\tiny\mbox{N}-1}$ symmetry. The $\tau$ dependence of the solutions to the BPS equation 
(\ref{PBScenter}) is then a pure phase and no 
tachyonic but only massless tangential fluctuations exist. These fluctuations 
lead to an instantaneous reheating at the phase boundary. The order parameter $|\Phi_k|$ jumps 
from zero to a finite value across the phase boundary. While the energy density of the 
ground-state jumps to zero at the phase boundary (BPS saturation), 
the ground-state pressure jumps to zero only after 
$|\Phi_k|=|\Phi^{\tiny\mbox{min}}_k|$ is reached. 

At finite N a flux quantum and a radial displacement $\Delta|\Phi_k|$ 
are generated in each tunneling process. This reduces the ground-state energy 
density locally and brings the field $\Phi_k$ closer to one of the 
minima of $V_C$. The closer $\Phi_k$ to a minimum the less likely a 
tunneling process since the 
associated energy gain is quickly reduced ($|\pd_{|\Phi_k|}V_C|$ decreases!). 
If the number of tunneling processes taking place until 
relaxation would be roughly independent of N then 
for small N many unit-root directions will be reached in a relaxation process 
and thus it is likely that the 
average value of $\Phi_k$ close to the phase boundary is 
close to zero. For large N only a small angular sector 
would be reached by tunneling and thus close to the phase 
boundary the average value of $\Phi_k$ is not close to zero: a strong discontinuity of $\Phi_k$ across the 
transition should be observed.

\begin{figure}
\begin{center}
\leavevmode
\leavevmode
\vspace{4.5cm}
\includegraphics{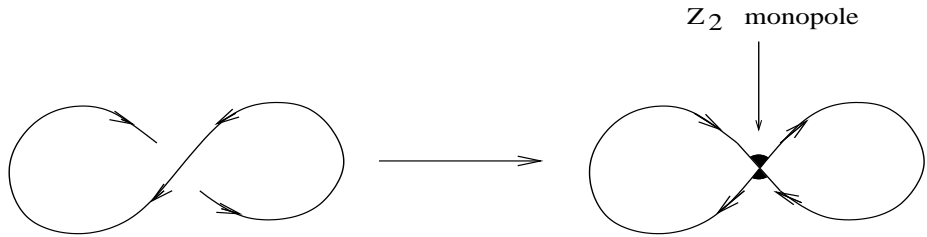}
\end{center}
\caption{The creation of an isolated $Z_2$ monopole by self-intersection 
of a $Z_2$ center vortex. \label{intersect}}      
\end{figure}

Closed center-vortex flux lines can self-intersect and in this way form 
isolated $Z_{\tiny\mbox{N}}$ monopoles which reverse the flux, see Fig.\,\ref{intersect} for the N=2 case.
Let us discuss the case N=2 more explicitly. 
During the phase transition the process leading to self intersection can be performed $L$ times 
on a nonintersecting vortex loop. This generates $L$ isolated $Z_2$ 
monopoles, each of mass $\sim \Lambda_C$, such that the mass $M_L$ 
of the $L$-monopole state is $M_L\sim L\Lambda_C$. Since the number of possible 
$L$-monopole states roughly grows as $L!$ 
(see \cite{BenderWu1976} for a more precise estimate) we conclude that the density of 
states $\rho(E)$ for the particle excitations in the center phase of an SU(2) theory 
is over-exponentially growing in energy $E$:
\eqb
\label{rhoenergydesn}
\rho(E)>\Lambda_C\exp\left[\frac{E}{\Lambda_C}\right]\,.
\eqe
For N$>$2 the situation is similar. So we conclude that there exists a 
highest temperature $T_{\tiny\mbox{H}}\sim \Lambda_C$ in the center 
phase of an SU(N) Yang-Mills theory: a situation which was 
anticipated for any strongly interacting four dimensional field theory a 
long time ago \cite{Hagedorn1965}, see also \cite{HongLiu2004} for a discussion of the Hagedorn transition in
an SU(N) matrix model. Moreover, we conclude that 
the center-magnetic phase transition can by no means be
thermal (the spatial homogeneity of the system is violated during the transition) 
and thus that the thermodynamical pressure may jump across the transition.
A single and a one-time self-intersecting center-vortex loop for N=2 are 
{\sl fermions} \cite{Hofmann20033}. This result is crucial for 
our understanding of the nature of leptons in the present Standard Model of 
Particle Physics. The time scale for the relaxation to one of the minima of $V_C$ 
is roughly given by $|(\Phi^{\tiny\mbox{vac}}_k)|^{-1}$. A quantitative investigation 
of this reheating process would need methods of nonequilibrium field theory, see 
for example \cite{nonequ}. In a thermal approach it would be interesting 
to check in a lattice simulation of the {\sl effective theory} 
in the center phase Eq.\,(\ref{actcenter}) whether a density of states 
as in Eq.\,(\ref{rhoenergydesn}) is indeed seen.

The fact that the pressure and the energy density of the ground state are 
precisely vanishing at the minima $\Phi^{\tiny\mbox{vac}}_k$ and that there 
are no radiative corrections to this situation clearly is of 
cosmological relevance.

\section{Scale matching\label{MCC}}

The scales $\La_E$ and $\La_M$, which determine the magnitudes of the 
adjoint Higgs field $\phi$ and the 
monopole condensates $\varphi_l$ at a given temperature in the electric and the magnetic phase, 
can be 
related by imposing the condition that the thermodynamical pressure $P$ be continuous 
across the thermal electric-magnetic phase transition. Disregarding the mismatch in the number of 
polarizations for some TLM modes in the electric 
phase and some dual gauge bosons in the magnetic phase, we derive 
\eqb
\label{laElaM}
\Lambda_E=(1/4)^{1/3}\Lambda_M\,\ \ \ \ (\mbox{N\ \ even})\,.
\eqe
For N=3 we have $\Lambda_E=(1/2)^{1/3}\Lambda_M$. The match between the 
magnetic and the center phase is less determined for the following reasons: 
(i) For N$>$3 the 
winding numbers of the fields $\Phi_i\,\ \ (i=\mbox{N}/2+1,\cdots,\mbox{N}-1)$ 
close to the center-magnetic phase boundary are dynamically generated during the phase transition 
and so cannot be derived from the boundary conditions for 
the electric phase. For N=2 and N=3 the winding numbers of $\Phi_1$ and 
$(\Phi_1,\Phi_2)$ are unity, and we can set 
$\Lambda_C=\Lambda^\prime_C$ in Eq.\,(\ref{vCright}). 
(ii) An analytical description based on the potential $V_C$ 
of the center-magnetic phase transition at finite N breaks down 
at the phase boundary, see Sec.\,(\ref{CVCM}).

In the electric and in the magnetic phase (recall that the latter confines fundamental test charges) 
we encounter the thermodynamical analogue to dimensional transmutation in perturbation theory: 
Assuming MGSB, a single, fixed mass scale determines the thermodynamics of the SU(N) Yang-Mills 
theory in these phases. At a given temperature 
this mass scale can be experimentally inferred from the mass 
spectrum of the gauge boson fluctuations. 

\section{Pressure, energy density, and entropy density\label{PEEN}}

\subsection{Numerical results}

In this section we present our numerical results for one-loop 
temperature evolutions of thermodynamical potentials through the 
electric and magnetic phase. For the actual computations we consider N=2,3 only. 

\noindent In the electric phase the pressure $P$ divided by $T^4$ is given as
\eqb
\label{pressureEP}
\frac{P}{T^4}=-\frac{(2\pi)^4}{\lambda_E^4}
\left[\frac{2\lambda_E^4}{(2\pi)^6}\left\{2\,(\mbox{N}-1)\bar{P}(0)+
3\sum_{k=1}^{\tiny\mbox{N(N-1)}}\bar{P}(a_k)\right\}+
\frac{\lambda_E}{2}\left(\frac{\mbox{N}}{2}+1\right)\mbox{N}\right]\,,\ \ (\mbox{N\ \ even})\,,
\eqe
where the function $\bar{P}(a)$ and the dimensionless mass parameters $a_k$ are 
defined in Eqs.\,(\ref{P(a)}) and (\ref{dimlessdef}), respectively, and MGSB is assumed. 
In the magnetic phase we have
\eqb
\label{pressureMP}
\frac{P}{T^4}=-\frac{(2\pi)^4}{\lambda_M^4}
\left[\frac{2\lambda_M^4}{(2\pi)^6}\left\{2\,(\frac{\mbox{N}}{2}-1)\bar{P}(0)+
3\sum_{l=1}^{\tiny\mbox{N}/2}\bar{P}(a_l)\right\}+
\frac{\lambda_M}{16}\mbox{N}(\mbox{N}+2)\right]\,,\ \ (\mbox{N\ \ even})\,.
\eqe
The dimensionless mass parameters $a_l$ are defined in 
Eqs.\,(\ref{massspecM}) and Eq.\,(\ref{defM})). 

For N=3 we have in the electric and the magnetic phase, respectively:
\eab
\label{pressure3EPpressure3MP}
\frac{P}{T^4}&=&-\frac{(2\pi)^4}{\lambda_E^4}
\left[\frac{2\lambda_E^4}{(2\pi)^6}\left\{4\bar{P}(0)+
3\left(4\,\bar{P}(a)+2\,\bar{P}(2a)\right)\right\}+
2\lambda_E\right]\,,\ \ (\mbox{N=3})\,,\nonumber\\ 
\frac{P}{T^4}&=&-\frac{(2\pi)^4}{\lambda_M^4}
\left[\frac{12\lambda_M^4}{(2\pi)^6}
\bar{P}(a)+\lambda_M\right]\,,\ \ (\mbox{N=3})\,.
\eae
The mass parameters $a_k$ and $a_l$ evolve with temperature according to the (inverted) 
solutions to Eqs.\,(\ref{eeq}), (\ref{eeq3}), (\ref{evolMeven}), and (\ref{evolM3}).

\begin{figure}
\begin{center}
\leavevmode
\leavevmode
\vspace{4.5cm}
\includegraphics{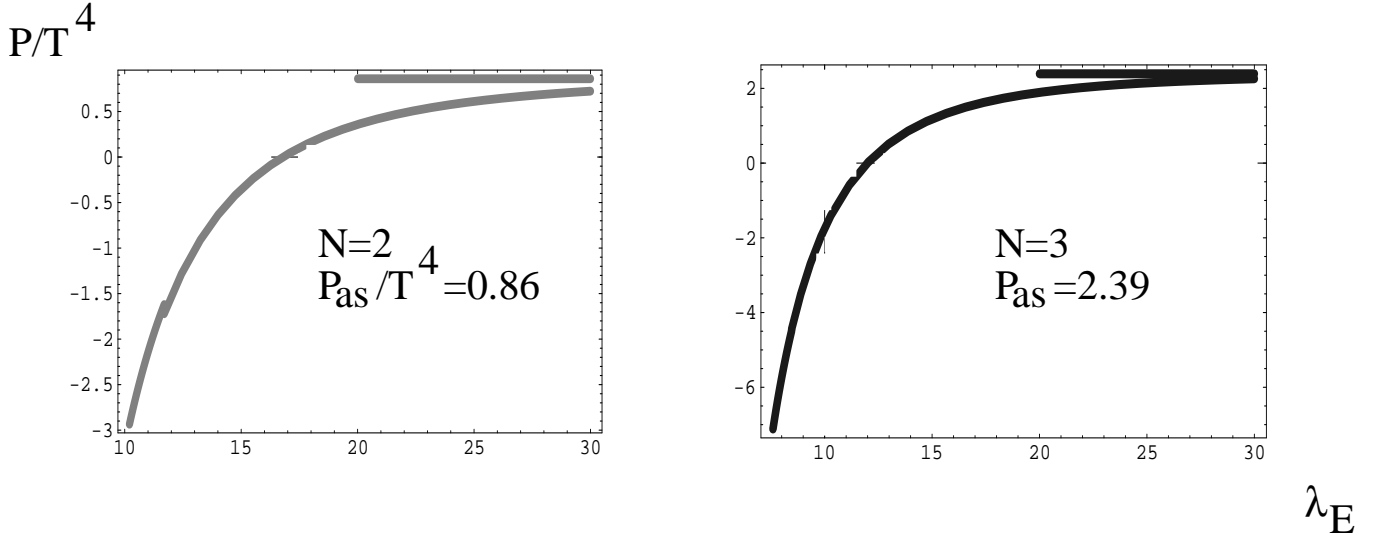}
\end{center}
\caption{$\frac{P}{T^4}$ as a function of temperature for N=2,3. The horizontal lines denote 
the respective asymptotic values. For N=2 we have $\lambda_{E,c}=11.65$ 
and for N=3 $\lambda_{E,c}=8.08$. \label{pressure}}      
\end{figure}

Our results for $\frac{P}{T^4}$ as 
a function of temperature in the electric and 
magnetic phase are shown in Fig.\,\ref{pressure}.
\begin{figure}
\begin{center}
\leavevmode
\leavevmode
\vspace{4.5cm}
\includegraphics{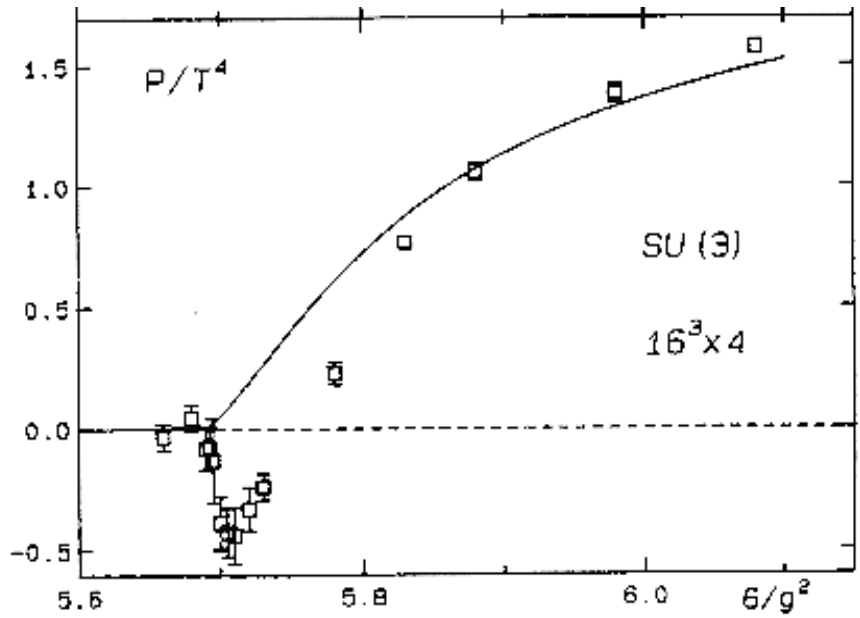}
\end{center}
\caption{$\frac{P}{T^4}$ as a function of temperature for N=3 as obtained on a $16^3\times4$ 
lattice using the differential method with a universal two-loop 
perturbative $\beta$ function \cite{Brown1988,Deng1988} and using the integral 
method (solid line) \cite{EngelsFingberg1990}. The figure is taken 
from \cite{EngelsFingberg1990}.  
\label{pressureLat}}      
\end{figure}
Notice that the pressure is negative 
in the electric phase close to $\lambda_{E,c}$ and 
even more so in the magnetic phase where the ground-state 
strongly dominates the thermodynamics.

\begin{figure}
\begin{center}
\leavevmode
\leavevmode
\vspace{4.5cm}
\includegraphics{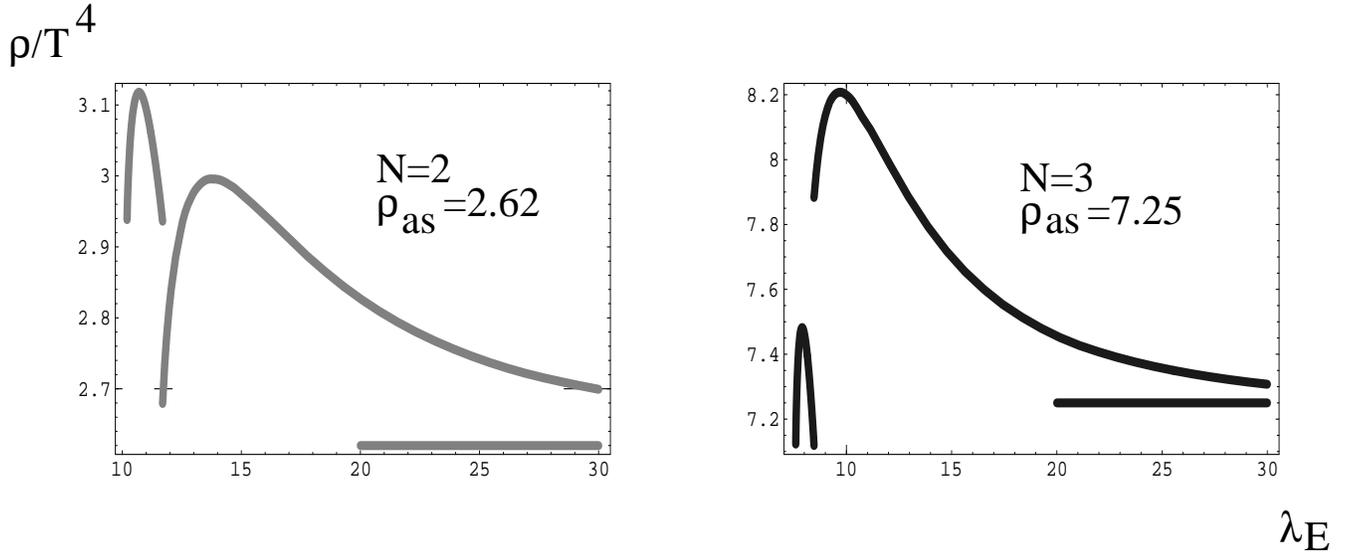}
\end{center}
\caption{$\frac{\rho}{T^4}$ as a function of temperature for N=2,3. The horizontal lines denote 
the respective asymptotic values. \label{rho}}      
\end{figure}

\noindent In the electric phase the energy density $\rho$ divided by $T^4$ is given as
\eqb
\label{rhoEP}
\frac{\rho}{T^4}=\frac{(2\pi)^4}{\lambda_E^4}
\left[\frac{2\lambda_E^4}{(2\pi)^6}\left\{2\,(\mbox{N}-1)\bar{\rho}(0)+
3\sum_{k=1}^{\tiny\mbox{N(N-1)}}\bar{\rho}(a_k)\right\}+
\frac{\lambda_E}{2}\left(\frac{\mbox{N}}{2}+1\right)\mbox{N}\right]\,,\ \ (\mbox{N\ \ even})\,,
\eqe
where the function $\bar{\rho}(a)$ is defined as
\eqb
\label{rhobar}
\bar{\rho}(a)\equiv \int_{0}^{\infty}dx\,x^2 \frac{\sqrt{x^2+a^2}}{\exp(\sqrt{x^2+a^2})-1}\,.
\eqe
In the magnetic phase we have
\eqb
\label{rhoMP}
\frac{\rho}{T^4}=\frac{(2\pi)^4}{\lambda_M^4}
\left[\frac{2\lambda_M^4}{(2\pi)^6}\left\{2\,(\frac{\mbox{N}}{2}-1)\bar{\rho}(0)+
3\sum_{l=1}^{\frac{\tiny\mbox{N}}{2}}\bar{\rho}(a_l)\right\}+
\frac{\lambda_M}{16}\mbox{N}(\mbox{N}+2)\right]\,,\ \ (\mbox{N\ \ even})\,.
\eqe
For N=3 we have in the electric and the magnetic phase, respectively:
\eab
\label{rho3EPrho3MP}
\frac{\rho}{T^4}&=&\frac{(2\pi)^4}{\lambda_E^4}
\left[\frac{2\lambda_E^4}{(2\pi)^6}\left\{4\bar{\rho}(0)+
3\left(4\,\bar{\rho}(a)+2\,\bar{\rho}(2a)\right)\right\}+
2\lambda_E\right]\,,\ \ (\mbox{N=3})\,,\nonumber\\ 
\frac{\rho}{T^4}&=&\frac{(2\pi)^4}{\lambda_M^4}
\left[\frac{12\lambda_M^4}{(2\pi)^6}
\bar{\rho}(a)+\lambda_M\right]\,,\ \ (\mbox{N=3})\,.
\eae
Our results for $\frac{\rho}{T^4}$ as 
a function of temperature in the electric and 
magnetic phase are shown in Fig.\,\ref{rho}. Slight discontinuities at $\lambda_{E,c}$ are seen. 
This is explained by the mismatch in the number of 
polarizations of fluctuating gauge bosons across the electric-magnetic transition - 
an approximation to scale-matching - 
and the fact that continuity in $P$ does not imply continuity in $\rho$. Again, 
the energy density is dominated by the
ground-state contribution in the electric phase close to the electric 
magnetic transition and even more so in the magnetic
phase. 
The entropy density $S$ is defined as the derivative of the 
pressure with respect to temperature:
\eqb
\label{Sdef}
S\equiv\frac{dP}{dT}\,.
\eqe
Using the thermodynamical relation $\rho=T\frac{dP}{dT}-P$, we may write
\eqb
\label{sdef}
\frac{S}{T^3}=\frac{1}{T^4}\left(\rho+P\right)\,.
\eqe
Our results for $S/T^3$ are shown in Fig.\,\ref{ST3}. 
The reasons for the slight discontinuities at $\lambda_{E,c}$ are the same as in the 
case $\frac{\rho}{T^4}$. The entropy density $S$ is a measure for the `mobility' of gauge 
boson excitations. That $S$ is zero at the critical 
temperature $\lambda_{M,c}$ for the center transition 
is explained by the fact that all dual gauge 
bosons acquire an infinite mass there and thus no fluctuating degrees 
of freedom are left in the thermodynamical balance.

\subsection{Comparison with the lattice\label{complat}}

An early lattice measurements of the energy density $\rho$ 
and the interaction measure $\Delta\equiv \rho-3P$ in a pure SU(2) gauge theory 
were reported on in \cite{EnKaSaMo1982}. In that work the critical temperature $T_c$ for 
the deconfinement transition was determined from the critical behavior of the 
Polyakov-loop expectation and the peak position 
of the specific heat using a Wilson action. The function $\Delta(T)$ was extracted by multiplying the 
lattice $\beta$ function with the difference of plaquette expectations at finite and zero temperature. 
This assures that $\Delta$ vanishes as $T\to 0$. What is subtracted at finite $T$ is, however, 
{\sl not} the value $\Delta(T=0)$ since the associated plaquette expectation is 
multiplied with the value of the $\beta$ function at {\sl finite} $T$. Apart from this 
approximation, the use of a perturbative $\beta$ function was assumed for all 
temperatures. 
\begin{figure}
\begin{center}
\leavevmode
\leavevmode
\vspace{5.0cm}
\includegraphics{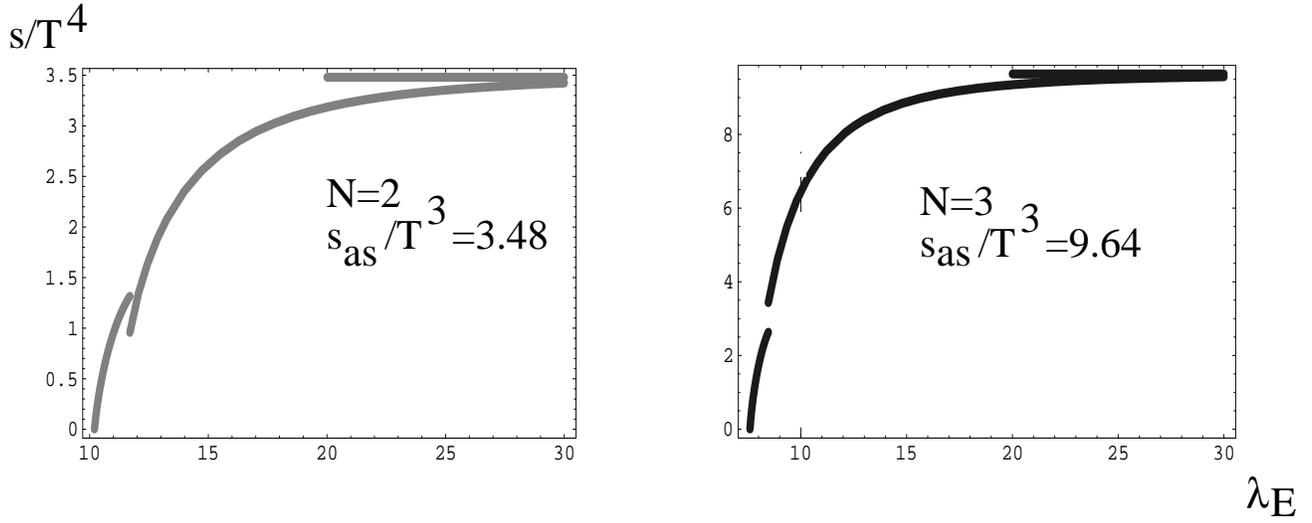}
\end{center}
\caption{$\frac{S}{T^3}$ as a function of temperature for N=2,3. The horizontal lines denote 
the respective asymptotic values. \label{ST3}}      
\end{figure}
The lattice results for $\rho$ differ drastically from 
our results for temperatures close the deconfinement, that is, the electric-magnetic 
transition. We obtain
\eqb
\label{rhooverrhoSBc}
\left.\frac{\rho}{\rho_{SB}}\right|_{T\sim T_{E,c}}\sim 1.49
\eqe
where $\rho_B\equiv\frac{\pi^2}{5}T^4$ denotes the Stefan-Boltzmann limit 
(ideal gas of massless gluons with two polarizations). On 
the lattice, this ratio is measured to be
smaller than unity. At $T\sim 5T_{E,c}$ we obtain 
\eqb
\label{rhooverrhoSB5}
\left.\frac{\rho}{\rho_{SB}}\right|_{T\sim 5\,T_{E,c}}\sim 1.34\,
\eqe
while the lattice measures a ratio of about 0.85. Our asymptotic 
value\footnote{The asymptotic temperature $\lambda_{E,as}=75$ is determined 
by the boundary condition $\lambda_E(0)=1000$ for solving the evolution equations 
(\ref{eeq}) and (\ref{eeq3}).} is 
\eqb
\label{rhooverrhoSBa}
\left.\frac{\rho}{\rho_{SB}}\right|_{T\sim 6.4\,T_{E,c}}\sim 1.33\,.
\eqe
Notice the latent heat released at the electric magnetic transition for N=3 
(Fig.\,\ref{rho}). 

Our result for the pressure $P$ indicates negative values for $T$ 
close to $T_{E,c}$ (see Fig.\, \ref{pressure}) - much in 
contrast to the positive values obtained 
in \cite{EnKaSaMo1982}. At $T\sim 5 T_{E,c}$ we obtain
\eqb
\label{PoverrhoSB5}
\left.\frac{P}{P_{SB}}\right|_{T\sim 5\,T_{E,c}}\sim 1.30\,
\eqe
while the lattice measures ($P$ is extracted 
from the measured values of 
$\Delta$ and $\rho$) a ratio of about 0.88. Our asymptotic 
value for $P$ is 
\eqb
\label{PoverrhoSBa}
\left.\frac{P}{P_{SB}}\right|_{T\sim 6.4\,T_{E,c}}\sim 1.32\,.
\eqe
The asymptotic values of Eqs.\,(\ref{rhooverrhoSBa}) and 
(\ref{PoverrhoSBa}) are very close to the ratio $R$ of the number of polarizations 
for massive TLH modes and massless TLM modes and the 
number of polarizations for massless gluon modes with two polarizations:
\eqb
\label{R}
R=\frac{2\times 3+1\times 2}{3\times 2}=\frac{4}{3}\sim 1.33\,.
\eqe
Indeed, at $\lambda_E\sim 75$ the mass 
parameter $a$, defined in Eq.\,(\ref{massP}), is 
$a\sim 2\pi\frac{5.5}{650}\sim 0.053$ and 
therefore the Boltzmann suppression of TLH modes is small. 
At extremely high temperatures a TLH mode `remembers' its massiveness 
at low temperatures in terms of an extra polarization coming from 
a tiny mass which, however, still solves the infrared problem 
of naive perturbation theory.  

For a comparison with N=3 lattice data we use the results obtained with a Wilson action in 
\cite{Bielfeld1996} on the lattice of the largest 
time extension, $N_\beta=8$. In the vicinity of the transition point $T_{E,c}$ 
the situation for both $\rho$ and $P$ is similar as for N=2: 
drastic differences between the lattice measurements and 
our results occur. Again, $P$ comes out negative in our calculation, 
contradicting the positive values obtained in 
\cite{Bielfeld1996}. It should be remarked at this point that a 
lattice simulation of $P$, which did not rely on the integral method (see below) 
as it was used in \cite{Bielfeld1996}, has seen negative 
pressure for $T$ not far above $T_{c,e}$ \cite{Deng1988}. Moreover, 
the most negative value of $P/T^4\sim -0.5$ obtained 
in \cite{Deng1988} very close to the phase transition coincides 
with our result at the electric-magnetic transition, 
see Figs.\,\ref{pressure} and \ref{pressureLat}.

\noindent At $T=5\,T_{E,c}$ we have
\eqb
\label{rhoN3}
\left.\frac{\rho}{\rho_{SB}}\right|_{T\sim 5\,T_{E,c}}\sim 1.38
\eqe
while the lattice measures a ratio of about 0.93. Our asymptotic 
value for $\rho$ is 
\eqb
\label{rhooverrhoSB3a}
\left.\frac{\rho}{\rho_{SB}}\right|_{T\sim 8.86\,T_{E,c}}\sim 1.37\,.
\eqe
For the pressure $P$ we obtain at $T=5\,T_{E,c}$:
\eqb
\label{PoverrhoSB35}
\left.\frac{P}{P_{SB}}\right|_{T\sim 5\,T_{E,c}}\sim 1.34\,
\eqe
while the lattice measures a ratio of about 0.97. Our asymptotic 
value for $P$ is 
\eqb
\label{PoverrhoSB3a}
\left.\frac{P}{P_{SB}}\right|_{T\sim 8.86\,T_{E,c}}\sim 1.36\,.
\eqe
Both asymptotic values in Eqs.\,(\ref{rhooverrhoSB3a}) and 
(\ref{PoverrhoSB3a}) are very close to the ratio of 
polarizations $R=\frac{11}{8}=1.375$.

According to Fig.\,(\ref{ST3}) the entropy 
density $\frac{S}{T^3}$ vanishes at $T_{M,c}$. In our approach 
this reflects the fact that at $T_{M,c}$ all gauge-field are thermodynamically decoupled 
because of their infinite mass. As a consequence, 
thermodynamics is entirely determined by the ground state. 
This is not observed in the lattice simulations \cite{EngelsKarschScheideler1999} 
for both N=2,3 where a continuous behavior 
of $\frac{S}{T^3}$ across the deconfinement transition at $T_{E,c}$ was obtained.

\begin{figure}
\begin{center}
\leavevmode
\leavevmode
\vspace{5.0cm}
\includegraphics{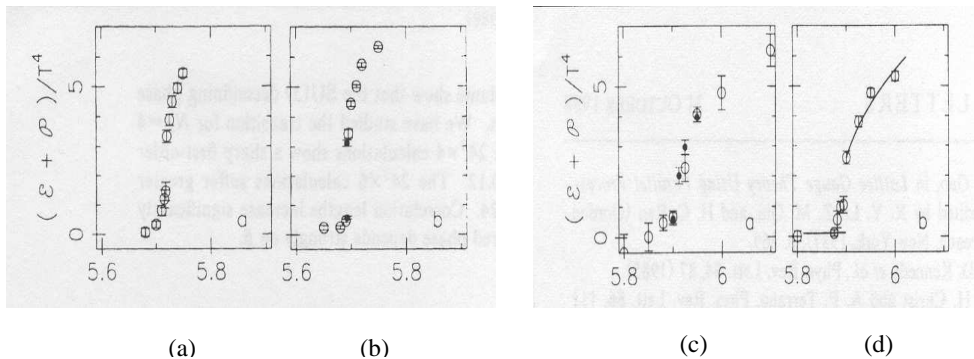}
\end{center}
\caption{$\frac{S}{T^3}$ as a function of $\beta$ 
obtained in SU(3) lattice gauge theory using the differential method and a 
perturbative beta function. The simulations were performed on (a) $16^3\times 4$, (b) 
$24^3\times 4$, (c) $16^3\times 6$ (open circles) and $20^3\times 6$ (closed circles), and 
(d) $24^3\times 6$ lattices. Using the $24^3\times 6$ lattice, the critical 
value of $\beta$ is between 5.8875 and 5.90. \label{ST3lat}}      
\end{figure}

In \cite{Brown1988} a discontinuous behavior of $\frac{S}{T^3}$ was observed for N=3 
using a Wilson action and a perturbative beta function. 
There is an excellent agreement of their 
result with our result for temperatures ranging $T_{E,c}$ down to $T_{M,c}$, 
compare Figs.\,\ref{ST3} and \ref{ST3lat}. 
The almost discontinuous 
behavior in Fig.\,(\ref{ST3}) is due to the large rise 
of the magnetic gauge coupling with decreasing temperature. 
According to Fig.\,(\ref{gevol}) the `duration' $\delta T\equiv\frac{T_{E,c}-T_{M,c}}{T_{E,c}}$ 
of the magnetic phase is only $\delta T\sim 0.1$. In a lattice simulation the resolution of 
such a small temperature interval depends very much on the choice of the beta function. 
In \cite{Brown1988} a universal perturbative beta function was used which may have lead to interprete 
the behavior of $\frac{S}{T^3}$ as discontinuous in 
dependence on temperature. Since $\frac{S}{T^3}$ is a 
quantity which is much less sensitive to the infrared physics 
than, say, $\frac{P}{T^4}$ (the direct 
contribution of the ground state is canceled out in $\frac{S}{T^3}$) the use of (the universal part of) a 
perturbative beta function 
in the lattice simulations \cite{Brown1988} may be justified. We stress at this point 
that lattice simulations based on lattice sizes of 1-3 times the inverse Yang-Mills scale 
are not capable of being sensitive to the infrared effects of the theory which 
have correlation lengths of the order of the gauge couplings $e$ and $g$ times the inverse Yang-Mills 
scale at decoupling (Standard Model physics in the electroweak sector 
suggest that $e_{\tiny\mbox{decoup}}\sim g_{\tiny\mbox{decoup}}\sim 10^6\,$!).

The alert reader would 
object that the thermodynamical relation
\eqb
\label{pdt}
dP=S\,dT\,,
\eqe
which implies that in a homogeneous thermal system the pressure has to be a 
monotonic function of temperature, is violated at the center transition ($T_{c,M}$), see 
Fig.\,\ref{pressure}, where with decreasing temperature 
the pressure quickly rises from a negative to a value close to zero
\footnote{This effect should be the 
more pronounced the higher N is \cite{LuciniTeperWenger2003}.}. What is the resolution of 
this puzzle? There are two anwers. First, on a microscopic level, the 
homogeneity of the system at the point where center
vortices start to condense is badly violated by the generation of 
(intersecting) center-vortex loops through discontinuous and local 
phase changes of the fields $\Phi_k(x)$ 
in Eq.\,(\ref{Wilsonloopx}). The derivation of Eq.\,(\ref{pdt}) 
from the partition function, however, relies on homogeneity. 
Second, assuming homgeneity, one can easily convinces oneself that 
the spectral density $\rho(E)$ in the center phase is 
exponentially increasing with energy $E$\footnote{The number
of center vortex loops with $L$ intersections increases stronger than 
factorially with $L$ and the energy of a vortex state with $L$ intersections 
is $\propto L\Lambda_C$, for the SU(2) case see \cite{BenderWu1976} where a counting of the vacuum
diagrams in a $\lambda\phi^4$ theory is carried out.}:
\eqb
\label{specdens}
\rho(E)\propto\exp[\frac{E}{T_{\tiny\mbox{H}}}]\,.
\eqe
Thus the partition function diverges at $T=T_{\tiny\mbox{H}}\sim T_{c,M}$, 
the homogeneous system would need an infinite amount of energy per volume 
to increase the temperature beyond $T_{\tiny\mbox{H}}$. We conclude that 
homogeneity is violated at $T=T_{\tiny\mbox{H}}$. We conclude that the SU(N) YM dynamics 
indeed predicts a violation of the thermodynamical relation in Eq.\,(\ref{pdt}) 
at $T=T_{\tiny\mbox{H}}\sim T_{c,M}$ \footnote{The author would like to thank 
D. T. Son for initiating this discussion.}.

What are the possible reasons for the qualitative difference between the results obtained 
in \cite{Bielfeld1996,EngelsKarschScheideler1999} and \cite{Brown1988,Deng1988}? To avoid the use 
of derivatives of the bare coupling, which are multiplying the sum of spatial 
and time plaquette averages in the 
differential method for the computation of the pressure, the integral method was introduced in 
\cite{EngelsFingberg1990}. Using a perturbative beta function 
in the differential method, negative values for the pressure for $T$ close to 
$T_c$ and a rapid approach of the $\rho$ and
$P$ to their respective ideal gas limits were obtained in \cite{Deng1988}. 
The rapid approach to the ideal gas limit in $\rho$ is also obtained in the present approach, 
compare Eqs.\,(\ref{rhooverrhoSB5}), 
(\ref{rhooverrhoSBa}) and Eqs.\,(\ref{rhoN3}),(\ref{rhooverrhoSB3a}). 
At the time when the results in 
\cite{Brown1988,Deng1988} appeared a negative pressure was regarded as 
unphysical and attributed to the use of a 
perturbative beta function. The integral method proposed in \cite{EngelsFingberg1990} 
solely generates positive pressure. This method {\sl assumes} the validity of the 
thermodynamical limit in the lattice simulation. Namely, 
for large spatial volume the thermodynamical relation
\eqb
\label{ptd}
P=T\frac{\pd\ln Z}{\pd V}
\eqe
valid for a finite volume $V$ is approximated by 
\eqb
\label{ptdL}
P=T\frac{\ln Z}{V}\,
\eqe
so that the pressure equals minus the free energy density.
In Eqs.\,(\ref{ptd}) and (\ref{ptdL}) $Z$ denotes the partition function. 
Based on Eq.\,(\ref{ptdL}) the derivative of the 
pressure with respect to the bare coupling $\beta=\frac{2\mbox{N}}{\bar{g}^2}$ 
can be expressed as an expectation over the sum of spatial and time plaquettes 
without the beta-function prefactor. Thus the pressure can be obtained 
by an integral over $\beta$ up to an unknown integration constant. The latter 
is chosen in such a way that the pressure vanishes at a temperature well 
below $T_c$. Instead of only integrating over minus the sum of spatial and 
time plaquette expectations twice 
the plaquette expectation at $T=0$ 
was added to the integrand in \cite{Bielfeld1996,EngelsKarschScheideler1999} 
to assure that the pressure vanishes at $T=0$. We would like to stress that 
this prescription does not follow from relation (\ref{ptdL}). 
Thus it seems to be natural that integral and differential methods 
generate qualitatively different results. The results for $P(T)$ obtained by the 
integral method show a rather large dependence on the cutoff and the time extent $N_\tau$ of the lattice 
\cite{Bielfeld1996}. We believe that this 
reflects the considerable deviation from the assumed 
thermodynamical limit for so-far available lattice sizes. The problem 
was addressed in \cite{EngelsKarschScheideler1999} where a 
correction factor $r$ was introduced to relate $P$ 
obtained with the integral method to $P$ obtained 
with the differential method. For a given $N_\tau$ 
the factor $r$ was determined from the pressure ratio at $\bar{g}=0$. 
Subsequently, this value of $r$ was used at finite coupling $\bar{g}$ to extract the spatial anisotropy 
coefficient $c_\sigma$ (essentially the beta function) 
by demanding the equality of the pressure obtained 
with the integral and the differential method. 
In doing so, twice the plaquette expectation at $T=0$ was, again, added to minus 
the sum of spatial and time plaquette expectations in the expression 
for the pressure using the differential method. It may be questioned 
whether a simple multiplicative correction $r$ does correctly 
account for finite-size and cutoff effects and, if yes, 
whether it is justified to determine $r$ in the limit 
of noninteracting gluons\footnote{The $c_\sigma$-values obtained in this way 
do not coincide with those obtained in \cite{Klassen1998}.}.  

To summarize, we observe a qualitative disagreement 
between the lattice results \cite{EnKaSaMo1982,Bielfeld1996} for $\rho$ and $P$, using the integral method, 
and the results of the present approach 
for temperatures close to $T_{E,c}$. At $T\sim 5\,T_{E,c}$ there is better agreement. 
Although for N=3 our negative values for $P$ in the vicinity of $T_{E,c}$ 
disagree with those obtained with the differential method 
(we obtain a modulus at $T_{E,c}$ which {\sl coincides} with that of the minimal value obtained by using a 
perturbative beta-function differential method \cite{Deng1988}). On the other hand, 
the entropy density obtained in \cite{Brown1988} with the differential method 
agrees well with our results. This is expected since the entropy 
is an infrared insensitive quantity.

\section{Conclusions and Outlook\label{CO}}

We have developed a nonperturbative approach to SU(N) Yang-Mills thermodynamics 
which is based on the (self-consistent) assumption 
that the theory `condenses' SU(2) embedded, BPS saturated 
topological fluctuations of trivial holonomy 
at an asymptotically high temperature. In \cite{HerbstHofmann2004} we have shown 
for the SU(2) case the redundancy of this assumption. 
The concept for the construction of an effective theory based 
on the above assumption is similar to that applied to the construction of 
the macroscopic field theory for superconductivity \cite{GinzburgLandau1950,Abrikosov1957}. 
We stress that the effects on nontrivial topology die 
off in a power-like fashion in the effective theory 
(as a function of temperature). Thus the perturbatively 
derived suppression of topologically nontrivial field 
configurations does take place in our effective theory as well.   

We have constructed a (uniquely determined) potential for the thermodynamics of an 
energy- and pressure-free adjoint Higgs background $\phi$ 
which macroscopically describes the collective effects due to noninteracting, 
trivial-holonomy calorons in the only deconfining phase of the 
theory which we call electric phase for obvious reasons. As a consequence, 
the ground state of the system is described by a 
BPS saturated solution to the field equation of the scalar 
sector and an associated pure-gauge configuration solving the macroscopic 
equation of motion for trivial-topology gauge-field fluctuations. 
The latter has macroscopic, nontrivial holonomy and 
thus describes the presence of isolated magnetic monopoles being generated 
from decaying nontrivial-holonomy calorons as a result of microscopic 
interactions between trivial-holonomy calorons. These interactions are mediated 
by the trivial-topology sector of the theory. The modulus of $\phi$ falls off as 
$\propto T^{-1/2}$ and the ground-state pressure is only linear in $T$ so that 
nonperturbative effects (apart from extra polarizations for excitations) 
are {\sl irrelevant} at asymptotically high temperatures.

Some of the topologically trivial gauge-field fluctuations 
are massive on tree-level due to the adjoint Higgs mechanism, and 
the underlying SU(N) gauge symmetry is
spontaneously broken accordingly. An evolution equation, describing the temperature 
dependence of the effective gauge coupling constant $e$, was obtained from the requirement of thermodynamical
self-consistency of the effective theory. The two fixed points of this 
evolution were identified. These fixed points 
predict the existence of a highest and a lowest
attainable temperature in the electric phase.  
Based on the evolution $e(T)$ a physical argument was given why caloron 
`condensation' in a grandly unifying theory must take place at a temperature close 
to the cutoff scale for validity of a local, four dimensional 
field-theory description. We have investigated some aspects of the loop expansion of 
thermodynamical potentials in the effective electric theory. Our conclusion 
is that the present nonperturbative approach resolves the infrared problems associated with the usual, 
perturbative loop expansion. We investigate the two-loop corrections 
to the pressure for N=2 in \cite{HerbstHofmannRohrer2004}. Two-loop contributions to the 
pressure are corrections 
to the one-loop contributions which 
range within the $0.1$\% level. Thus as far a bulk thermodynamical quantities are concerned 
the gauge-boson fluctuations in the electric fields are 
practically {\sl noninteracting} at the expense of 
some of them being thermal quasiparticles.  

The downward temperature evolution of the effective gauge coupling $e$ has an attractor and 
thus the IR-UV decoupling observed in the underlying theory due to 
renormalizability is recovered in the effective theory. The temperature evolution $e(T)$ 
predicts a transition, driven by the 
condensation of magnetic monopoles, to a phase with less gauge symmetry 
(magnetic phase, confining heavy fundamental test charges). This transition 
is the deconfinement-confinement transition identified in lattice simulations. Due to the 
typical correlation length in the monopole condensate being of the order of $e_{\tiny\mbox{decoup}}\times
\Lambda^{-1}$, where $\Lambda$ denotes the Yang-Mills scale and 
$e_{\tiny\mbox{decoup}}\gg 1$, it is very hard 
(in practice impossible) for lattice simulations performed within the magnetic phase to 
predict thermodynamical quantities that are 
infrared sensitive. 

The macroscopic ground-state structure of the 
magnetic phase is determined in analogy to the electric phase, and, 
as a consequence, some of the residual (dual) 
gauge-field fluctuations acquire mass by the Abelian Higgs mechanism. The evolution of the 
magnetic gauge coupling constant $g$, again being a consequence of 
thermodynamical self-consistency, predicts 
the existence of a highest and a lowest attainable temperature 
also for the magnetic phase. At the lowest temperature a transition to the center phase, 
where center-vortex loops are condensed into the ground state, takes place. Assuming maximal gauge-symmetry 
breaking in the electric phase, {\sl all} gauge-field modes decouple thermodynamically at the 
transition point, and at this point the equation of state is maximally negative, $P=-\rho$. 

The magnetic-center transiton is of 
the Hagedorn type and thus nonthermal. A remarkable feature of the center phase is that 
ground-state pressure and energy density are {\sl precisely} zero after a period of rapid 
reheating has taken place. This follows from the shape of the effective potential $V_C$ for the 
vortex-condensate fields $\Phi_k$ which implements the spontaneous 
breakdown of the local center symmetry $Z_{\mbox{N,mag}}$. 

In the limit N$\to\infty$ analytical access to the center-vortex dynamics is granted by the 
thermodynamical and quantum mechanical stability of the 
classical solutions to the BPS equations for the center-vortex fields. 
For finite N this stability is lost 
due to the presence of tachyonic modes if the center-vortex condensates $\Phi_k$ 
are away from the minima of their potential. Once the 
minima are reached, no fluctuations in $\Phi_k$ exist for any $\mbox{N}\ge 2$, 
and thus the result of a vanishing 
ground-state energy and pressure, which is based on the 
classical analysis, is strictly reliable. 

For N=2,3 the present approach predicts a Stefan-Boltzmann 
like behavior (with additional polarizations) of the thermodynamical potentials pressure, energy density, 
and entropy density at temperatures of about $10\,T_{E,c}$. 
Throughout the magnetic phase we predict a {\sl negative} 
equation of state which is in contradiction to lattice results for N=2,3 
using the integral method. For the (infrared insensitive) entropy density at N=3 
we obtain excellent agreement with lattice data generated with the 
differential method and a perturbative beta function.

There are many applications of the approach presented in this paper. In \cite{Hofmann20032} we have 
proposed that a strongly interacting gauge theory underlying the leptonic sector of the Standard Model 
should be based on the following gauge group   
\eqb
\label{SIGUE}
SU(2)_{\tiny\mbox{CMB}}\times SU(2)_{e}\times SU(2)_{\mu}\times SU(2)_{\tau}\cdots\,.
\eqe
In addition, mixing angles for the gauge bosons of one factor with 
those of the other factors at temperatures much higher than the Yang-Mills 
scale of the last factor should be supplied. In Eq.\,(\ref{SIGUE}) the Yang-Mills scale of the first factor is 
roughly given (but can be precisely determined 
\cite{Hofmann20032}) by the temperature $T_{\tiny\mbox{CMB}}$ 
of the cosmic microwave background $\sim 10^{-4}\,\mbox{eV}$. The 
Yang-Mills scales of the other factors are roughly given by the mass of the corresponding charged leptons. 
While the CMB-scale theory is in its magnetic phase 
very close to the magnetic-electric transition (only there is the dual gauge boson, the photon, massless) 
the other theories are in their center phases generating the stable leptons as solitons (neutrino ... single
center-vortex loop, charged lepton ... center-vortex loop with one self-intersection.). 
The two polarization states of these solitons 
arise as a consequence of the spontaneously broken, local $Z_{2,\tiny\mbox{mag}}$ symmetry. The photon, 
which is the fluctuating degree of freedom in the magnetic phase of 
the CMB-scale theory, would couple to the electric charges and the 
magnetic moments of these leptons because the gauge dynamics subject to Eq.\,(\ref{SIGUE}) was 
embedded into the gauge dynamics subject to a higher gauge group SU(N) at temperatures 
larger than the mass of the heaviest charged lepton. The same reasoning goes through 
for the coupling of the photon to quarks if we allow for SU(3) factors in Eq.\,(\ref{SIGUE}). 
Given the mixing angles it is possible to compute the 
electric charge of each soliton from the plateau value of the gauge-coupling evolution in the 
electric phase in each factor theory. Since the CMB-scale theory is in its magnetic phase at the magnetic-electric transition 
(a very small magnetic gauge coupling constant $g$) the ground state of the universe at present
is slightly superconducting: a possible explanation for 
intergalactic magnetic fields. The ground-state energy density due to 
the CMB-scale theory is about 1\% of the gravitationally observed value \cite{WMAP2003,Hofmann20032}, no 
contribution. Recall, that no ground-state energy density of pressure is generated by 
SU(N) Yang-Mills theories in their center phases. We believe that the missing part can be 
linked to a CP violating, additional 
term in the Yang-Mills action \cite{Wilczek2004}, see also \cite{AlexanderMoffat2004} 
for another intelligent (albeit incomplete) way of addressing 
the cosmological constant `problem'. That the temperature of the 
Universe is stabilized at $T=T_{\tiny\mbox{CMB}}$ follows from the behavior of the energy density $\rho$ at the
electric-magnetic transition, see Fig.\,\ref{rho}. This is an extraordinary useful fact 
since it allows for our mere existence \footnote{Biology would be unthinkable 
with a massive photon.}. The decoupled $W^\pm$ bosons of 
SU(2)$_{\tiny\mbox{CMB}}$ are stable since they cannot decay into the matter that would 
arise if SU(2)$_{\tiny\mbox{CMB}}$ would go into its center phase (a disaster for 
entropy generating individuals). They are an extraordinarily viable candidate for clustering dark matter (the stuff 
responsible for the anomalous rotation curves of galaxies).   

The $Z_0$ and the $W_{\pm}$ 
bosons of the Standard Model would be interpreted as the thermodynamically 
decoupled dual and TLH gauge-boson 
fluctuations of the $SU(2)_{e}$ factor in Eq.\,(\ref{SIGUE}). One would expect to 
see heavy gauge bosons $Z^\prime_0$ and $W^\prime_{\pm}$, arising 
from the factor $SU(2)_{\mu}$ in Eq.\,(\ref{SIGUE}), at about $\frac{m_\mu}{m_e}\sim 200$ 
times the mass of the $Z_0$ and the $W_{\pm}$ bosons, respectively. There is no {\sl fluctuating} 
Higgs-field in this (stepwise) description of electroweak symmetry breaking. Obviously, 
a lot of the extraordinarily precisely checked features of 
the Standard Model can not be derived at the present stage of development, 
for example the absence of flavor-changing neutral currents 
on tree-level. Moreover, it is 
questionable that scattering processes like $e^+e^-\to e^+e^-$, say, at the $Z_0$ resonance 
can be well understood in a thermodynamical framework. 
It is, however, conspicuous that the total cross section
of this process deviates substantially from the QED prediction for $\sqrt{s}\sim m_e, m_\mu$ 
\cite{QEDtests}. The apparent structurelessness of a charged lepton as measured 
for momentum transfers away from its mass 
may be understood by the over-exponentially rising density of (instable) 
states in the center phase of the SU(2)$_e$, SU(2)$_\mu$, ... Yang-Mills theories in Eq.\,(\ref{SIGUE}).   

To make contact with ultra-relativistic heavy ion collision our approach to pure SU(3) Yang-Mills 
theory would have to be extended to include fundamentally charged fermions. The assumption of 
rapid thermalization, which underlies the (very successful) hydrodynamical approach to 
the early stages of an ultrarelativistic heavy-ion collision 
\cite{ShuryakTeaney2001}, would be explained 
by rigid correlations in the magnetic phase (magnetic monopole condensate) 
or the electric phase (caloron `condensate') if the ground-state structure 
of a pure SU(3) gauge theory would not 
significantly be altered by the presence of quarks. 

To describe thermalized Quantum Chromodynamics 
one would introduce quarks as fundamental fermionic fields 
$\psi_i$ where $i$ is a flavor index and the color index is implicit. 
Quarks may couple to the caloron `condensate' $\phi$ 
in the electric phase via Yukawa terms
\eqb
\label{Yuk}
\kappa\sum_i \bar{\psi}_i\phi\psi_i
\eqe
and to topologically trivial gauge-field fluctuation $\delta a_\rho$ 
via the usual covariant derivative. Due to 
Eq.\,(\ref{Yuk}) quarks acquire mass dynamically. The ground-state structure in this approach 
would be the same as the one of a pure SU(3) 
Yang-Mills theory. In addition to gauge-field loops there 
would be quark loops in the expansion of the thermodynamical potentials. 
Thermodynamical self-consistency would imply a system of two coupled 
evolution equations whose solutions would predict $e(T)$ and $\kappa(T)$ and can be used to compute 
the temperature dependence of the thermodynamical potentials. Presumably, the dynamical quark 
masses would become large close to the transition to the magnetic phase. 
This would mean that chiral
symmetry is dynamically broken. As a consequence, quarks would decouple 
thermodynamically at the phase boundary and be replaced by (relatively strongly interacting) 
chiral Goldstone modes in the magnetic phase. The latter could overcompensate 
the negative pressure generated by the ground state of 
condensed magnetic monopoles. An extension of this approach to the case 
of finite quark chemical potential $\mu_q$ should be relatively straight forward.  
A more fundamental approach, were quarks arise as topological solitons in the 
center phases of various SU(3) Yang-Mills 
theories, would be much more difficult.    

Due to the dominance of the ground state in the magnetic phase a 
gauge theory for cosmological inflation based on SU(N) Yang-Mills 
thermodynamics would be a natural application. This would 
be a gauge-theory realization of warm inflation \cite{Berera}. 
Density perturbations generated in the magnetic phase would be dominated by 
thermal fluctuations. Along these lines an attempt to construct 
a gauge theory for warm inflation was made in \cite{HofmannKeil2002}.  

If the entire matter of the Universe would be described in terms of a `mother' SU(N) Yang-Mills 
theory, which break into SU(K) factors at $M_P$, 
then the energy-momentum tensor of the ground state would vanish 
identically for all those `daughter' theories that are in their center 
phases now. {\sl No cosmological constant is generated by the latter.}

\section*{Acknowledgments}
The author would like to thank B. Garbrecht, H. Gies, Th. Konstandin, T. Prokopec, H. Rothe, K. Rothe, 
M. Schmidt, I.-O. Stamatescu, and W. Wetzel for very helpful, continuing discussions. 
Important support for numerical calculations was provided by J. Rohrer and is 
thankfully acknowledged. 
Very useful discussions with P. van Baal, E. Gubankova, J. Moffat, J. Polonyi, D. Rischke, and F. Wilczek and 
illuminating conversations with D. B\"odeker, R. Brandenberger, G. Dunne, Ph. de Forcrand, A. Guth, 
F. Karsch, A. Kovner, M. Laine, H. Liu, C. Nunez, R. D. Pisarski, K. Rajagopal, K. Redlich, D. T. Son, 
A. Vainshtein, J. Verbaarschot, and F. Wilczek are gratefully acknowledged. 
The warm hospitality of the Center for 
Theoretical Physics at M.I.T, where part of this research was carried out 
(sponsored by Deutsche Forschungsgemeinschaft), is thankfully acknowledged.\\  
This paper is dedicated to my family.

\bibliographystyle{prsty}

\end{document}